\documentclass[aip,jcp,reprint,longbibliography,amsmath,amssymb,floatfix]{revtex4-2}

\usepackage{graphicx}
\usepackage{siunitx}
\usepackage{bm}
\usepackage{tikz}
\usepackage{tabularx}
\usepackage{mathtools}

\usepackage[acronym]{glossaries}

\usepackage[colorlinks=true, allcolors=blue]{hyperref}

\newcommand{\operator}[1]{\bm{\mathrm{#1}}}
\renewcommand{\tensor}[2]{\operator{#1}^{(#2)}}
\newcommand{\tensortop}[3]{\operator{#1}^{(#2)#3}}
\newcommand{\telement}[3]{\mathrm{#1}^{(#2)}_{#3}}
\newcommand{\foperator}[1]{\bm{\mathcal{#1}}}
\newcommand{\ftensor}[2]{\foperator{#1}^{(#2)}}
\newcommand{\ftelement}[3]{\mathcal{#1}^{(#2)}_{#3}}
\renewcommand{\vector}[1]{\bm{\mathrm{#1}}}
\newcommand{\velement}[2]{\mathrm{#1}_{#2}}
\renewcommand{\dot}[2]{({#1} \cdot {#2})}
\newcommand{\bigdot}[2]{\big({#1} \cdot {#2}\big)}
\newcommand{\threedot}[3]{\big({#1} \cdot {#2} \cdot {#3}\big)}
\newcommand{\cross}[3]{\{{#1} \!\times {#2}\}^{(#3)}}
\newcommand{\bigcross}[3]{\big\{{#1} \!\times {#2}\big\}^{(#3)}}
\newcommand{\elcross}[4]{\{{#1} \!\times {#2}\}^{(#3)}_{#4}}
\newcommand{\bigelcross}[4]{\big\{{#1} \!\times {#2}\big\}^{(#3)}_{#4}}
\newcommand{\iiij}[6]{\begin{pmatrix}#1&#2&#3\\#4&#5&#6\end{pmatrix}}
\newcommand{\vij}[6]{\begin{Bmatrix}#1&#2&#3\\#4&#5&#6\end{Bmatrix}}
\newcommand{\ixj}[9]{\begin{Bmatrix}#1&#2&#3\\#4&#5&#6\\#7&#8&#9\end{Bmatrix}}
\newcommand{\cmat}[1]{\langle l||\tensor{c}{#1}||l\rangle}
\newcommand{\ident}{\mathbb{I}}
\newcommand{\itbra}[1]{\langle\mathit{#1}|}
\newcommand{\itket}[1]{|\mathit{#1}\rangle}
\newcommand{\Ndiff}{N_{\mathit{diff}}}
\newcommand{\Adiff}{A_{\mathit{diff}}}
\newcommand{\Asame}{A_{\mathit{same}}}
\newcommand{\Nswap}{N_{\mathit{swap}}}
\newcommand{\Neval}{N_{\mathit{eval}}}
\newcommand{\symiiij}{\mbox{3-j}}
\newcommand{\symvij}{\mbox{6-j}}
\newcommand{\symixj}{\mbox{9-j}}
\newcommand{\ameli}{\texttt{AMELI}{}}

\newacronym{doi}{DOI}{Digital Object Identifier}
\newacronym{api}{API}{Application Programming Interface}
\newacronym{coo}{COO}{Coordinate Format}
\newacronym{cfi}{CFI}{Crystal Field Intensity}
\newacronym{cfp}{CFP}{Coefficient of Fractional Parentage}

\makeatletter
\def\@email#1#2{%
 \endgroup
 \patchcmd{\titleblock@produce}
  {\frontmatter@RRAPformat}
  {\frontmatter@RRAPformat{\produce@RRAP{*#1\href{mailto:#2}{#2}}}\frontmatter@RRAPformat}
  {}{}
}%
\makeatother

\newcommand{\drawhistogram}[1]{
    \begin{tikzpicture}[line width=\lwidth]
        \def\data{#1}
        \pgfmathsetmacro{\n}{dim(\data)}           
        \pgfmathsetmacro{\lastidx}{int(\n-1)}      
        \pgfmathsetmacro{\binwidth}{\imgwidth/\n}  

        \draw[gray!40] (0,0) rectangle (\imgwidth, \imgheight);

        \ifnum\lastidx>0
            \foreach \i in {1,...,\lastidx} {
                \draw[gray!40] ({\i * \binwidth}, 0) -- ({\i * \binwidth}, \imgheight);
            }
        \fi

        \fill[red!20] (0,0) -- (0, {\data[0] * \imgheight}) 
            \foreach \x in {1,...,\lastidx} {
                -- ({\x * \binwidth}, {\data[\x-1] * \imgheight}) 
                -- ({\x * \binwidth}, {\data[\x] * \imgheight})
            }
            -- (\imgwidth, {\data[\lastidx] * \imgheight}) -- (\imgwidth, 0) -- cycle;
            
        \draw[red, line width={1.5 * \lwidth}] 
            (0, {\data[0] * \imgheight}) 
            \foreach \x in {1,...,\lastidx} {
                -- ({\x * \binwidth}, {\data[\x-1] * \imgheight}) 
                -- ({\x * \binwidth}, {\data[\x] * \imgheight})   
            }
            -- (\imgwidth, {\data[\lastidx] * \imgheight}); 
    \end{tikzpicture}
}

\begin{document}

\preprint{AMELI/JCP/1.1}

\title[]{AMELI: Angular Matrix Elements of Lanthanide Ions}

\author{Reinhard Caspary}
\email[]{reinhard.caspary@phoenixd.uni-hannover.de}
\affiliation{Cluster of Excellence PhoenixD, Leibniz Universität Hannover, Germany}

\date{\today}
\begin{abstract}
Matrix elements of spherical tensor operators are fundamental to analyzing lanthanide spectra in both amorphous and crystalline host materials.
This work presents a comprehensive framework for calculating angular matrix elements using a Slater determinant basis and their subsequent transformation to the traditional $LS$-coupling scheme using the classification introduced by Racah~\cite{Racah.1942,Racah.1942b,Racah.1943,Racah.1949}.
While computationally demanding, this direct product-state approach is more universally applicable than conventional methods and remains well within modern desktop computing capabilities.
We provide a concise set of general rules to calculate angular matrix elements for virtually any spherical tensor operator within an $f^N$ configuration.
Because these matrices are mathematical constants independent of the host environment, they need only be calculated once.
A comprehensive set of calculated matrix elements for unit and angular momentum operators, alongside perturbation Hamiltonians, is made available in the open-access repository \ameli.
By utilizing exact arithmetic, \ameli\ eliminates the numerical artifacts and rounding errors inherent to conventional floating-point representations.
This takes full advantage of the selection rules and symmetry properties of each operator, resulting in a very compact data format due to high sparsity of the matrices and the small number of unique non-zero elements.
While the evaluation of final physical observables requires subsequent numerical diagonalization, this foundational repository is intended to replace legacy tables currently used for semi-empirical calculations.
Extensive quantitative comparisons to classic tables from Judd and Carnall are presented, and application examples are demonstrated using the open-source Python reference implementation \texttt{YALIP}.
\end{abstract}
\maketitle 

\section{Introduction}
\label{sec.introduction}

It is well known that solutions and solid materials containing lanthanide ions (also called rare earths) exhibit a series of sharp absorption and emission lines spanning from the infrared to the ultraviolet regions.
Moreover, the spectra show a minor dependence on the host material only.
These properties make lanthanides particularly interesting for laser applications.
A similar behavior is observed in actinide ions, and everything discussed here applies to them as well.
However, because of their radioactivity, the spectroscopic properties of most actinides receive significantly less attention.
Therefore, they will not be mentioned explicitly further in this discussion.

The line spectra of the lanthanides are associated with transitions within their partially filled electronic $4f$ shell.
This shell has a radial distribution that spatially remains well within that of the fully occupied $5s$ and $5p$ shells, which are at lower energy.
Consequently, the surrounding host material and its chemical bonds to the lanthanide ion alter the wave functions of the $4f$ shell only slightly.
Therefore, lanthanide ions behave very much like free ions even when they are integrated into matter.
As a result, the energy-level spectrum, and thus the positions of the absorption and emission lines, are remarkably similar for lanthanides in different materials.

It is important to note that the same argument does not apply to the relative intensities of the absorption and emission lines of the lanthanides.
This is because radiative transitions within $4f^N$ configurations are predominantly of electric dipole nature.
Such transitions require the initial and final states to have different parities and are, therefore, forbidden within a pure $4f^N$ configuration.
The observed electric dipole transitions occur only because of the weak interaction with excited configurations of opposite parity induced by the surrounding host material.
Consequently, these transitions of the lanthanides are very weak and depend on the material.
Indeed, they are so weak that their magnitude is comparable with that of magnetic dipole transitions, which are not forbidden by parity.
However, the selection rules for magnetic dipole transitions permit only a limited number of these transitions.

\subsection{Computational Approaches}

The mathematical description of lanthanide electronic structures is historically divided between the rigorous group-theoretical treatment of the free ion (angular matrices) and the phenomenological modeling of the ligand environment (radial integrals).

The group-theoretical foundations of this work rest upon the seminal tensor algebra of Racah~\cite{Racah.1942,Racah.1942b,Racah.1943,Racah.1949}, which was subsequently refined and specialized for the lanthanide $f^N$ configuration by Judd~\cite{Judd.1963} and expanded for crystal field interactions by Wybourne~\cite{Wybourne.1965}.
The comprehensive categorization of the perturbation Hamiltonians and their application to complex $f^N$ spectra reached its definitive form in the work of Goldschmidt~\cite{Goldschmidt.1978}.

This framework allows for a highly granular description of electronic interactions, typically categorized into several types of perturbation Hamiltonians.
While basic models often limit themselves to first-order Coulomb ($\operator{H}_1$) and spin-orbit ($\operator{H}_2$) interactions~\cite{Hehlen.2013}, or include two-electron operators of the second-order Coulomb interaction ($\operator{H}_3$)~\cite{Brik.2019}, a complete semi-empirical description requires the inclusion of more effects.
A much more comprehensive treatment, detailed by Goldschmidt~\cite{Goldschmidt.1978}, incorporates the three-electron operators of the second-order Coulomb interaction ($\operator{H}_4$), spin-spin and spin-other-orbit first-order interactions ($\operator{H}_5$), and electrostatic spin-orbit second-order effects ($\operator{H}_6$).
This set of six types of perturbation Hamiltonians has since then been accepted as reference.

Two implementations of this full set marked the time at which even personal computer systems had reached the processing power required for brute-force approaches in the product space based on Slater determinants: The models of Edvardsson and Åberg in 2001~\cite{Edvardsson.2001} and the one from the doctoral dissertation~\cite{Caspary.2002,Caspary.2005} of the author of this work in 2002.
However, the former implementation could not connect to the common $LS$-state nomenclature in the field of lanthanide spectra while the later provided a transformation matrix from product states to $LS$-coupling.

Peijzel et al.\ also used all six types of perturbation Hamiltonians~\cite{Peijzel.2005}, but their publication lacks details about the underlying implementation.
Despite the theoretical clarity provided by Goldschmidt~\cite{Goldschmidt.1978} and several full-set implementations, the operators $\operator{H}_4-\operator{H}_6$ are still often perceived as ``very complicated'' even in recent literature~\cite{Brik.2019}.

\subsection{Applications}

There exists a large scientific community of experimentalists dedicated to investigating the optical properties of lanthanide ions incorporated in a multitude of different materials.
These experimentalists are experts in the recording of absorption, excitation, and emission spectra of their materials, but they face a significant challenge regarding the evaluation of their spectra.
Application models, such as laser simulations, require absolute values for transition strengths, whereas emission spectroscopy typically yields only unscaled line shapes.
The theory of the calculation of all the radiative excited state absorption or emission transitions relevant to their applications from wave functions and dipole elements is available~\cite{Reisfeld.1975,Walsh.2006}, but the respective quantum-mechanical calculations pose a significant obstacle.

Because of the mathematical difficulty in generating the respective angular matrices from scratch, the spectroscopic community to a large degree still depends on legacy resources to skip the direct calculation of energy levels.
For identifying energy positions, databases like the one provided by Martin et al.~\cite{Martin.1978} remains the standard.
For transition intensities, researchers predominantly use the squared reduced matrix elements tabulated by Carnall~\cite{Carnall.1968,Carnall.1968b,Carnall.1968c,Carnall.1968d, Carnall.1978} for aquo lanthanides and $\mathrm{LaF_3}$ as host material.
These tables effectively act as a black box, allowing researchers to apply the Judd-Ofelt theory~\cite{Judd.1962,Ofelt.1962,Walsh.2006} without engaging with the underlying angular momentum algebra of tensor operators.

The local environment of the lanthanide ion in matter is handled by diverse computational strategies as weak perturbation.
Semi-empirical approaches like the INDO/S method~\cite{Kotzian.1995} or overlap-based models (SAAO)~\cite{Sa.2000, Malta.2002} provide a bridge between molecular orbitals and atomic states.
At the most rigorous end of the spectrum, the multireference ab initio methods used by Freidzon et al.~\cite{Freidzon.2018} allow for the calculation of intermediate states even in low-symmetry environments without empirical scaling, although at a significant computational cost.

The practical application of these energy-level calculations becomes most apparent in the study of radiative transitions.
Since the 40th anniversary of the Judd-Ofelt theory \cite{Judd.2003,Ofelt.2003,Smentek.2003,Naguleswaran.2003}, the focus has shifted toward refining intensity parameters.
Tools like \verb"LUMPAC"~\cite{Dutra.2013,Dutra.2014} and \verb"JOYspectra"~\cite{Moura.2021} incorporate both static~\cite{Reid.1983} and dynamic coupling~\cite{Reid.1984, Reid.2005} to model energy transfer and hypersensitive transitions.
Meanwhile, automated fitting utilities such as \verb"JOES"~\cite{Ciric.2019} and the interface \verb"LOMS.cz"~\cite{Hrabovsky.2025} provide accessible ways to extract the three Judd-Ofelt parameters, while still relying on the underlying transition matrix elements which largely trace back to the 1970s literature.

\subsection{The Case for an Exact Matrix Repository}

This work addresses some of the fundamental weak points of the current situation in the field.
At present, researchers must rely on legacy tables of matrix elements that are scattered across the literature in disparate formats and of varying quality.
While some sources provide exact analytical expressions, like those provided by Nielson and Koster~\cite{Nielson.1963}, many report values in floating-point representation with inconsistent numerical precision~\cite{Hansen.1996}.
Consequently, researchers who choose to compute these matrix elements independently must still validate their results against existing compilations.
This process invariably introduces numerical discrepancies that require rigorous analysis to determine whether they stem from genuine physical or mathematical deviations or merely rounding errors inherent to floating-point precision~\cite{LizarazoFerro.2026}.
When genuine discrepancies arise, isolating the underlying error, whether located in the new calculations or within the legacy tables, is an exceptionally resource-intensive and time-consuming endeavor.

Several factors explain the current absence of a universal software package for lanthanide calculations.
Chief among these is that existing software suites have historically been developed for, and optimized around, specific applications.
Extracting matrix elements from these specialized frameworks for alternative use cases can be exceedingly complex, and often not all required elements are available.
Furthermore, because these values are produced in floating-point formats, their precision remains intrinsically tied to the specific computational pipeline used to generate them.
Faced with the challenge of integrating incompatible legacy codebases, researchers frequently opt to develop another proprietary, custom-tailored software~\cite{LizarazoFerro.2026}. 
As a result, the lifecycle of lanthanide software remains highly fragmented and short-lived.

The present work argues that the primary need of the scientific community is not another specialized lanthanide application, but rather a comprehensive repository of operator matrix elements computed and stored with absolute mathematical precision.
The open-access repository \ameli~\cite{Caspary.2026b} therefore provides a large set of angular matrix elements of tensor operators for every $f^N$ configuration calculated and stored using exact arithmetic.
Trust in this repository will naturally grow as it undergoes continuous peer validation.
Crucially, unlike the current paradigm, utilizing exact arithmetic leaves no margin for numerical improvement in the future.
This positions the repository to serve as a definitive, modern reference for lanthanide calculations, effectively superseding all legacy tables.

A major advantage of computing the operator matrix elements using exact arithmetic, rather than standard floating-point representations, is the unambiguous determination of selection rules.
In conventional numerical frameworks, accumulated rounding errors frequently convert analytically vanishing matrix elements into small, non-zero floating-point artifacts.
Conversely, genuinely weak transitions can be lost entirely due to underflow limitations.
Because selection rules fundamentally require a strict distinction between an absolute algebraic zero and an arbitrarily small value, floating-point representations are inherently inadequate for this task.
By utilizing exact arithmetic, the \ameli\ repository ensures that every vanishing matrix element corresponds strictly to an exact, mathematically rigorous selection rule.

Together with the matrices, the \ameli\ repository for each configuration provides a transformation matrix between the common state representation in $LS$-coupling and the product state space of Slater determinants, again using an exact numerical representation.
This allows users to take full advantage of the globally synchronized phases of the product states.
For cases of rotational state symmetry like in amorphous hosts, calculations such as energy-level or Judd-Ofelt fits can even be carried out directly in a reduced $LS$-space, because the phases of the $LS$-states in each $J$-multiplet have been synchronized.

The step from pure basis states to intermediate coupling marks the point where exact arithmetic becomes unfeasible due to the necessity of numerical matrix diagonalization, requiring the use of floating-point approximations in the application software.
Consequently, \ameli\ does not provide matrix elements in intermediate coupling.
Nevertheless, the repository renders all the printed tables of reduced matrix elements in intermediate coupling obsolete which have been used for applications of the Judd-Ofelt theory so far.
Lanthanide spectroscopists can instead utilize a suitable set of published radial integrals to calculate all states and reduced matrix elements in intermediate coupling using basic numerical matrix operations.
Ideally, a full energy-level fit based on measured energy-level positions would be used to determine the specific radial integrals for a certain host material~\cite{Kozak.2005}.

Experimentalists studying lanthanides in crystalline hosts can take even more advantage of the \ameli\ repository.
It contains even-rank crystal field operators in the Wybourne nomenclature for convenience~\cite{Wybourne.1965,Fiorucci.2025}.
Energy-level fits to lanthanide spectra incorporating these operators naturally account for $J$-mixing, yielding eigenstates that are a superposition of different $J$-multiplets alongside the $LS$-term mixing intrinsic to intermediate coupling.

\subsection{Methodological Advancements and Overview}

The \ameli\ software used to calculate the set of angular matrices is published as open source software~\cite{Caspary.2026}, but it is primarily intended for documentation and not for its integration into other software projects.
Instead, matrices from the repository~\cite{Caspary.2026b} should be used directly.
The open-source Python package \texttt{YALIP}~\cite{Caspary.2026c} serves as reference implementation for applications based on \ameli.
\texttt{YALIP} currently supports three types of applications:
(1) Raw access to matrix elements from the repository,
(2) Numerical calculation of energy levels and transition strengths using radial integrals and Judd-Ofelt parameters, and
(3) Fitting of radial integrals and Judd-Ofelt parameters to measured absorption spectra and their intensities.

The primary conceptual advance of this work is that it provides the first software implementation and data repository of a comprehensive set of angular tensor operators in exact arithmetic instead of floating-point approximations of operator matrix elements.
The analytic determination of these matrix elements so far has been a highly manual endeavor reliant upon the operator symmetries in $LS$-coupling.
This work demonstrates the fully automated calculation in exact arithmetic for the first time.
Central to this advancement is the computation of matrix elements in the space of uncoupled product states, which until now has only been used for numerical calculations~\cite{Edvardsson.2001,Caspary.2002}.

The following text provides the mathematical background required to calculate every angular tensor operator matrix from scratch, including even the mentioned ``very complicate'' perturbation Hamiltonians.
Particular emphasis is placed on aspects in which it deviates from traditional methods.
This work does not describe the first calculation of matrix elements based on Slater determinants~\cite{Edvardsson.2001,Caspary.2002}, but it provides an algorithmic refinement by not just mentioning the rules, but developing a detailed data structure for the identification of potentially non-zero matrix elements, see Section~\ref{sec.elements}.
Although the computational effort for the preparation of this data structure is high, it reduces the number of calculation steps for each subsequent operator matrix significantly.

A conceptual advance of the \ameli\ approach is that the element-wise calculation of matrix elements is reduced to the absolute minimum of three elementary tensor operators as building blocks for general one-, two-, or three-electron operators, see Section~\ref{sec.tensors}.
Every practical many-electron tensor operator like an angular momentum or Hamilton operator is derived from these three elementary operators by basic high-level matrix operations.
This minimizes the required software code in the sensitive area of element-wise calculation in which debugging can be very difficult.
Note that Ref.~\onlinecite{Caspary.2002} still used ten different elementary tensor operators instead.

The algorithm to determine a transformation matrix between determinantal product states and $LS$-coupling was first introduced in the dissertation of the author~\cite{Caspary.2002} for the numerical computation of matrix elements in the software \texttt{Lanthanide}~\onlinecite{Caspary.2005}.
However, Section~\ref{sub:transform} provides a much more detailed description and a formalized algorithmic refinement of this procedure.

While the signs of $LS$-states within each $J$-multiplet have previously been synchronized using a complex iterative algorithm based on the Wigner-Eckart theorem~\cite{Caspary.2002}, this work introduces a conceptual advancement based on a ladder operator, see Section~\ref{sub:phase}.
This new method also corrects an error of the software~\onlinecite{Caspary.2005} which failed to apply the sign synchronization to the $\tau$-doublets in the configurations from $f^5$ to $f^9$.

Since arithmetic values of matrix elements leave no room for further improvement, a suitable approach for long-term storage of the results is detailed in Section~\ref{sec.data}, resulting in significant improvements in packaging, curation, and reproducibility.
The goal is to keep these digital datasets available for a longer period of time than the magnetic tapes on which the first calculated matrix elements were available in the 1960s.
Another important aspect of modern research data management is reusability.
Instead of a raw number array each matrix is stored in the \ameli\ repository\cite{Caspary.2026b} in a self-contained data container together with extensive meta data.

For the first time this text provides the details of the calculation workflow from the electron configuration to the final matrix elements in $LS$-coupling in Section~\ref{sec.workflow} as an important algorithmic refinement.
It is intended as a guideline for users who want to study the software code and probably expand it for other tensor operators or implement it in a different software environment.

The text finishes with an extensive comparison to legacy tables from Judd and Carnall in Section~\ref{sec.comparison} and a discussion of two application examples in Section~\ref{sec.examples}.
For a publication that aims to update fundamental elements of well established procedures, it is very important to demonstrate its compatibility with existing results.

Section~\ref{sec.comparison} also discusses mathematical validity tests for the matrix elements.
A particularly interesting type of test is provided by the well-known fact that different interactions in $f^N$ configurations share the same symmetries and thus the same angular matrix elements.
This effect is known as parameter absorption or screening, and some radial integrals resulting from semi-empirical energy-level fits cover the effects of several different perturbations.
Knowledge of such linear relations between certain perturbation operators can serve as a useful testing tool to identify potential calculation errors as detailed in Section~\ref{sub:elementwise}.
For two of these linear combinations this text gives arithmetic expressions for the first time, namely the Eqs.~\eqref{eq:t5} and \eqref{eq:p0}.

\section{Product States}
\label{sec.product}

This section briefly reviews some well-known fundamentals of many-electron systems in order to establish a consistent terminology and nomenclature for the remainder of the paper.
It also emphasizes some basic properties that will become essential as we continue.

\subsection{Central Field Approximation}

The energy levels of a many-electron atom or ion are given by the eigenvalues $E_a$ of the total Hamilton operator $\operator{H}$:
\begin{equation}
  \label{eq:H}
  \operator{H} |\Psi_a\rangle = E_a |\Psi_a\rangle
\end{equation}
with the eigenstate $|\Psi_a\rangle$. Eigenvalues and eigenstates are obtained by diagonalizing the energy matrix with a matrix element between the states $a$ an $b$ given by:
\begin{equation}
  \label{eq:Helement}
  \langle\Psi_b| \operator{H} |\Psi_a\rangle
  = \int \Psi_b(\vector{r})\, \operator{H}\, \Psi_a^\ast(\vector{r})
  \,d^3\vector{r}
  \ .
\end{equation}

For an ion with a nucleus of charge number $Z$ and $N$ electrons in
partly filled shells the electrostatic Hamiltonian is given by~\cite{Wybourne.1965,CohenTannoudji.1977}
\begin{equation}
  \label{eq:Hcoulumb}
  \operator{H}' = -\sum_{i}\frac{\vector{p}_i^{\,2}}{2m_e}
  +\frac{e^2}{4\pi\epsilon_0} 
  \left(-\sum_{i}\frac{Z}{r_i} + \sum_{i<j}\frac{1}{r_{ij}}\right)
\end{equation}
with the momentum operator $\vector{p}_i$ of electron~$i$, the electron mass~$m_e$, the electron charge~$e$, the dielectric constant~$\epsilon_0$, the distance $r_i$ of electron~$i$ from the nucleus, and the distance $r_{ij}=|\vector{r}_j-\vector{r}_i|$ between electrons $i$ and $j$.
The first term describes the kinetic energy of the electrons, the second term is the Coulomb energy of the electrons in the central field of the nucleus, and the third term takes into account the repulsive Coulomb forces between all electron pairs.
The prime in $\operator{H}'$ reminds us that this is not the total Hamiltonian.

The usual approach to obtain the eigenvalues and eigenstates of a many-electron system is based on the central field approximation.
A Hamiltonian $\operator{H}_0$ with an arbitrary central potential $U(r)$ for the electrons is used as zeroth order approximation.
It is represented as a sum of operators $\operator{h}_0$ acting on single electrons~\cite{Wybourne.1965,CohenTannoudji.1977}:
\begin{equation}
  \label{eq:H0}
  \operator{H}_0 =
  \sum_{i} \left(-\frac{\vector{p}_i^{\,2}}{2m_e}+U(r_i)\right)
  = \sum_{i} \operator{h}_{0,i}
  \ .
\end{equation}

The eigenfunction of each of these hydrogen-like one-electron operators $\operator{h}_0$ is~\cite{Wybourne.1965,CohenTannoudji.1977,CohenTannoudji.1977b}
\begin{equation}
  \label{eq:psi}
  \psi_{\alpha}(\vector{r}) =
  R_{nl}(r) Y^l_{m_l}(\theta,\phi)\chi_{m_s}
\end{equation}
with the radial function~$R_{nl}(r)$, the spherical
harmonic~$Y^l_{m_l}(\theta,\phi)$, and the spin function~$\chi_{m_s}$
defined by $\chi_i\chi_j=\delta_{ij}$. The abbreviation
$\alpha=nlm_lm_s$ represents the set of quantum numbers of an electron: shell~($n$), orbital angular momentum~($l$ and $m_l$), and spin~($m_s$).

\subsection{Product States}
\label{sub:ProductStates}

Any product of $N$ one-electron wave functions~\eqref{eq:psi}
\begin{equation}
  \label{eq:Psi}
  \Psi = \psi_{1,\alpha_1}
  \psi_{2,\alpha_2}
  \cdots\psi_{N,\alpha_N}
\end{equation}
can serve as a mathematically valid eigenvector for Eq.~\eqref{eq:Helement} and split it into the product of a radial and an angular integral.
Because of the characteristics of the radial function in Eq.~\eqref{eq:psi} the radial integral depends only on the quantum numbers $n$ and $l$ of the involved electrons.
These numbers are identical for all electrons inside a $f^N$ configuration and therefore the radial integral is essentially a common factor for all angular matrix elements in Eq.~\eqref{eq:Helement}.

In order to become a physically valid eigenstate, an eigenvector~\eqref{eq:Psi} must be transformed in such a way that it respects the Pauli principle which requires electronic eigenfunctions to be antisymmetric with regard to the exchange of any pair of electrons. The solution to this requirement is the well known Slater determinant~\cite{Slater.1929,Wybourne.1965,CohenTannoudji.1977b}
\begin{equation}
  \label{eq:determinant}
  \Psi_{\alpha_1\alpha_2\ldots\alpha_N} = 
  \frac{1}{\sqrt{N!}}
  \begin{vmatrix}
    \psi_{1,\alpha_1} &
    \psi_{1,\alpha_2} &
    \cdots &
    \psi_{1,\alpha_N} \\
    \psi_{2,\alpha_1} &
    \psi_{2,\alpha_2} &
    \cdots &
    \psi_{2,\alpha_N} \\
    \vdots & \vdots & \ddots & \vdots \\
    \psi_{N,\alpha_1} &
    \psi_{N,\alpha_2} &
    \cdots &
    \psi_{N,\alpha_N}
  \end{vmatrix}
  \ ,
\end{equation}
which provides an antisymmetric linear combination of all electron permutations. We use the abbreviation
\begin{equation}
  \label{eq:Psi-short}
  \Psi_{\alpha_1\alpha_2\ldots\alpha_N} = 
  \left\{\alpha_1\alpha_2\ldots\alpha_N\right\}
\end{equation}
for such antisymmetric product wave functions.

\subsection{Perturbation Theory}
\label{sub:perturbation}

The pure central field nature of the operator $\operator{H}_0$ results in rotational symmetry without any angular dependency.
All eigenstates of a lanthanide ion are thus degenerated with respect to this operator.
In an energy-level fit its effect is therefore taken into account by a fitting parameter acting as a global energy offset for all levels.
A second important consequence is that we are free to use any electron-state coupling scheme for energy-level calculations which suits our calculation approach best.

The impact of any interaction between the electrons and/or the core of the lanthanide ion on the energy-level spectrum which is not covered by the central field approximation in Eq.~\eqref{eq:H0} turns out to be small compared to the zeroth order energy.
Each interaction can therefore be treated as perturbation $\operator{H}_p$ to the zeroth order Hamiltonian:
\begin{equation}
    \operator{H}_\mathit{corr} = \operator{H}_0 + \operator{H}_p
    \ .
\end{equation}

According to the well known perturbation theory in quantum mechanics the elements of the first order correction matrix are~\cite{CohenTannoudji.1977b}
\begin{equation}
    \label{eq:Hp1st}
    [H_{p}]_{ab} = \langle\Psi_a| \operator{H}_p |\Psi_b\rangle
\end{equation}
with the zeroth order wave functions $\Psi_a$ and $\Psi_b$ of the final and the initial state, respectively.
Valid states are thus Slater determinants~\eqref{eq:Psi-short} or any unitary linear transformation of them into another coupling scheme.
In a semi-empirical energy-level fit only the material-independent angular part of this matrix element is actually calculated while the material-dependent radial integral is used as fitting parameter.

While the first order correction in Eq.~\eqref{eq:Hp1st} takes only interactions inside the $f^N$ base configuration into account, the second order correction accounts for interactions of the base configuration with each state $i$ of each excited configuration~$c$ by the matrix elements~\cite{CohenTannoudji.1977b}:
\begin{equation}
    \label{eq:Hp2nd}
    [H_{p}]_{ab} = \sum_{c,i}
    \frac{\langle\Psi_a| \operator{H}_p |\Psi^c_i\rangle 
    \langle\Psi^c_i| \operator{H}_p |\Psi_b\rangle}%
    {E_0 - E_c}
\end{equation}
with the wave function $\Psi^c_i$ of state $i$ in the excited configuration $c$ and the barycenter energies $E_0$ and $E_c$ of the base configuration $f^N$ and the excited configuration~$c$, respectively.

Although Eq.~\eqref{eq:Hp2nd} seems to be far too complex for useful calculations, an equivalent operator~\cite{Rajnak.1963,Racah.1967}
\begin{equation}
    \label{eq:Hp-eff}
    \operator{H}'_p = \sum_{c,i}
    \frac{\operator{H}_p|\Psi^c_i\rangle \langle\Psi^p_i|\operator{H}_p}{E_0 - E_c}
\end{equation}
was actually determined for each relevant interaction.
Such an effective operator acts solely inside the base configuration:
\begin{equation}
    \label{eq:Hp2st-alt}
    [H_{p}]_{ab} = \langle\Psi_a| \operator{H}'_p |\Psi_b\rangle
    \ .
\end{equation}

Based on these effective operators, the treatment of intra-configuration (1st order) and inter-configuration (2nd order) interactions is essentially identical and we use the same notation $\operator{H}_p$ without prime for either a first order or an effective second order interaction Hamiltonian.
The energy-level spectrum of a lanthanide ion is therefore calculated by diagonalizing the total interaction matrix with the elements
\begin{equation}
    \label{eq:Hab}
    [H]_{ab} = \sum_{p} R_p
    \langle\Phi_a|\operator{H}_p|\Phi_b\rangle
    \ ,
\end{equation}
the radial integral $R_p$, and the angular parts $\Phi_a$ and $\Phi_b$ of the zero order wave functions $\Psi_a$ and $\Psi_b$.

\section{Matrix Elements}
\label{sec.elements}

The conceptual simplicity of the Slater determinant comes with severe costs, which in the past restricted their application to very low numbers of electrons only.
Reason is that for an $N$ electron system each determinant consists of $N!$ summands, resulting in $(N!)^2$ additions just for a single matrix element.

The traditional way to calculate matrix elements of many-electron systems therefore uses $LS$-coupling.
This reduces the number of calculations drastically, but these calculations follow optimized operator-dependent rules based on involved quantum-mechanical and group-theoretical considerations~\cite{Judd.1963}.

Operator matrices of product states in contrast are not well shaped, but they are sparse as well.
Lists including all potentially non-zero matrix elements can be prepared by following some operator-independent basic rules which we shall exploit here.
Modern computers are well able to handle the remaining number of calculations and the necessary operations follow simple and generic rules.

While the traditional approach was ideal for manual computation and early computer systems, the rules for product state matrix elements are perfect for automated processing on current standard computers.

\subsection{Example}
\label{sub:Example}

The example of a matrix element for a two-electron system illustrates how potentially non-zero matrix elements are identified.
A one-electron operator in a two-electron configuration is the sum of two essentially identical elementary one-electron operators one acting on the first and the other acting on the second electron:
\begin{equation}
  \label{eq:exF}
  \operator{F} = \operator{f}_1 + \operator{f}_2
  \ .
\end{equation}

The product state or Slater determinant for a two-electron configuration with the sets of electron quantum numbers $\alpha_a$ and $\alpha_b$ is
\begin{equation}
  \label{eq:alpha_ab}
  |\{\alpha_a\alpha_b\}\rangle =
  \frac{1}{\sqrt{2}}\big(
  |\alpha_a\alpha_b\rangle -
  |\alpha_b\alpha_a\rangle
  \big)
\end{equation}
where the first set in $|\alpha_i\alpha_j\rangle$ contains the quantum numbers of the first electron and the second set is attributed to the second electron.

The full expression of a general matrix element of a one-electron operator in a two-electron configuration is then given by
\begin{multline}
  \label{eq:MatF-1}
  \langle\{\alpha_a\alpha_b\}|\operator{F}|\{\alpha_c\alpha_d\}\rangle =
  \\
  = {\textstyle \frac12} \big[
  \langle\alpha_a\alpha_b|\operator{f}_1|\alpha_c\alpha_d\rangle
  -\langle\alpha_a\alpha_b|\operator{f}_1|\alpha_d\alpha_c\rangle -
  \\
  -\langle\alpha_b\alpha_a|\operator{f}_1|\alpha_c\alpha_d\rangle
  +\langle\alpha_b\alpha_a|\operator{f}_1|\alpha_d\alpha_c\rangle +
  \\
  +\langle\alpha_a\alpha_b|\operator{f}_2|\alpha_c\alpha_d\rangle
  -\langle\alpha_a\alpha_b|\operator{f}_2|\alpha_d\alpha_c\rangle -
  \\
  -\langle\alpha_b\alpha_a|\operator{f}_2|\alpha_c\alpha_d\rangle
  +\langle\alpha_b\alpha_a|\operator{f}_2|\alpha_d\alpha_c\rangle
  \big]
  \ .
\end{multline}

There is one summand contained in this equation for every permutation of the sets of quantum numbers and every elementary operator $\operator{f}_i$.
Operator $\operator{f}_1$ acts on the first electron only and
$\operator{f}_2$ on the second one, therefore the last four summands must be identical to the first four, resulting in the shorter expression
\begin{multline}
  \label{eq:MatF-2}
  \langle\{\alpha_a\alpha_b\}|\operator{F}|\{\alpha_c\alpha_d\}\rangle =
  \\
  =
  \langle\alpha_a\alpha_b|\operator{f}_1|\alpha_c\alpha_d\rangle
  -\langle\alpha_a\alpha_b|\operator{f}_1|\alpha_d\alpha_c\rangle -
  \\
  -\langle\alpha_b\alpha_a|\operator{f}_1|\alpha_c\alpha_d\rangle
  +\langle\alpha_b\alpha_a|\operator{f}_1|\alpha_d\alpha_c\rangle
  \ .
\end{multline}

Since $\operator{f}_1$ acts only on the first electron, the quantum numbers of the second sets must be identical for any summand to be non-zero:
\begin{multline}
  \label{eq:MatF-3}
  \langle\{\alpha_a\alpha_b\}|\operator{F}|\{\alpha_c\alpha_d\}\rangle =
  \\
  =
  \langle\alpha_a|\operator{f}_1|\alpha_c\rangle \delta(\alpha_b,\alpha_d)
  -\langle\alpha_a|\operator{f}_1|\alpha_d\rangle \delta(\alpha_b,\alpha_c) -
  \\
  -\langle\alpha_b|\operator{f}_1|\alpha_c\rangle \delta(\alpha_a,\alpha_d)
  +\langle\alpha_b|\operator{f}_1|\alpha_d\rangle \delta(\alpha_a,\alpha_c)
  \ .
\end{multline}

From the Kronecker deltas in this expression we can deduce the only two cases in which the matrix element might be non-zero. The first case is that the final state $|\{\alpha_a\alpha_b\}\rangle$ and the initial state $|\{\alpha_c\alpha_d\}\rangle$ share exactly one common set of quantum numbers. If, for example $\alpha_d=\alpha_b$, the matrix element is
\begin{equation}
  \label{eq:MatF-4a}
  \langle\{\alpha_a\alpha_b\}|\operator{F}|\{\alpha_c\alpha_b\}\rangle
  = \langle\alpha_a|\operator{f}_1|\alpha_c\rangle
  \ .
\end{equation}

In the second case the initial and the final state share both sets of quantum numbers. If, for example $\alpha_c=\alpha_a$ and $\alpha_d=\alpha_b$, the matrix element is
\begin{equation}
  \label{eq:MatF-4b}
  \langle\{\alpha_a\alpha_b\}|\operator{F}|\{\alpha_a\alpha_b\}\rangle
  = \langle\alpha_a|\operator{f}_1|\alpha_a\rangle
  +\langle\alpha_b|\operator{f}_1|\alpha_b\rangle
  \ .
\end{equation}

\subsection{Preparation}
\label{sub:Preparation}

The general rules to identify potentially non-zero matrix elements are all based on the identification of identical sets of quantum numbers in their initial and final states.

As it can be seen from the results of the example in Subsection~\ref{sub:Example}, the sign of the respective matrix element depends on the order of the sets.
If we would have $\alpha_d=\alpha_a$, for example, the right hand side of Eq.~\eqref{eq:MatF-4a} would be $-\langle\alpha_b|\operator{f}_1|\alpha_c\rangle$ with a negative sign.

In order to keep track of the correct signs it is therefore necessary to define a common order for the list of all potential sets of quantum numbers for a given configuration.
The actual choice doesn't matter, but it is advisable to stay with the standard order introduced by Condon and Shortley~\cite{Condon.1935}. For the single-electron configuration $f^1$ and the nomenclature $m_l^\pm=(m_l,m_s=\pm\frac12)$ the 14 states in standard order are
\begin{multline}
  \label{eq:StdOrder}
  +3^+, +3^-,
  +2^+, +2^-,
  +1^+, +1^-,
   0^+,  \\ 0^-,
  -1^+, -1^-,
  -2^+, -2^-,
  -3^+, -3^-
  \ .
\end{multline}
For any configuration $f^N$ the list of all product states is the lexicographically sorted list of $N$-combinations taken from this list and arranged in standard order.
The total number of states is
\begin{equation}
  \begin{pmatrix}2\,(2l+1)\\N\end{pmatrix} =
  \begin{pmatrix}14\\N\end{pmatrix}
  \ .
\end{equation}

Note that the actual quantum numbers are irrelevant for the following algorithm until the elementary operators are evaluated in the very last step.
This is an important advantage of our approach.
The representation of $\alpha_i$ in the \ameli\ code is just an index into the list of single-electron states~\eqref{eq:StdOrder} with a value in the range of 0 to 13 for the $f^N$ configurations.

For every diagonal matrix element initial and final state are identical, they share the same sorted list of indices $[\alpha_1,\alpha_2,\ldots,\alpha_N]$ and the number of non-matching indices $\Ndiff=0$.
All diagonal elements are potentially non-zero for every operator.

For non-diagonal matrix elements we have $\Ndiff>0$.
We will see in Subsection~\ref{sub:Calculation} that for the identification of all potentially non-zero matrix elements we need to keep track of all matrix elements for with $0\le\Ndiff\le3$.

The value of $\Ndiff$ does not change when initial and final state are exchanged.
Therefore only the lower or the upper non-diagonal triangle needs to be evaluated to generate a list of all potentially non-zero matrix elements, regardless of the symmetry of the matrix on which the list will be applied.

In order to simplify the determination of $\Ndiff$, certain datasets $D_i$ with $i=1,2,3$ are prepared for each state.
For each dataset we split the sorted list of indices $A=[\alpha_1,\alpha_2,\ldots,\alpha_N]$ into pairs of two lists $\Asame,\Adiff$.
The list $\Adiff$ contains $i$ sorted indices picked from $A$, and $\Asame$ contains the remaining sorted indices of $A$.
All possible pairs $\Asame, \Adiff$ are collected for $D_1$, $D_2$, and $D_3$.
Together with each pair we also store $\Nswap$, the number of neighbor indices which need to be exchanged in $A$ to transform it into the concatenated list of $\Asame$ and $\Adiff$.

Using the datasets $D_1$, $D_2$, and $D_3$, $\Ndiff$ can be determined for any matrix element using the following algorithm:
If it is a diagonal element, $\Ndiff=0$.
Otherwise search for a common list $\Asame$ in the datasets $D_1$ of initial and final state.
If unsuccessful, search in $D_2$ and finally in $D_3$.
As soon as the search is successful, the value of $\Ndiff=i$ is given by the index of the respective $D_i$ and the matrix element might be non-zero for an $n$-electron operator with $n=1,\ldots,i$.
If no search is successful, the respective matrix element is zero for every operator.

Whenever identified, a potentially non-zero matrix element is registered in certain lists~$M_n$ of matrix elements for operators with the particle-rank $n=1,\ldots,\Ndiff$.
The following information is stored in each list: the state indices of initial and final state, the common index list $\Asame$, the index lists $\Adiff(\mathit{inital})$ and $\Adiff(\mathit{final})$, and the total number of swap operations $\Nswap=\Nswap(\mathit{initial})+\Nswap(\mathit{final})$.

Note that the lists $M_1$, $M_2$, and $M_3$ are material-independent and operator-indepedent. Once generated, they can be used infinitely to calculate the matrix elements of arbitrary one-, two-, and three-electron operators.

\subsection{Calculation}
\label{sub:Calculation}

Now we need to discuss the calculation of all matrix elements based on the lists $M_n$ of potentially non-zero elements.
The rules for this calculation depend on the number $n$ of electrons the operator is acting on (its particle-rank) as well as the number $\Ndiff$ of non-identical sets of electron quantum numbers for the respective matrix element.

We use the following abbreviation for the angular part of a general product state:
\begin{equation}
    |\Phi\rangle = | \{\alpha_1\alpha_2\ldots\alpha_N\} \rangle
    \ .
\end{equation}

In the following formulation of calculation rules we will use $\alpha_i$ for indices to single-electron states~\eqref{eq:StdOrder} which are identical in the initial and final product state (elements of $\Asame$) and $\beta_i$ or $\beta'_i$ for indices which appear only in the initial or in the final product state, respectively (elements of $\Adiff)$.

A one-electron operator~$\operator{F}$ in an $N$-electron configuration is the sum of $N$ essentially identical elementary one-electron operators $\operator{f}_i$ acting on electron~$i$:
\begin{equation}
  \label{eq:opF}
  \operator{F} = \sum_i \operator{f}_i
  \ .
\end{equation}
For $\Ndiff=0$, the matrix element is~\cite{Judd.1963,Goldschmidt.1978,Condon.1935}
\begin{equation}
  \label{eq:elementF-0}
  \langle \Phi' | \operator{F} | \Phi \rangle = 
  \sum_i \langle \alpha_i | \operator{f} | \alpha_i \rangle
\end{equation}
and for $\Ndiff=1$, it is
\begin{equation}
  \label{eq:elementF-1}
  \langle \Phi' | \operator{F} | \Phi \rangle = 
  \langle \beta'_1 | \operator{f} | \beta_1 \rangle
  \ .
\end{equation}

A two-electron operator~$\operator{G}$ in an $N$-electron configuration is the sum of $N(N-1)/2$ essentially identical elementary two-electron operators $\operator{g}_{ij}$ acting on the pair of electrons~$i,j$:
\begin{equation}
  \label{eq:opG}
  \operator{G} = \sum_{i<j} \operator{g}_{ij}
  \ .
\end{equation}
For $\Ndiff=0$, the matrix element is~\cite{Judd.1963,Goldschmidt.1978}
\begin{equation}
  \label{eq:elementG-0}
  \langle \Phi' | \operator{G} | \Phi \rangle = 
  \sum_{i<j} \langle \{\alpha_i\alpha_j\} | \operator{g} 
  | \{\alpha_i\alpha_j\} \rangle
  \ ,
\end{equation}
for $\Ndiff=1$, it is
\begin{equation}
  \label{eq:elementG-1}
  \langle \Phi' | \operator{G} | \Phi \rangle = 
  \sum_i \langle \{\alpha_i\beta'_1\} | \operator{g} 
  | \{\alpha_i\beta_1\} \rangle
  \ ,
\end{equation}
and for $\Ndiff=2$
\begin{equation}
  \label{eq:elementG-2}
  \langle \Phi' | \operator{G} | \Phi \rangle = 
  \langle \{\beta'_1\beta'_2\} | \operator{g} | \{\beta_1\beta_2\} \rangle
  \ .
\end{equation}

A three-electron operator~$\operator{H}$ in an $N$-electron configuration is the sum of $N(N-1)(N-2)/6$ essentially identical elementary three-electron operators $\operator{h}_{ijk}$ acting on the triple of electrons~$i,j,k$:
\begin{equation}
  \label{eq:opH}
  \operator{H} = \sum_{i<j<k} \operator{h}_{ijk}
  \ .
\end{equation}
For $\Ndiff=0$, the matrix element is~\cite{Edvardsson.2001}
\begin{equation}
  \label{eq:elementH-0}
  \langle \Phi' | \operator{H} | \Phi \rangle = 
  \sum_{i<j<k} \langle \{\alpha_i\alpha_j\alpha_k\} | \operator{h} 
  | \{\alpha_i\alpha_j\alpha_k\} \rangle
  \ ,
\end{equation}
for $\Ndiff=1$, it is
\begin{equation}
  \label{eq:elementH-1}
  \langle \Phi' | \operator{H} | \Phi \rangle = 
  \sum_{i<j} \langle \{\alpha_i\alpha_j\beta'_0\} | \operator{h} 
  | \{\alpha_i\alpha_j\beta_0\} \rangle
  \ ,
\end{equation}
for $\Ndiff=2$
\begin{equation}
  \label{eq:elementH-2}
  \langle \Phi' | \operator{H} | \Phi \rangle = 
  \sum_i \langle \{\alpha_i\beta'_1\beta'_2\} | \operator{h} 
  | \{\alpha_i\beta_1\beta_2\} \rangle
  \ ,
\end{equation}
and for $\Ndiff=3$
\begin{equation}
  \label{eq:elementH-3}
  \langle \Phi' | \operator{H} | \Phi \rangle = 
  \langle \{\beta'_1\beta'_2\beta'_3\} | \operator{h} | \{\beta_1\beta_2\beta_3\} \rangle
  \ .
\end{equation}

Note that three-electron operators can not result directly from elementary physical interactions, which never involve more than two interaction partners.
However, they appear as effective operators for perturbations in second order.

It should also be noted that the \ameli\ package implements the rules above in a single abstract algorithm.
The particle-rank~$n$ of the elementary operator is treated as a free parameter, which may also exceed the current maximum of $n=3$, should it be required for certain tensor operators in the future.

Each matrix element calculated according to these rules finally must be multiplied by $-1$ if the total number of swap operations $\Nswap$ is odd.
Diagonal elements ($\Ndiff=0$) keep their sign, because no swap operation is involved.

Most of our operators will be scalar products and therefore symmetric. For asymmetric operators each calculation of a non-diagonal matrix element must be repeated a second time with initial and final state swapped.

We summarize that the rules for the calculation of a matrix element reduce the number of evaluations of the respective $n$-electron elementary operator $\Neval$ from $(N!)^2$ for every element to 
\begin{equation}
  \Neval = \begin{pmatrix}N-\Ndiff\\n-\Ndiff\end{pmatrix} (n!)^2
\end{equation}
only for potentially non-zero elements.
For lanthanides the largest number of evaluations is 10296 for the diagonal elements ($\Ndiff=0$) of a three-electron operator in the $f^{13}$ configuration.

The number of evaluations could be reduced even further when symmetries of the specific elementary operator would be taken into account.
If the value of a two-electron operator $g_{ij}$ for example does not depend on the order of the electrons, only half of the evaluations are necessary.
Such considerations, however, would make the algorithm operator-dependent.

Reminder: Angular matrices are ma\-te\-ri\-al-in\-de\-pend\-ent. 
The angular matrix of an operator is an immutable property of the respective electron configuration.
Once determined, it can be stored and never needs to be calculated again.
This makes further operator-specific optimizations unnecessary.

\section{Spherical Tensor Operators}
\label{sec.tensors}

So far we have prepared electronic product states, split off the radial integrals, and developed rules to identify which elementary one-, two-, or three-electron operator matrix elements need to be evaluated for the calculation of the angular matrix elements of any many-electron operator in the space of product states.
Now we are going to introduce the many-electron operators in concrete terms as spherical tensor operators and reduce them to elementary unit tensor operators that can be evaluated using the tools from the last section.

In his seminal series of four papers \emph{Theory of Complex Spectra}, published between 1942 and 1949, G.~Racah introduced the concept of a spherical tensor operator to the theory of many-electron systems, which we will use here~\cite{Racah.1942,Racah.1942b,Racah.1943,Racah.1949}.
We will give a very brief introduction to the theory of spherical tensor operators in order to establish a consistent nomenclature and certain expressions, which we will need later.
For further details the reader is referred in particular to the book \emph{Operator Techniques in Atomic Spectroscopy} from B.~R.~Judd~\cite{Judd.1963}.

In the traditional theory of many-electron systems the symmetry properties of spherical tensor operators are exploited for the evaluation of Hamiltonians in an extended $LS$-coupling scheme.
This approach made it possible to greatly simplify the calculation of the operators, but it required each operator to be analyzed separately and manually with regard to its symmetries in order to determine an individually optimized calculation strategy.
We are going to use spherical tensor operators in a different manner, but we will come back to this coupling scheme in Section~\ref{sec.coupling}.

Here we will define two elementary unit tensor operators acting in the orbital and spin space, respectively.
Our goal is to express each many-electron operator in terms of one-, two-, or three-electron tensor products of these two elementary unit tensor operators.
It turns out that just one mixed tensor operator for each number of involved electrons is sufficient to express any many-electron operator.

This means that only these three elementary mixed tensor operators actually need to be evaluated element-wise and the respective expressions will be given in Subsection~\ref{sub:unit}.
The matrix elements of all many-electron operators are derived from these results using basic matrix operations.
Highly efficient implementations of these matrix operations exist for virtually every modern computer system and programming language.

\subsection{Basic Properties}

A spherical tensor operator~$\tensor{A}{k}$ of rank~$k$ with the components~$\telement{A}{k}{q}$ and $q=-k,\ldots,+k$ is defined by its commutation relations with the components of an angular momentum vector operator~$\vector{J}$ acting in the respective space:
\begin{equation}
  \label{eq:Commutator}
  \begin{split}
    \displaystyle
    [\velement{J}{z}, \telement{A}{k}{q}] &= q\, \telement{A}{k}{q}
    \\
    \displaystyle
    [\velement{J}{x}\pm i\velement{J}{y}, \telement{A}{k}{q}] &= \sqrt{k(k+1)-q(q\pm1)}\,
    \telement{A}{k}{q\pm1}
    \ .
  \end{split}
\end{equation}
The theory of tensor operators is a generalization of the theory of vector operators from Condon and Shortley~\cite{Condon.1935}.
Vector operators can be derived from tensor operators of rank~$1$ using the equations~\cite{Judd.1963}
\begin{equation}
  \label{eq:VectorOp}
  \begin{split}
    \velement{A}{x} &=  \displaystyle
    \frac{-1}{\sqrt2}\, \big(\telement{A}{1}{-1} - \telement{A}{1}{+1}\big)
    \\
    \velement{A}{y} &= \displaystyle
    \frac{i}{\sqrt2}\, \big(\telement{A}{1}{-1} + \telement{A}{1}{+1}\big)
    \\
    \velement{A}{z} &= \displaystyle
    \telement{A}{1}{0}
    \ .
  \end{split}
\end{equation}

The following expression defines the scalar product of tensor operators~\cite{Judd.1963}
\begin{equation}
  \label{eq:ScalarProduct-1}
  \dot{\tensor{A}{k}}{\tensor{B}{k}} =
  \sum_q (-1)^q\,  \telement{A}{k}{q} \telement{B}{k}{-q}
  \ .
\end{equation}
This definition keeps the value of the vectorial scalar product
\begin{equation}
  \dot{\vector{A}}{\vector{B}} =
  \dot{\tensor{A}{1}}{\tensor{B}{1}}
  \ .
\end{equation}

The angular behavior of every spherical tensor operator is identical and given by the Wigner-Eckart theorem~\cite{Wigner.1959,Eckart.1930,Judd.1963}:
\begin{multline}
  \label{eq:WignerEckart}
  \langle j' m' | \telement{A}{k}{q} | j m \rangle = \\
  = (-1)^{j'-m'}
  \iiij{j'}{k}{j}{-m'}{q}{m}
  \langle j' || \tensor{A}{k} || j \rangle
\end{multline}
with the eigenstate $|jm\rangle$ of the angular momentum operator $\operator{j}$ acting in the state space of the tensor operator.
The six numbers in parentheses are a Wigner \symiiij~symbol and the symbol $\langle j' || \tensor{A}{k} || j \rangle$ is called a reduced matrix element, which evaluates to a scalar.
The equation tells us that a tensor operator is completely defined by its reduced matrix elements.

The general tensor product of two tensor operators $\tensor{A}{k_1}$ and $\tensor{B}{k_2}$ results in a new tensor operator~$\tensor{Q}{k}$.
This product is called a mixed tensor operator~\cite{Judd.1963}:
\begin{equation}
  \label{eq:MixedTensor}
  \tensor{Q}{k} =
  \cross{\tensor{A}{k_1}}{\tensor{B}{k_2}}{k}
  \ .
\end{equation}

Due to the definition of the spherical tensor operator in relation to an angular momentum operator in Eq.~\eqref{eq:Commutator}, the mixing of tensor operators is closely related to the coupling of angular momenta.
This becomes obvious from the equation for the component of a mixed tensor operator~\cite{Judd.1963}:
\begin{multline}
  \label{eq:MixedElement.CG}
  \telement{Q}{k}{q} = 
  \sum_{q_1,q_2}
  \langle k_1,q_1,k_2,q_2 | k_1, k_2, k, q \rangle
  \telement{A}{k_1}{q_1}
  \telement{B}{k_2}{q_2}
  \ ,
\end{multline}
which contains the Clebsch-Gordan coupling coefficient in angled brackets.
Using the \symiiij~symbol instead~\cite{Judd.1963}, this expression translates to
\begin{multline}
  \label{eq:MixedElement}
  \telement{Q}{k}{q} = 
  (-1)^{k+q} \sqrt{2k+1} \\
  \times
  \sum_{q_1,q_2}
  \iiij{k_1}{k}{k_2}{q_1}{-q}{q_2}
  \telement{A}{k_1}{q_1}
  \telement{B}{k_2}{q_2}
  \ ,
\end{multline}
where the \symiiij~symbol actually reduces the double sum to a single one, because its value is zero if the sum of its lower elements not equals zero.
Therefore, for $k_i \leq k_j$ the sum is carried out over $q_i=-k_i,\ldots,+k_i$, while the second parameter is fixed to $q_j=q-q_i$.

Swapping the factors of a tensor product may involve a sign flip according to the properties of the \symiiij~symbol:
\begin{multline}
  \label{eq:ReOrder}
  \cross{\tensor{A}{k_1}}{\tensor{B}{k_2}}{k} =
  \\ =
  (-1)^{k_1+k_2+k}\,
  \cross{\tensor{B}{k_2}}{\tensor{A}{k_1}}{k}
  \ .
\end{multline}

An important special case is the component of a mixed scalar tensor operator:
\begin{equation}
  \label{eq:MixedScalar}
  \elcross{\tensor{A}{k}}{\tensor{B}{k}}{0}{0} = 
  \frac{(-1)^{k}}{\sqrt{2k+1}}
  \sum_{q}
  (-1)^{q}
  \telement{A}{k}{q}
  \telement{B}{k}{-q}
  \ ,
\end{equation}
where we used the relation
\begin{equation}
  \label{eq:3j-spec1}
  \iiij{j'}{0}{j}{-m'}{0}{m} = \delta(j',j)\delta(m',m)\frac{(-1)^{j-m}}{\sqrt{2j+1}}
\end{equation}

When we compare the equations \eqref{eq:ScalarProduct-1} and \eqref{eq:MixedScalar}, we obtain the connection between the scalar product and the mixed scalar tensor:
\begin{equation}
  \label{eq:ScalarProduct-2}
  \dot{\tensor{A}{k}}{\tensor{B}{k}} =
  (-1)^k \sqrt{2k+1}\,
  \elcross{\tensor{A}{k}}{\tensor{B}{k}}{0}{0} 
  \ .
\end{equation}

The reduced matrix element of the mixed tensor operator is often used for the calculation of its matrix elements.
Without restriction of the operating space, it is related to the reduced matrix elements of its tensor factors by the general expression~\cite{Judd.1963}
\begin{multline}
  \label{eq:MixedRed}
  \langle j' || \tensor{Q}{q} || j \rangle =
  (-1)^{j'+k+j}\,
  \sqrt{2k+1}\,
  \sum_{j''}
  \vij{k_1}{k}{k_2}{j}{j''}{j'}
  \\ \times
  \langle j' || \tensor{A}{k_1} || j'' \rangle\,
  \langle j'' || \tensor{B}{k_2} || j \rangle
  \ , 
\end{multline}
with a \symvij~symbol in curly brackets.

The re-coupling of a triple tensor product is equivalent to the re-coupling of three angular momenta~\cite{Judd.1963}:
\begin{multline}
  \label{eq:ReCouple.ABC}
  \bigcross%
  {\cross{\tensor{A}{k_1}}{\tensor{B}{k_2}}{k_{12}}}%
  {\tensor{C}{k_3}}{k} =
  \\
  = \sum_{k_{23}}
  \langle(k_1k_2)k_{12},k_3,k\,|\,k_1,(k_2k_3)k_{23},k\rangle
  \\
  \times \bigcross%
  {\tensor{A}{k_1}}%
  {\cross{\tensor{B}{k_2}}{\tensor{C}{k_3}}{k_{23}}}{k}
\end{multline}
and the re-coupling coefficient in angled brackets can be expressed in terms of a \symvij~symbol:
\begin{multline}
  \label{eq:ReCouple.6j}
  \langle(k_1k_2)k_{12},k_3,k\,|\,k_1,(k_2k_3)k_{23},k\rangle =
  (-1)^{k_1+k_2+k_3+k}
  \\ \times
  \sqrt{(2k_{12}+1)(2k_{23}+1)}\,
  \vij{k_1}{k_2}{k_{12}}{k_3}{k}{k_{23}}
  \ .
\end{multline}

The evaluation of the scalar triple mixed tensor product using the equations \eqref{eq:MixedScalar} and \eqref{eq:MixedElement} delivers
\begin{multline}
  \label{eq:ReCouple.ABC.zero}
  \bigcross%
  {\cross{\tensor{A}{k_1}}{\tensor{B}{k_2}}{k_3}}%
  {\tensor{C}{k_3}}{0} =
  (-1)^{k_3-k_1}
  \\ \times
  \sqrt{\frac{2k_3+1}{2k_1+1}}\,
  \bigcross%
  {\tensor{A}{k_1}}%
  {\cross{\tensor{B}{k_2}}{\tensor{C}{k_3}}{k_1}}{0}
  \ .
\end{multline}
The same evaluation also shows that according to Eq.~\eqref{eq:ScalarProduct-2} the scalar product does not depend on the coupling order:
\begin{multline}
  \label{eq:ReCouple.ABC.scalar}
  \bigdot{\cross{\tensor{A}{k_1}}{\tensor{B}{k_2}}{k_3}}{\tensor{C}{k_3}} =
  \\
  = \bigdot{\tensor{A}{k_1}}{\cross{\tensor{B}{k_2}}{\tensor{C}{k_3}}{k_1}} =
  \\
  = \sum_{q_1,q_2,q_3}
  \iiij{k_1}{k_2}{k_3}{q_1}{q_2}{q_3}\,
  \telement{A}{k_1}{q_1}\,
  \telement{B}{k_2}{q_2}\,
  \telement{C}{k_3}{q_3}
  \ .
\end{multline}
In the literature this is called a triple scalar product~\cite{Rajnak.1963,Judd.1966}:\begin{multline}
  \label{eq:TripleDot}
  \threedot{\tensor{A}{k_1}}{\tensor{B}{k_2}}{\tensor{C}{k_3}} = \\
  = \sum_{q_1,q_2,q_3}
  \iiij{k_1}{k_2}{k_3}{q_1}{q_2}{q_3}\,
  \telement{A}{k_1}{q_1}\,
  \telement{B}{k_2}{q_2}\,
  \telement{C}{k_3}{q_3}
  \ ,
\end{multline}
which is used for an effective three-electron operator below.

Another important re-coupling case is the pair-wise re-coupling of a 4-element mixed tensor operator~\cite{Judd.1963}:
\begin{multline}
  \label{eq:ReCouple.ABCD}
  \bigcross%
  {\cross{\tensor{A}{k_1}}{\tensor{B}{k_2}}{k_{12}}}%
  {\cross{\tensor{C}{k_3}}{\tensor{D}{k_4}}{k_{34}}}{k} =
  \\
  = \sum_{k_{13},k_{24}}
  \langle(k_1k_2)k_{12},(k_3k_4)k_{34},k\,|\,(k_1k_3)k_{13},(k_2k_4)k_{24},k\rangle
  \\
  \times \bigcross%
  {\cross{\tensor{A}{k_1}}{\tensor{C}{k_3}}{k_{13}}}%
  {\cross{\tensor{B}{k_2}}{\tensor{D}{k_4}}{k_{24}}}{k}
  \ ,
\end{multline}
where the re-coupling coefficient expressed in terms of a \symixj~symbol is
\begin{multline}
  \label{eq:ReCouple.9j}
  \langle(k_1k_2)k_{12},(k_3k_4)k_{34},k\,|\,(k_1k_3)k_{13},(k_2k_4)k_{24},k\rangle =
  \\ =
  \sqrt{(2k_{12}+1)(2k_{34}+1)(2k_{13}+1)(2k_{24}+1)}
  \\ \times
  \ixj{k_1}{k_2}{k_{12}}{k_3}{k_4}{k_{34}}{k_{13}}{k_{24}}{k}
  \ .
\end{multline}
Using the identity relation for $k=0$
\begin{multline}
  \label{eq:9j.scalar}
  \ixj{k_1}{k_2}{k_{12}}{k_3}{k_4}{k_{12}}{k_{13}}{k_{13}}{0} = 
  (-1)^{k_2+k_3+k_{12}+k_{13}}
  \\ \times
  \frac{1}{\sqrt{(2k_{12}+1)(2k_{13}+1)}}
  \vij{k_1}{k_2}{k_{12}}{k_4}{k_3}{k_{13}}
  \ ,
\end{multline}
we obtain the simplified expression for the important scalar case~\cite{Judd.1963}:
\begin{multline}
  \label{eq:ReCouple.ABCD.zero}
  \bigcross%
  {\cross{\tensor{A}{k_1}}{\tensor{B}{k_2}}{k_{12}}}%
  {\cross{\tensor{C}{k_3}}{\tensor{D}{k_4}}{k_{12}}}{0} =
  \\
  = \sum_{k_{13}}
  (-1)^{k_2+k_3+k_{12}+k_{13}}
  \\ \times
  \sqrt{(2k_{12}+1)(2k_{13}+1)}
  \vij{k_1}{k_2}{k_{12}}{k_4}{k_3}{k_{13}}
  \\
  \times \bigcross%
  {\cross{\tensor{A}{k_1}}{\tensor{C}{k_3}}{k_{13}}}%
  {\cross{\tensor{B}{k_2}}{\tensor{D}{k_4}}{k_{13}}}{0}
  \ .
\end{multline}

Of special importance for the scope of this work are the three cases $(k_1,k_2,k_{12})=(0,0,0)$, $(0,1,1)$, and $(1,1,2)$.
Insertion into the algebraic definition of the \symvij~symbol delivers
\begin{multline}
  \label{eq:6j.special}
  \vij{k_1}{k_2}{k_{12}}{k_4}{k_3}{k_{13}} = 
  \vij{0}{0}{0}{k}{k}{k} =
  \vij{0}{1}{1}{k}{k+1}{k} =
  \\ =
  \vij{0}{1}{1}{k}{k-1}{k} =
  \vij{1}{1}{2}{k-1}{k+1}{k} =
  \\ =
  \frac{1}{\sqrt{(2k_{12}+1)(2k_{13}+1)}}
  \ .
\end{multline}

This shows that the pair-wise re-coupling in these cases does not change the magnitude of the operator. It keeps its sign for even $k_{12}$ and flips it for odd $k_{12}$.

\subsection{Elementary Unit Tensor Operators}
\label{sub:unit}

We define an elementary unit tensor operator $\tensor{u}{k}$ acting on one electron in the orbital space:
\begin{equation}
\label{eq:def.u}
  \langle l' || \tensor{u}{k} || l \rangle = \delta(l',l)
  \ ,
\end{equation}
with the Kronecker delta symbol $\delta(l',l)$, and an elementary unit tensor operator $\tensor{t}{k}$ acting on one electron in the spin space:
\begin{equation}
\label{eq:def.t}
  \langle s' || \tensor{t}{k} || s \rangle = \delta(s',s) = 1
  \ ,
\end{equation}
using the fact that the spin quantum number of every electron is $s'=s=1/2$.

In tensor operator expressions you come across mixed tensor operators acting in the same angular space and on the same electron, which can be reduced to a single tensor operator using Eq.~\eqref{eq:MixedRed}.
For elementary unit tensor operators in the orbital space of an $l^N$ configuration we get
\begin{multline}
  \label{eq:reduce-uu}
  \cross{\tensor{u}{k_1}}{\tensor{u}{k_2}}{k} =
  \\ =
  (-1)^{2l+k}
  \sqrt{2k+1}
  \vij{k_1}{k}{k_2}{l}{l}{l}\,
  \tensor{u}{k}
\end{multline}
and for the scalar product this equation reduces to
\begin{equation}
  \label{eq:reduce-uu-scalar-1}
  \dot{\tensor{u}{k}}{\tensor{u}{k}} =
  \frac{1}{\sqrt{2l+1}}\ 
  \tensor{u}{0}
  \ ,
\end{equation}
where we used the relation
\begin{equation}
  \label{eq:6j-spec1}
  \vij{0}{k}{k}{l}{l}{l} =
  \frac{(-1)^{2l+k}}{\sqrt{(2k+1)(2l+1)}}
  \ .
\end{equation}

Using the Wigner-Eckart theorem~\eqref{eq:WignerEckart} and the relation~\eqref{eq:3j-spec1} we obtain the matrix element of the elementary scalar unit tensor operator 
\begin{equation}
  \label{eq:u00}
  \langle m'_l | \telement{u}{0}{0} | m_l \rangle
  = \frac{\delta(m'_l,m_l)}{\sqrt{2l+1}}
  \ .
\end{equation}
Inserting into Eq.~\eqref{eq:reduce-uu-scalar-1} delivers the value of the scalar product 
\begin{equation}
  \label{eq:reduce-uu-scalar}
  \dot{\tensor{u}{k}}{\tensor{u}{k}} =
  \frac{1}{2l+1}
  \ .
\end{equation}

The same consideration in the spin space of an $l^N$ configuration delivers the matrix element of the elementary scalar unit tensor operator
\begin{equation}
  \label{eq:t00}
  \langle m'_s | \telement{t}{0}{0} | m_s \rangle
  = \frac{\delta(m'_s,m_s)}{\sqrt{2s+1}} = \frac{\delta(m'_s,m_s)}{\sqrt{2}}
\end{equation}
and the scalar product
\begin{equation}
  \label{eq:reduce-tt-scalar}
  \dot{\tensor{t}{k}}{\tensor{t}{k}} =
  \frac{1}{2}
  \ .
\end{equation}

According to the rules in Subsection~\ref{sub:Calculation} we need to calculate operator matrices by evaluating all potentially non-zero matrix elements individually based on preprocessed lists.
Even in many-electron configurations this breaks down to the evaluation of elementary one-, two-, or three-electron operators.

Every many-electron operator can be expressed by such elementary operators and we will do that in the following subsections.
However, for the sake of reliability of the evaluation code it is helpful to keep the number of types of elementary operators in the software implementation small and shift the differentiation between different many-electron operators from the level of individual matrix elements to the level of matrix operations.

\begin{widetext}
Therefore, the \ameli\ package uses the following elementary mixed tensor operator in the orbital and spin angular space acting on one electron:
\begin{equation}
  \label{eq:ElementaryOne}
  \ftensor{A}{k} =
  \cross{\tensor{u}{k_1}}{\tensor{t}{k_2}}{k}
  \ ,
\end{equation}
which allows us to describe any operator acting on a single electron by its three rank-parameters $k_1$, $k_2$, $k$, and its tensor component $q$.
The operator~$\ftensor{A}{k}$ is named \verb"Unit_UT" in the software code.
The evaluation of the matrix element of the tensor component~$q$ of this operator starts with Eq.~\eqref{eq:MixedElement}:
\begin{equation}
  \itbra{a} \elcross{\tensor{u}{k_1}}{\tensor{t}{k_2}}{k}{q} \itket{b} = 
  \sum_{q_1,q_2}
  (-1)^{k + q}
  \sqrt{2k+1}
  \iiij{k_1}{k}{k_2}{q_1}{-q}{q_2}
  \itbra{a} \telement{u}{k_1}{q_1} \itket{b}
  \itbra{a} \telement{t}{k_2}{q_2} \itket{b}
  \ ,
\end{equation}
where we use the abbreviation $|a\rangle=|l_a m_{l_a} s_a m_{s_a}\rangle$ for a general single electron state.

For the evaluation of the two unit tensor operator matrix elements we use the Wigner-Eckart theorem~\eqref{eq:WignerEckart} and insert the reduced matrix elements from 
Eqs.\ \eqref{eq:def.u} and \eqref{eq:def.t}:
\begin{multline}
  \itbra{a} \elcross{\tensor{u}{k_1}}{\tensor{t}{k_2}}{k}{q} \itket{b} = 
  \sum_{q_1,q_2}
  (-1)^{k + q + l_a + s_a - m_{l_a} - m_{s_a}}
  \\ \times
  \delta(l_a, l_b)\,\delta(s_a, s_b)\,
  \sqrt{2k+1}
  \iiij{k_1}{k}{k_2}{q_1}{-q}{q_2}
  \iiij{l_a}{k_1}{l_b}{-m_{l_a}}{q_1}{m_{l_b}}
  \iiij{s_a}{k_2}{s_b}{-m_{s_a}}{q_2}{m_{s_b}}
  \ .
\end{multline}

When we take the properties of the \symiiij~symbols into account, which fix $q_1=m_{l_a}-m_{l_b}$ and $q_2=m_{s_a}-m_{s_b}$, the double sum reduces to a single term:
\begin{multline}
  \itbra{a} \elcross{\tensor{u}{k_1}}{\tensor{t}{k_2}}{k}{q} \itket{b} = 
  (-1)^{k + q + l_a + s_a - m_{l_a} - m_{s_a}}\,
  \delta(l_a, l_b)\,
  \sqrt{2k+1}
  \\ \times
  \iiij{k_1}{k}{k_2}{m_{l_a}-m_{l_b}}{-q}{m_{s_a}-m_{s_b}}
  \iiij{l_a}{k_1}{l_b}{-m_{l_a}}{m_{l_a}-m_{l_b}}{m_{l_b}}
  \iiij{s_a}{k_2}{s_b}{-m_{s_a}}{m_{s_a}-m_{s_b}}{m_{s_b}}
  \ ,
\end{multline}
where we also removed the second delta symbol, since we always have $s_a=s_b=1/2$.
From the properties of the \symiiij~symbol we also conclude that the matrix element of all tensor components except $q=m_{l_a}-m_{l_b}+m_{s_a}-m_{s_b}$ must vanish.
This brings us to the final expression implemented in the \ameli\ package:
\begin{multline}
  \label{eq:EvalOne}
  \itbra{a} \elcross{\tensor{u}{k_1}}{\tensor{t}{k_2}}{k}{q} \itket{b} = 
  (-1)^{k + l_a + s_a - m_{l_b} - m_{s_b}}\,
  \delta(l_a, l_b)\,
  \delta(q,m_{l_a}-m_{l_b}+m_{s_a}-m_{s_b})\,
  \sqrt{2k+1}
  \\ \times
  \iiij{k_1}{k}{k_2}{m_{l_a}-m_{l_b}}{-q}{m_{s_a}-m_{s_b}}
  \iiij{l_a}{k_1}{l_b}{-m_{l_a}}{m_{l_a}-m_{l_b}}{m_{l_b}}
  \iiij{s_a}{k_2}{s_b}{-m_{s_a}}{m_{s_a}-m_{s_b}}{m_{s_b}}
  \ .
\end{multline}

The elementary mixed tensor operator in the orbital and spin angular space acting on two electrons used by the \ameli\ package is
\begin{equation}
  \label{eq:ElementaryTwo}
  \ftensor{B}{k} = 
  \bigcross{\ftensor{A}{k_{12}}_1}{\ftensor{A}{k_{34}}_2}{k} =
  \bigcross%
  {\cross{\tensor{u}{k_1}_1}{\tensor{t}{k_2}_1}{k_{12}}}%
  {\cross{\tensor{u}{k_3}_2}{\tensor{t}{k_4}_2}{k_{34}}}{k}
  \ ,
\end{equation}
with the mixed tensor operator of rank~$k_{12}$ acting on electron~1 and the one of rank~$k_{34}$ acting on electron~2.
The operator~$\ftensor{B}{k}$ is named \verb"Unit_UTUT" in the software code.
It allows to express every many-electron operator acting on two electrons by its seven rank-parameters $k_1$, $k_2$, $k_{12}$, $k_3$, $k_4$, $k_{34}$, $k$, and its tensor component $q$.

Note that we can reduce this full-featured operator to simpler cases by setting certain ranks $k_1$ to $k_4$ to zero and compensate with scalar factors according to 
Eqs.\ \eqref{eq:u00} and \eqref{eq:t00}.
We could actually even use this operator as a replacement for the elementary one-electron operator~\eqref{eq:ElementaryOne} by setting $k_3=k_4=k_{34}=0$, but this would come with a significantly increased computing effort and it would require special software code for the calculation of operators in the $f^1$ configuration.

The evaluation of the matrix element of the tensor component~$q$ of the operator~$\ftensor{B}{k}$ starts with Eq.~\eqref{eq:MixedElement} again:
\begin{equation}
  \label{eq:ElementaryTwo-1}
  \itbra{ab}
  \elcross{\ftensor{A}{k_{12}}_1}{\ftensor{A}{k_{34}}_2}{k}{q}
  \itket{cd} = 
  \sum_{q_{12},q_{34}}
  (-1)^{k + q}
  \sqrt{2k+1}
  \iiij{k_{12}}{k}{k_{34}}{q_{12}}{-q}{q_{34}}
  \itbra{a} \ftelement{A}{k_{12}}{q_{12}} \itket{c}
  \itbra{b} \ftelement{A}{k_{34}}{q_{34}} \itket{d}
  \ .
\end{equation}

The two matrix elements at the end of this equation are given by Eq.~\eqref{eq:EvalOne}:
\begin{multline}
  \itbra{a} \ftelement{A}{k_{12}}{q_{12}} \itket{c} = 
  \itbra{a} \elcross{\tensor{u}{k_1}}{\tensor{t}{k_2}}{k_{12}}{q_{12}} \itket{c} = 
  \\ =
  (-1)^{k_{12} + l_a + s_a - m_{l_c} - m_{s_c}}\,
  \delta(l_a, l_c)\,
  \delta(q_{12}, m_{l_a} - m_{l_c} + m_{s_a} - m_{s_c})\,
  \sqrt{2k_{12}+1}
  \\ \times
  \iiij{k_1}{k_{12}}{k_2}{m_{l_a}-m_{l_c}}{-q_{12}}{m_{s_a}-m_{s_c}}
  \iiij{l_a}{k_1}{l_c}{-m_{l_a}}{m_{l_a}-m_{l_c}}{m_{l_c}}
  \iiij{s_a}{k_2}{s_c}{-m_{s_a}}{m_{s_a}-m_{s_c}}{m_{s_c}}
  \ ,
\end{multline}
\begin{multline}
  \itbra{b} \ftelement{A}{k_{34}}{q_{34}} \itket{d} = 
  \itbra{b} \elcross{\tensor{u}{k_3}}{\tensor{t}{k_4}}{k_{34}}{q_{34}} \itket{d} = 
  \\ =
  (-1)^{k_{34} + l_b + s_b - m_{l_d} - m_{s_d}}\,
  \delta(l_b, l_d)\,
  \delta(q_{34}, m_{l_b} - m_{l_d} + m_{s_b} - m_{s_d})\,
  \sqrt{2k_{34}+1}
  \\ \times
  \iiij{k_3}{k_{34}}{k_4}{m_{l_b}-m_{l_d}}{-q_{34}}{m_{s_b}-m_{s_d}}
  \iiij{l_b}{k_3}{l_d}{-m_{l_b}}{m_{l_b}-m_{l_d}}{m_{l_d}}
  \iiij{s_b}{k_4}{s_d}{-m_{s_b}}{m_{s_b}-m_{s_d}}{m_{s_d}}
  \ .
\end{multline}

When we insert these expressions into Eq.~\eqref{eq:ElementaryTwo-1} the delta symbols for $q_{12}$ and $q_{34}$ reduce the double sum to a single term and due to the properties of the \symiiij~symbol in \eqref{eq:ElementaryTwo-1} all components of the mixed tensor operator vanish except of $q=q_{12}+q_{34}$.
Therefore, the final expression implemented in the \ameli\ package is:
\begin{multline}
  \label{eq:EvalTwo}
  \itbra{ab}
  \bigelcross%
  {\cross{\tensor{u}{k_1}_i}{\tensor{t}{k_2}_i}{k_{12}}}%
  {\cross{\tensor{u}{k_3}_j}{\tensor{t}{k_4}_j}{k_{34}}}{k}{q}
  \itket{cd} = 
  \\
  = (-1)^{k+q+k_{12}+k_{34}+l_a+l_b+s_a+s_b-m_{l_c}-m_{l_d}-m_{s_c}-m_{s_d}}
  \\ \times
  \sqrt{(2k+1) (2k_{12}+1) (2k_{34}+1)}\,
  \delta(l_a, l_c)\,
  \delta(l_b, l_d)\,
  \delta(q,m_{l_a}+m_{l_b}-m_{l_c}-m_{l_d}+m_{s_a}+m_{s_b}-m_{s_c}-m_{s_d})
  \\ \times
  \iiij{k_{12}}{k}{k_{34}}%
  {m_{l_a}-m_{l_c}+m_{s_a}-m_{s_c}}{-q}{m_{l_b}-m_{l_d}+m_{s_b}-m_{s_d}}
  \\ \times
  \iiij{k_1}{k_{12}}{k_2}%
  {m_{l_a}-m_{l_c}}{-m_{l_a}+m_{l_c}-m_{s_a}+m_{s_c}}{m_{s_a}-m_{s_c}}
  \iiij{l_a}{k_1}{l_c}{-m_{l_a}}{m_{l_a}-m_{l_c}}{m_{l_c}}
  \iiij{s_a}{k_2}{s_c}{-m_{s_a}}{m_{s_a}-m_{s_c}}{m_{s_c}}
  \\ \times
  \iiij{k_3}{k_{34}}{k_4}%
  {m_{l_b}-m_{l_d}}{-m_{l_b}+m_{l_d}-m_{s_b}+m_{s_d}}{m_{s_b}-m_{s_d}}
  \iiij{l_b}{k_2}{l_d}{-m_{l_b}}{m_{l_b}-m_{l_d}}{m_{l_d}}
  \iiij{s_b}{k_3}{s_d}{-m_{s_b}}{m_{s_b}-m_{s_d}}{m_{s_d}}
  \ .
\end{multline}

Instead of defining a generic three electron operator acting in the orbital and spin space, the \ameli\ package uses the triple scalar product of elementary unit tensor operators in the orbital space acting on three electrons.
This pragmatic choice is based on the fact that we will need only this elementary three-electron operator for one of the perturbation Hamiltonians:
\begin{equation}
  \label{eq:ElementaryThree}
  \ftensor{C}{0} =
  \threedot{\tensor{u}{k_1}_1}{\tensor{u}{k_2}_2}{\tensor{u}{k_3}_3}
  \ .
\end{equation}

The operator~$\ftensor{C}{k}$ is named \texttt{Unit\_UUU} in the software code.
The calculation of the matrix elements of this operator starts with Eq.~\eqref{eq:TripleDot}:
\begin{equation}
  \itbra{abc}
  \threedot{\tensor{u}{k_1}_1}{\tensor{u}{k_2}_2}{\tensor{u}{k_3}_3}
  \itket{def} = 
  \sum_{q_1,q_2,q_3}  
  \iiij{k_1}{k_2}{k_3}{q_1}{q_2}{q_3}
  \itbra{a} \telement{u}{k_1}{q_1} \itket{d}\,
  \itbra{b} \telement{u}{k_2}{q_2} \itket{e}\,
  \itbra{c} \telement{u}{k_3}{q_3} \itket{f}\,
\end{equation}
and then the Wigner-Eckart theorem~\eqref{eq:WignerEckart} is applied to each of the three matrix elements in the sum.
The triple sum is reduced to a single term due to the fact that the \symiiij~symbols are zero if the sum of their lower elements is not equal zero, which leads to the final expression used in the \ameli\ package:
\begin{multline}
  \label{eq:EvalThree}
  \itbra{abc}
  \threedot{\tensor{u}{k_1}_1}{\tensor{u}{k_2}_2}{\tensor{u}{k_3}_3}
  \itket{def} = 
  (-1)^{l_a + l_b + l_c - m_{l_a} - m_{l_b} - m_{l_c}}\,
  \\ \times
  \delta(l_a,l_d)\,
  \delta(l_b,l_e)\,
  \delta(l_c,l_f)\,
  \delta(m_{s_a},m_{s_d})\,
  \delta(m_{s_b},m_{s_e})\,
  \delta(m_{s_c},m_{s_f})
  \iiij{k_1}{k_2}{k_3}{m_{l_a}-m_{l_d}}{m_{l_b}-m_{l_e}}{m_{l_c}-m_{l_f}}
  \\ \times
  \iiij{l_a}{k_1}{l_d}{-m_{l_a}}{m_{l_a}-m_{l_d}}{m_{l_d}}
  \iiij{l_b}{k_2}{l_e}{-m_{l_b}}{m_{l_b}-m_{l_e}}{m_{l_e}}
  \iiij{l_c}{k_3}{l_f}{-m_{l_c}}{m_{l_c}-m_{l_f}}{m_{l_f}}
  \ .
\end{multline}
\end{widetext}

\subsection{Coulomb and Unit Tensor Operators}

All tensor operators which appear in the theory of many-electron systems can be reduced to three basic one-electron operators defined by their reduced matrix elements:
The Coulomb interaction or crystal field operator~\cite{Judd.1963}
\begin{equation}\begin{split}
  \label{eq:def.C}
  \langle l' || \tensor{c}{k} || l \rangle &=
  (-1)^{l'}\sqrt{(2l'+1)(2l+1)}\iiij{l'}{k}{l}{0}{0}{0} \\
  \tensor{C}{k} &= \sum_i \tensor{c}{k}_i
  \ ,
\end{split}\end{equation}
the unit tensor operator in the orbital space~\cite{Judd.1963}
\begin{equation}\begin{split}
  \label{eq:def.U}
  \langle l' || \tensor{u}{k} || l \rangle &= \delta(l',l) \\
  \tensor{U}{k} &= \sum_i \tensor{u}{k}_i
  \ ,
\end{split}\end{equation}
and the unit tensor operator in the spin space~\cite{Judd.1963}
\begin{equation}\begin{split}
  \label{eq:def.T}
  \langle s' || \tensor{t}{k} || s \rangle &= \delta(s',s) = 1 \\
  \tensor{T}{k} &= \sum_i \tensor{t}{k}_i
  \ .
\end{split}\end{equation}

For single-shell configurations $l^N$, not only the spin quantum number $s=1/2$ is the same for all states, but also the quantum number~$l$ of the orbital angular momentum.
This allows us to replace the Coulomb operator by the orbital unit tensor operator:
\begin{equation}
  \label{eq:ck}
  \tensor{c}{k} = (-1)^{l}(2l+1)\iiij{l}{k}{l}{0}{0}{0} \tensor{u}{k}
  \ .
\end{equation}

In order to define the many-electron unit tensor operator~$\tensor{U}{k}$ in terms of our general elementary one-electron operator~\eqref{eq:ElementaryOne} we set $k_1=k$, and $k_2=0$.
The latter reduces the effect of the operator~$\operator{t}$ to a factor $1/\sqrt{2}$, given by Eq.~\eqref{eq:t00}.
When we compensate for this factor, Eq.~\eqref{eq:def.U} translates into the final expression for the \ameli\ software:
\begin{equation}
  \label{eq:U}
  \tensor{U}{k} = \sqrt{2}\, 
  \sum_i \cross{\tensor{u}{k}_i}{\tensor{t}{0}_i}{k}
  \ .
\end{equation}

The parameters for the many-electron unit tensor operator~$\tensor{T}{k}$ are $k_1=0$, and $k_2=k$.
Eq.~\eqref{eq:u00} delivers the compensation factor $\sqrt{2l+1}$ to translate Eq.~\eqref{eq:def.T} into
\begin{equation}
  \label{eq:T}
  \tensor{T}{k} = \sqrt{2l+1}\, 
  \sum_i \cross{\tensor{u}{0}_i}{\tensor{t}{k}_i}{k}
  \ .
\end{equation}

Scalar products of unit tensor operators operators are very common.
They result in the combination of a one-electron and a two-electron operator. The squared operator $\tensor{U}{k}$ is
\begin{equation}
  \dot{\tensor{U}{k}}{\tensor{U}{k}} =
  \sum_i \dot{\tensor{u}{k}_i}{\tensor{u}{k}_i} +
  2\sum_{i<j} \dot{\tensor{u}{k}_i}{\tensor{u}{k}_j}
  \ .
\end{equation}

The one-electron operator according to Eq.~\eqref{eq:reduce-uu-scalar} is a state-independent scalar.
Expansion of the two-electron scalar product to an mixed tensor operator of the type~\eqref{eq:ElementaryTwo} by application of equations \eqref{eq:ScalarProduct-2} and \eqref{eq:t00} results in the compensation factor $(-1)^k(2s+1)\sqrt{2k+1}$.
The final form for the \ameli\ software is therefore
\begin{multline}
  \label{eq:UU}
  \dot{\tensor{U}{k}}{\tensor{U}{k}} =
  \frac{N}{2l+1}\ident +
  (-1)^k\, 4\sqrt{2k+1}
  \\ \times
  \sum_{i<j} \bigelcross%
  {\cross{\tensor{u}{k}_i}{\tensor{t}{0}_i}{k}}%
  {\cross{\tensor{u}{k}_j}{\tensor{t}{0}_j}{k}}{0}{0}
  \ ,
\end{multline}
with the identity matrix $\ident$ and $2s+1=2$.

The same considerations for the squared operator $\tensor{T}{k}$ result in the following expression used by the \ameli\ software:
\begin{multline}
  \label{eq:TT}
  \dot{\tensor{T}{k}}{\tensor{T}{k}} =
  \frac{N}{2}\ident +
  (-1)^k\, 2(2l+1)\sqrt{2k+1}
  \\ \times
  \sum_{i<j} \bigelcross%
  {\cross{\tensor{u}{0}_i}{\tensor{t}{k}_i}{k}}%
  {\cross{\tensor{u}{0}_j}{\tensor{t}{k}_j}{k}}{0}{0}
  \ .
\end{multline}

The mixed scalar product is also important:
\begin{equation}
  \dot{\tensor{U}{k}}{\tensor{T}{k}} =
  \sum_i \dot{\tensor{u}{k}_i}{\tensor{t}{k}_i} +
  2\sum_{i<j} \dot{\tensor{u}{k}_i}{\tensor{t}{k}_j}
  \ .
\end{equation}

The one-electron operator in this equation is obviously not state-independent.
Conversion of the scalar products to mixed tensor operators according to Eq.~\eqref{eq:ScalarProduct-2} delivers a common factor $(-1)^k \sqrt{2k+1}$:
\begin{multline}
  \dot{\tensor{U}{k}}{\tensor{T}{k}} =
  (-1)^k \sqrt{2k+1}
  \\ \times
  \bigg[
  \sum_i \elcross{\tensor{u}{k}_i}{\tensor{t}{k}_i}{0}{0} +
  2\sum_{i<j} \elcross{\tensor{u}{k}_i}{\tensor{t}{k}_j}{0}{0}
  \bigg]
\end{multline}
and the conversion of the two-electron operator in the second term to our general elementary two-electron operator~\eqref{eq:ElementaryTwo} results in another factor $\sqrt{(2s+1)(2l+1)}$.
The final form used by the \ameli\ package is then:
\begin{multline}
  \label{eq:UT}
  \dot{\tensor{U}{k}}{\tensor{T}{k}} =
  (-1)^k \sqrt{2k+1}
  \\ \times
  \bigg[
  \sum_i \elcross{\tensor{u}{k}_i}{\tensor{t}{k}_i}{0}{0} +
  2\sqrt{2(2l+1)}
  \\ \times
  \sum_{i<j} \bigelcross%
  {\cross{\tensor{u}{k}_i}{\tensor{t}{0}_i}{k}}%
  {\cross{\tensor{u}{0}_j}{\tensor{t}{k}_j}{k}}{0}{0}
  \bigg]
  \ .
\end{multline}

\subsection{Angular Momentum Operators}

The relationship between angular momentum operators and unit tensor operators is very close.
The rank of angular momentum operators is always one because of their vectorial nature.
Expressed in terms of the unit tensor operator the elementary orbital angular momentum operator is 
\begin{equation}
  \label{eq:def.L}
  \tensor{l}{1} / \hbar =
  \sqrt{l(l+1)(2l+1)}\,
  \tensor{u}{1}
  \ ,
\end{equation}
and the elementary spin operator
\begin{equation}
  \label{eq:def.S}
  \tensor{s}{1} / \hbar =
  \sqrt{s(s+1)(2s+1)}\,\tensor{t}{1} =
  \sqrt{3/2}\,\tensor{t}{1}
  \ ,
\end{equation}
with the reduced Planck constant $\hbar=h/2\pi$.
It is obviously advisable to define angular momentum operators in units of $\hbar$ in a software implementation. This results in integer or half-integer eigenvalues in terms of the respective quantum numbers.

The definition of the many-electron operator $\operator{L}$ used by the \ameli\ software is based on Eq.~\eqref{eq:U}:
\begin{equation}
  \label{eq:L}
  \tensor{L}{1} / \hbar =
  \sum_i \tensor{l}{1}_i / \hbar =
  \sqrt{l(l+1)(2l+1)}\,
  \tensor{U}{1}
\end{equation}
and Eq.~\eqref{eq:T} delivers the definition of the operator $\operator{S}$:
\begin{equation}
  \label{eq:S}
  \tensor{S}{1} / \hbar =
  \sum_i \tensor{s}{1}_i / \hbar =
  \sqrt{3/2}\,\tensor{T}{1}
  \ .
\end{equation}

The tensor operator of the total angular momentum is obtained as the sum of the 
Eqs.\ \eqref{eq:L} and \eqref{eq:S}:
\begin{equation}
  \label{eq:J}
  \tensor{J}{1} = \tensor{L}{1}+\tensor{S}{1}
  \ .
\end{equation}

The operator $\operator{L}^2$ is commonly used.
Within the \ameli\ software its definition is based on Eq.~\eqref{eq:L}:
\begin{equation}
  \label{eq:LL}
  \dot{\tensor{L}{1}}{\tensor{L}{1}} / \hbar^2 =
  l(l+1)(2l+1)\,
  \dot{\tensor{U}{1}}{\tensor{U}{1}}
\end{equation}
and in the same way the operator $\operator{S}^2$ is based on Eq.~\eqref{eq:S}:
\begin{equation}
  \label{eq:SS}
  \dot{\tensor{S}{1}}{\tensor{S}{1}} / \hbar^2 =
  \frac32\,
  \dot{\tensor{T}{1}}{\tensor{T}{1}}
  \ .
\end{equation}

The scalar product of orbital and spin angular momentum $\operator{L}\operator{S}$ appears for example as part of the squared total angular momentum operator $\operator{J}^2$:
\begin{multline}
  \label{eq:JJ}
  \dot{\tensor{J}{1}}{\tensor{J}{1}} =
  \\ =
  \dot{\tensor{L}{1}}{\tensor{L}{1}} +
  \dot{\tensor{S}{1}}{\tensor{S}{1}} +
  2\dot{\tensor{L}{1}}{\tensor{S}{1}}
  \ .
\end{multline}
The tensor expression $\operator{L}\operator{S}$ used by the \ameli\ software is based on Eqs.\ \eqref{eq:L} and \eqref{eq:S} again:
\begin{equation}
  \label{eq:LS}
  \dot{\tensor{L}{1}}{\tensor{S}{1}} / \hbar^2 =
  \sqrt{\frac{3l(l+1)(2l+1)}{2}}
  \dot{\tensor{U}{1}}{\tensor{T}{1}}
  \ .
\end{equation}

\subsection{Intra-Configuration Hamiltonians}

As explained in Subsection~\ref{sub:perturbation}, the energy-level spectrum of a $f^N$ configuration is defined by the linear combination of angular perturbation Hamiltonians with the respective radial integrals as weight factors, see Eq.~\eqref{eq:Hab}.
Until the 1970s, a comprehensive set of electric and magnetic electronic perturbation operators originating from the Coulomb as well as the spin-orbit interactions in first and second order were established~\cite{Goldschmidt.1978} and since then accepted as reference.
The full set of these operators consists of the six types in Tab.~\ref{tab:perturbation}.
The operators $\operator{H}_1$ to $\operator{H}_6$ are ordered by their relative magnitude.

We are now going to express all six types of Hamiltonians in terms of elementary unit tensor operators, starting with the first-order intra-configuration Hamiltonians $\operator{H}_1$, $\operator{H}_2$, and $\operator{H}_5$ here, and proceeding in Subsection~\ref{sub:2ndOrder} with the second-order inter-configuration Hamiltonians $\operator{H}_3$, $\operator{H}_4$, and $\operator{H}_6$.

\begin{table}
  \caption{\label{tab:perturbation} Free ion interaction Hamiltonians for the configuration~$f^N$ of Coulomb type (electric) or relativistic spin-orbit type (magnetic) with operator name, radial integrals,  perturbation order, and number of involved electrons (particle-rank)}
  \renewcommand{\arraystretch}{1.2}
  \begin{tabular}{|c|c|c|c|}
    \hline
    Operator & Radial Integrals & Order & Electrons \\
    \hline
    $\operator{H}_1$ & $F^2,F^4,F^6$ & 1 & 2\\
    $\operator{H}_2$ & $\zeta$ & 1 & 1 \\
    $\operator{H}_3$ & $\alpha,\beta,\gamma$ & 2 & 2 \\
    $\operator{H}_4$ & $T^2,T^3,T^4,T^6,T^7,T^8$ & 2 & 3 \\
    $\operator{H}_5$ & $M^0,M^2,M^4$ & 1 & 2 \\
    $\operator{H}_6$ & $P^2,P^4,P^6$ & 2 & 2 \\
  \hline
  \end{tabular}
\end{table}

The Coulomb interaction between the electrons delivers by far the largest contribution to the separation of energy levels~\cite{Wybourne.1965}.
The part of the full Coulomb operator missing in the free ion Hamiltonian in Eq.~\eqref{eq:H0} is $\operator{H}'_1=\operator{H}'-\operator{H}_0$:
\begin{equation}
  \label{eq:H1p}
  \operator{H}'_1 =
  -\sum_{i} U(r_i)
  + \frac{e^2}{4\pi\epsilon_0}
  \left(
  -\sum_{i}\frac{Ze}{r_i}
  +\sum_{i<j}\frac{1}{r_{ij}}
  \right)
  \ .
\end{equation}

The single electron terms of this operator are pure radial terms without any angular dependence.
Their contribution to all states is therefore the same and they deliver only a spectral offset, which we will ignore.
An energy-level fit takes this contribution into account by fixing the ground-state energy or the barycenter energy of the whole spectrum.
The remaining vector operator is~\cite{Wybourne.1965}
\begin{equation}
  \label{eq:H1.vector}
  \operator{H}_1 =
  \frac{e^2}{4\pi\epsilon_0}
  \sum_{i<j}\frac{1}{r_{ij}}
  \ .
\end{equation}

In terms of elementary spherical tensor operators this translates into~\cite{Wybourne.1965}
\begin{equation}
  \label{eq:H1.tensor}
  \operator{H}_1 =
  \frac{e^2}{4\pi\epsilon_0}
  \sum_k
  \sum_{i<j}
  \frac{r_<^k}{r_>^{k+1}}\,
  \dot{\tensor{c}{k}_i}{\tensor{c}{k}_j}
  \ ,
\end{equation}
where the abbreviation $r_<$ stands for the smaller and $r_>$ the larger one of $r_i$ and $r_j$.

After evaluating the radial Slater integrals~\cite{Slater.1960,Slater.1960b}
\begin{equation}
  \label{eq:SlaterIntegral}
  F^k =
  \frac{e^2}{4\pi\epsilon_0}
  \int\!\!\!\int 
  \frac{r_<^k}{r_>^{k+1}}\,
  R_{nl}^2(r_i) R_{nl}^2(r_j)\,
  dr_i dr_j
\end{equation}
for first order perturbations, we are left with a set of angular Coulomb operators
\begin{equation}
  \label{eq:fk}
  \operator{f}_k =
  \sum_{i<j}
  \dot{\tensor{c}{k}_i}{\tensor{c}{k}_j}
  \ ,
\end{equation}
which in terms of elementary unit tensor operators translates to
\begin{equation}
  \label{eq:fk-unit-1}
  \operator{f}_k =
  \cmat{k}^2
  \sum_{i<j}
  \dot{\tensor{u}{k}_i}{\tensor{u}{k}_j}
  \ .
\end{equation}

According to Eq.~\eqref{eq:def.C}, the reduced matrix element of the elementary Coulomb operator is
\begin{equation}
  \cmat{k} = (-1)^{l}(2l+1)\iiij{l}{k}{l}{0}{0}{0}
  \ .
\end{equation}

We use equations \eqref{eq:ScalarProduct-2} and \eqref{eq:t00} to expand the scalar product and obtain the final expression for the \ameli\ software
\begin{multline}
  \label{eq:fk-unit}
  \operator{f}_k =
  (-1)^{k}\,2\sqrt{2k+1}\,
  \cmat{k}^2
  \\ \times
  \sum_{i<j} \bigelcross%
  {\cross{\tensor{u}{k}_i}{\tensor{t}{0}_i}{k}}%
  {\cross{\tensor{u}{k}_j}{\tensor{t}{0}_j}{k}}{0}{0}
  \ .
\end{multline}

The \symiiij~symbol in this equation restricts $k$ to even values from $0$ to $2l$.
We also ignore $k=0$, since it results in a common offset for all states.
The Coulomb interaction inside the $f^N$ configuration in first order is therefore determined by three angular operators with $k=2,4,6$.
The eigenstates of the operator
\begin{equation}
  \operator{H}_1 = \sum_{k=2,4,6} F^k \operator{f}_k
\end{equation}
are actually $LS$-states.
 
The spin-orbit interaction is also contributing substantially to the splitting of energy levels and must never be omitted in an energy-level fit.
The vector operator of the spin-orbit interaction is~\cite{Judd.1963}
\begin{equation}
  \label{eq:SpinOrbit.vector}
  \operator{H}_2 =
  \frac{1}{2m_e^2c^2}
  \sum_i
  \dot{\vector{s}_i}%
  {(\vector{\nabla}_i U(r_i)\times\vector{p}_i)}
  \ ,
\end{equation}
which after the evaluation of the gradient translates into the spherical tensor operator~\cite{Judd.1963}
\begin{equation}
  \label{eq:SpinOrbit.tensor}
  \operator{H}_2 =
  \frac{1}{2m_e^2c^2}
  \sum_i
  \frac{1}{r_i}\,
  \frac{d}{d r_i} U(r_i)\,
  \dot{\tensor{s}{1}_i}{\tensor{l}{1}_i}
  \ .
\end{equation}

After evaluating the radial integral
\begin{equation}
  \label{eq:zeta}
  \zeta =
  \frac{\hbar^2}{2m_e^2c^2}
  \int
  \frac{1}{r}\,
  \frac{d}{dr} U(r)\,
  R_{nl}^2(r)\,
  dr
\end{equation}
for first order perturbations, we are left with the angular spin-orbit operator
\begin{equation}
  \label{eq:z-1}
  \operator{z} =
  \frac{1}{\hbar^2}
  \sum_i
  \dot{\tensor{s}{1}}{\tensor{l}{1}}
  \ .
\end{equation}

Since $\tensor{s}{1}$ and $\tensor{l}{1}$ are proportional to the elementary unit
tensor operators $\tensor{t}{1}$ and $\tensor{u}{1}$ respectively and the order does not matter, we get from Eqs.\ \eqref{eq:def.L} and \eqref{eq:def.S}:
\begin{equation}
  \label{eq:z-2}
    \operator{z} =
    \sqrt{\frac{3l(l+1)(2l+1)}{2}}\,
    \sum_i
    \dot{\tensor{u}{1}}{\tensor{t}{1}}
\end{equation}
and the final expression for the \ameli\ software using Eq.~\eqref{eq:ScalarProduct-2}:
\begin{equation}
  \label{eq:z}
    \operator{z} =
    -3\sqrt{\frac{l(l+1)(2l+1)}{2}}\,
    \sum_i
    \elcross{\tensor{u}{1}}{\tensor{t}{1}}{0}{0}
  \ .
\end{equation}

The spin-orbit operator is mixing spin and orbital angular momentum and is thus not diagonal in the $LS$-state space. The characteristic quantity of eigenstates of the operator
\begin{equation}
  \operator{H}_2 = \zeta \operator{z}
\end{equation}
is the quantum number~$J$ of the total angular momentum.
This quantum number is conserved by all Hamilton operators in Tab.~\ref{tab:perturbation}, because they are all scalar operators.
The $J$-symmetry is only broken when external fields or crystal fields introduce a directional reference frame.

The perturbation Hamiltonian $\operator{H}_5$ covers the interaction of the spin of one electron with either the spin or the orbital angular momentum of another electron.
The former is called spin-spin interaction, the later spin-other-orbit interaction:
\begin{equation}
  \label{eq:H5}
  \operator{H}_5 = \operator{H}_{ss} + \operator{H}_{soo}
  \ .
\end{equation}
This operator is very weak and, in contrast to $\operator{H}_1$ and $\operator{H}_2$, only considered in first perturbation order.

We will see that although the angular parts of both interactions are different, they share the same radial integral.
We start with the vector operator expression of the spin-spin interaction~\cite{Horie.1953,Judd.1963,Goldschmidt.1978}:
\begin{equation}
  \label{eq:Hss.vector}
    \operator{H}_{ss}
    =
    \frac{1}{4\pi\epsilon_0}\,
    \frac{2\beta_m^2}{\hbar^2 c^2}
    \sum_{i \neq j}
    \bigg[
    \frac{\dot{\vector{s}_i}{\vector{s}_j}}{r_{ij}^{3}}
    -
    \frac{3 \dot{\vector{r}_{ij}}{\vector{s}_i}
      \dot{\vector{r}_{ij}}{\vector{s}_j}}{r_{ij}^5}
  \bigg]
\end{equation}
with the Bohr magneton $\beta_m=e\hbar/2m_e$ and the speed of light~$c$. This equation may be
transformed to a tensor operator form as shown by Judd~\cite{Judd.1963}:
\begin{multline}
  \label{eq:Hss.tensor}
  \operator{H}_{ss}
  =
  -\frac{1}{4\pi\epsilon_0}\,
  \frac{2\beta_m^2}{\sqrt{5}\hbar^2 c^2}
  \sum_k (-1)^k \sqrt{\frac{(2k+5)!}{(2k)!}} \\
  \times \sum_{i<j}
  \frac{r_<^k}{r_>^{k+3}}
  \bigdot{\cross{\tensor{c}{k}_i}{\tensor{c}{k+2}_j}{2}}%
  {\cross{\tensor{s}{1}_i}{\tensor{s}{1}_j}{2}}
  \ .
\end{multline}

After evaluating the Marvin integrals~\cite{Marvin.1947} for a $l^N$ configuration
\begin{equation}
  \label{eq:MarvinIntegral}
  M^k =
  \frac{1}{4\pi\epsilon_0}\,
  \frac{\beta_m^2}{2 c^2}
  \int\!\!\!\int 
  \frac{r_<^k}{r_>^{k+3}}
  R_{nl}^2(r_i) R_{nl}^2(r_j)\,
  dr_i dr_j
  \ ,
\end{equation}
we obtain the angular part of the spin-spin operator
\begin{multline}
  \operator{m}_{k,ss} =
  (-1)^{k+1} \frac{4}{\hbar^2} \sqrt{\frac{(2k+5)!}{5(2k)!}} \\
  \times \sum_{i<j}
  \bigdot{\cross{\tensor{c}{k}_i}{\tensor{c}{k+2}_j}{2}}%
  {\cross{\tensor{s}{1}_i}{\tensor{s}{1}_j}{2}}
\end{multline}
and in terms of unit tensor operators according to equations \eqref{eq:def.C} and \eqref{eq:S}:
\begin{multline}
  \operator{m}_{k,ss}
  =
  -12 \cmat{k} \cmat{k+2}\\
  \times \sqrt{\frac{(k+1)(k+2)(2k+1)(2k+3)(2k+5)}{5}} \\
  \times \sum_{i<j}
  \bigdot{\cross{\tensor{u}{k}_i}{\tensor{u}{k+2}_j}{2}}%
  {\cross{\tensor{t}{1}_i}{\tensor{t}{1}_j}{2}}
  \ .
\end{multline}

We convert the scalar product to a tensor product by application of Eq.~\eqref{eq:ScalarProduct-2}, which gives a factor $\sqrt{5}$ and re-order the resulting mixed tensor operator pair-wise without changing its value according to Eqs.\ \eqref{eq:ReCouple.ABCD.zero} and \eqref{eq:6j.special}.
This leads to the final expression for the \ameli\ software:
\begin{multline}
  \label{eq:mkss}
  \operator{m}_{k,ss}
  =
  -12 \cmat{k} \cmat{k+2}\\
  \times \sqrt{(k+1)(k+2)(2k+1)(2k+3)(2k+5)} \\
  \times \sum_{i<j}
  \bigelcross%
  {\cross{\tensor{u}{k}_i}{\tensor{t}{1}_i}{k+1}}%
  {\cross{\tensor{u}{k+2}_j}{\tensor{t}{1}_j}{k+1}}{0}{0}
  \ ,
\end{multline}
where the \symiiij~symbol in the reduced matrix element of $\tensor{c}{k+2}$ restricts the rank $k$ to the even values between 0 and $2l-2$ inside the $l^N$ configuration.

The vector operator expression of the spin-other-orbit operator is~\cite{Judd.1963,Goldschmidt.1978}
\begin{equation}
  \label{eq:Hsoo.vector}
    \operator{H}_{soo}
    =
    \frac{1}{4\pi\epsilon_0}\,
    \frac{2\beta_m^2}{\hbar}
    \sum_{i \neq j}
    \bigdot{[\vector{\nabla}_{\!i}\frac{1}{r_{ij}}
      \!\times\!
      \vector{p}_i]}%
    {[\vector{s}_i \!+\! 2\vector{s}_j]}
    \ .
\end{equation}

The general tensor operator expression given by Goldschmidt~\cite{Goldschmidt.1978} is rather large, but inside an $l^N$ configuration only two terms are not vanishing:
\begin{multline}
  \label{eq:Hsoo.tensor}
  \operator{H}_{soo} =
  \frac{1}{4\pi\epsilon_0}\,
  \frac{2\beta_m^2}{\sqrt{3}\hbar^2 c^2} \sum_k (-1)^k (2k+1)
  \\
  \times \sum_{i < j} \big( \big[
  \frac{r_<^{k-2}}{r_>^{k+1}} \sqrt{2k-1}
  \, \cross{\tensor{c}{k}_j}{\cross{\tensor{c}{k}_i}{\tensor{l}{1}_i}{k-1}}{1}
  \\
  -\frac{r_<^{k}}{r_>^{k+3}} \sqrt{2k+3}
  \, \cross{\tensor{c}{k}_j}{\cross{\tensor{c}{k}_i}{\tensor{l}{1}_i}{k+1}}{1}
  \big]
  \\
  \cdot \big[
  \tensor{s}{1}_i + 2\tensor{s}{1}_j
  \big] \big)
  \ .
\end{multline}

The substitution $k \to k+2$ in the first of the two tensor terms shows that the radial integral of this interaction is again the Marvin integral from Eq.~\eqref{eq:MarvinIntegral}.
The angular part of the spin-other-orbit operator is then
\begin{multline}
  \operator{m}_{k,soo}
  =
  (-1)^k\frac{4}{\hbar^2} \sqrt{\frac{2k+3}{3}}
  \\
  \times \sum_{i < j} \big( \big[
  (2k+5)\,
  \cross{\tensor{c}{k+2}_j}{\cross{\tensor{c}{k+2}_i}{\tensor{l}{1}_i}{k+1}}{1}
  \\[-0.333\baselineskip]
  -
  (2k+1)\,
  \cross{\tensor{c}{k}_j}{\cross{\tensor{c}{k}_i}{\tensor{l}{1}_i}{k+1}}{1}
  \big]
  \\
  \cdot \big[
  \tensor{s}{1}_i + 2\tensor{s}{1}_j
  \big] \big)
  \ .
\end{multline}

Inside a $l^N$ configuration the rank $k$ is again limited to even values between 0 and $2l-2$ and we get the following unit tensor operator expression with positive sign factor:
\begin{multline}
  \operator{m}_{k,soo}
  =
  4 \sqrt{\frac{l(l+1)(2l+1)(2k+3)}{2}}
  \\
  \times
  \sum_{i<j} \big( \big[
  (2k+5)\, \cmat{k+2}^2
  \\
  \times \cross{\tensor{u}{k+2}_j}%
  {\cross{\tensor{u}{k+2}_i}{\tensor{u}{1}_i}{k+1}}{1}
  \\
  -(2k+1)\, \cmat{k}^2
  \\
  \times \cross{\tensor{u}{k}_j}%
  {\cross{\tensor{u}{k}_i}{\tensor{u}{1}_i}{k+1}}{1}
  \big]
  \\
  \cdot \big[
  \tensor{t}{1}_i + 2\tensor{t}{1}_j
  \big] \big)
  \ ,
\end{multline}

Both tensor terms in this expression contain a tensor product of two orbital unit tensor operators acting on the same electron $i$.
Such products can be evaluated as shown in Eq.~\eqref{eq:reduce-uu} and the result is
\begin{multline}
  \operator{m}_{k,soo}
  =
  -\sqrt{2(2k+3)}
  \\
  \times
  \sum_{i<j} \big( \big[
  \sqrt{(2l+k+3)(2l-k-1)(k+2)(2k+5)}
  \\
  \times \cmat{k+2}^2 \cross{\tensor{u}{k+1}_i}{\tensor{u}{k+2}_j}{1}
  \\[0.333\baselineskip]
  +
  \sqrt{(2l+k+2)(2l-k)(k+1)(2k+1)}
  \\
  \times \cmat{k}^2 \cross{\tensor{u}{k+1}_i}{\tensor{u}{k}_j}{1}
  \big]
  \\[0.333\baselineskip]
  \cdot \big[
  \tensor{t}{1}_i + 2\tensor{t}{1}_j
  \big] \big)
  \ .
\end{multline}

Expanding the scalar product of two sums of tensors results in four terms:
\begin{multline}
  \operator{m}_{k,soo}
  =
  -\sqrt{2(2k+3)}
  \\
  \times \Bigg[
  \cmat{k}^2 \sqrt{(2l+k+2)(2l-k)(k+1)(2k+1)}
  \\
  \times \Big[
  \sum_{i<j}
  \bigdot{\cross{\tensor{u}{k+1}_i}{\tensor{u}{k}_j}{1}}{\tensor{t}{1}_i}
  \\
  + 2\sum_{i<j}
  \bigdot{\cross{\tensor{u}{k+1}_i}{\tensor{u}{k}_j}{1}}{\tensor{t}{1}_j}
  \Big]
  \\
  +\cmat{k+2}^2 \sqrt{(2l+k+3)(2l-k-1)(k+2)(2k+5)}
  \\
  \times \Big[
  \sum_{i<j}
  \bigdot{\cross{\tensor{u}{k+1}_i}{\tensor{u}{k+2}_j}{1}}{\tensor{t}{1}_i}
  \\
  + 2 \sum_{i<j}
  \bigdot{\cross{\tensor{u}{k+1}_i}{\tensor{u}{k+2}_j}{1}}{\tensor{t}{1}_j}
  \Big]
  \Bigg]
  \ ,
\end{multline}
which allows us to use the symmetry properties of the scalar product and swap the indices $i$ and $j$ in the first and third tensor term:
\begin{multline}
  \operator{m}_{k,soo}
  =
  -\sqrt{2(2k+3)}
  \\
  \times \bigg[
  \cmat{k}^2 \sqrt{(2l+k+2)(2l-k)(k+1)(2k+1)}
  \\
  \times \Big[
  \sum_{i<j}
  \bigdot{\cross{\tensor{u}{k}_i}{\tensor{u}{k+1}_j}{1}}{\tensor{t}{1}_j}
  \\
  + 2\sum_{i<j}
  \bigdot{\cross{\tensor{u}{k+1}_i}{\tensor{u}{k}_j}{1}}{\tensor{t}{1}_j}
  \Big]
  \\
  +\cmat{k+2}^2 \sqrt{(2l+k+3)(2l-k-1)(k+2)(2k+5)}
  \\
  \times \Big[
  \sum_{i<j}
  \bigdot{\cross{\tensor{u}{k+2}_i}{\tensor{u}{k+1}_j}{1}}{\tensor{t}{1}_j}
  \\
  + 2 \sum_{i<j}
  \bigdot{\cross{\tensor{u}{k+1}_i}{\tensor{u}{k+2}_j}{1}}{\tensor{t}{1}_j}
  \Big]
  \bigg]
  \ .
\end{multline}

In the final step, we convert the scalar product to a scalar tensor product, which according to Eq.~\eqref{eq:ScalarProduct-2} gives a factor $-\sqrt{3}$.
We get another factor $\sqrt{2}$ by expanding the tensor products to mixed tensor operators of the type~\eqref{eq:ElementaryTwo} using Eq.~\eqref{eq:t00} and yet another factor of $+1$ from Eq.~\eqref{eq:ReCouple.ABCD.zero} when we reorder the four-fold tensor operators with odd rank pair-wise.
This results in the final expression for the \ameli\ software:
\begin{multline}
  \label{eq:mksoo}
  \operator{m}_{k,soo}
  =
  2\sqrt{3(2k+3)}
  \\
  \times \bigg[
  \cmat{k}^2 \sqrt{(2l+k+2)(2l-k)(k+1)(2k+1)}
  \\
  \times \Big[
  \sum_{i<j}
  \bigelcross%
  {\cross{\tensor{u}{k}_i}{\tensor{t}{0}_i}{k}}%
  {\cross{\tensor{u}{k+1}_j}{\tensor{t}{1}_j}{k}}{0}{0}
  \\
  + 2\sum_{i<j}
  \bigelcross%
  {\cross{\tensor{u}{k+1}_i}{\tensor{t}{0}_i}{k+1}}%
  {\cross{\tensor{u}{k}_j}{\tensor{t}{1}_j}{k+1}}{0}{0}
  \Big]
  \\
  +\cmat{k+2}^2 \sqrt{(2l+k+3)(2l-k-1)(k+2)(2k+5)}
  \\
  \times \Big[
  \sum_{i<j}
  \bigelcross%
  {\cross{\tensor{u}{k+2}_i}{\tensor{t}{0}_i}{k+2}}%
  {\cross{\tensor{u}{k+1}_j}{\tensor{t}{1}_j}{k+2}}{0}{0}
  \\
  + 2 \sum_{i<j}
  \bigelcross%
  {\cross{\tensor{u}{k+1}_i}{\tensor{t}{0}_i}{k+1}}%
  {\cross{\tensor{u}{k+2}_j}{\tensor{t}{1}_j}{k+1}}{0}{0}
  \Big]
  \bigg]
  \ .\hspace{-1em}
\end{multline}

The total angular integral of the operator~$\operator{H}_5$ in Eq.~\eqref{eq:H5} is the sum of the angular integrals of the spin-spin and spin-other-orbit interaction:
\begin{equation}
  \label{eq:mk}
  \operator{m}_k = \operator{m}_{k,ss} + \operator{m}_{k,soo}
  \ .
\end{equation}

\subsection{Inter-Configuration Hamiltonians}
\label{sub:2ndOrder}

Second-order perturbations take the influence of excited electron configurations into account.
It may seem at first that this leads to an unlimited number of possibilities, but systematic investigations showed that only a small number of configurations can actually deliver contributions to second order Hamiltonians.
Since fundamental physical interactions always connect two partners, potential configurations are restricted to those differing to $f^N$ in no more than two electrons.
Furthermore, the interacting configuration must have the same parity as the respective $f^N$ configuration in order to contribute to a scalar Hamiltonian.

Tab.~\ref{tab:CIconf} contains all five types of potentially interacting configurations in the general case of an $l^N$ configuration as discussed in the literature~\cite{Rajnak.1963,Racah.1967,Goldschmidt.1978}.
The first three configurations (a)-(c) differ by a single electron and the other two by two electrons, both taken (d) or given (e) by the base shell $l$.
Rajnak and Wybourne explain in Ref.~\onlinecite{Rajnak.1963} that the three additional types of configurations also differing by two electrons, but in which no or one electron is taken or given by the base shell, do not deliver other contributions than shifting the energy-level spectrum as a whole.

\begin{table}
    \caption{\label{tab:CIconf} All types of configurations which may interact with $l^N$. On the left of $l$ are electrons removed from former closed shells ($N'=4l'+2$) and on the right there are electrons in former empty shells ($N'=0$).}
    \renewcommand{\arraystretch}{1.2}
    \begin{tabular}{cc}
      (a) & $l'^{4l'+1}l^{N+1}$ \\
      (b) & $l'^{4l'+1}l^Nl''$ \\
      (c) & $l^{N-1}l'$ \\
      (d) & $l'^{4l'}l^{N+2}$ and $l'^{4l'+1}l''^{4l''+1}l^{N+2}$ \\
      (e) & $l^{N-2}l'^2$ and $l^{N-2}l'l''$
    \end{tabular}
\end{table}

The sum $\operator{H}_1+\operator{H}_2$ of the dominating Coulomb and spin-orbit  interactions inserted into Eq.~\eqref{eq:Hp2nd} of second order perturbation leads to three terms~\cite{Goldschmidt.1978} of the types $\operator{H}_1\operator{H}_1/\Delta E$, $\operator{H}_1\operator{H}_2/\Delta E$, and $\operator{H}_2\operator{H}_2/\Delta E$.
The first pure Coulomb operator was introduced by Trees~\cite{Trees.1951}.
It is represented by an effective two-electron operator~$\operator{H}_3$ and an effective three-electron operator~$\operator{H}_4$ and the second mixed Coulomb and spin-orbit operator delivers the effective two-electron operator~$\operator{H}_6$.
It was shown that the angular dependence of the last pure spin-orbit operator in second order is equivalent to the first order spin-orbit operator~\cite{Rajnak.1964}.
Its effect is thus already taken into account by $\operator{H}_2$.

In $f^N$ configurations the operator $\operator{H}_3$ consists of three terms~\cite{Rajnak.1963}
\begin{equation}
  \label{eq:H3}
  \operator{H}_3 = \alpha\operator{L}^2/\hbar^2 + \beta \operator{C}_2(\text{G}_2) + \gamma \operator{C}_2(\text{SO}(7))
  \ ,
\end{equation}
with the radial integrals $\alpha$, $\beta$, and $\gamma$.

The first term contains the squared total orbital angular momentum operator~$\operator{L}^2$, whose tensor expression is given in Eq.~\eqref{eq:LL}.

The angular operator of the second term is the quadratic Casimir operator of the special group $\text{G}_2$, which is given by~\cite{Racah.1949,Judd.1963}
\begin{equation}
  \label{eq:H3b}
  \operator{C}_2(\text{G}_2) =
  \frac14\,
  \sum_{k=1,5}
  \sqrt{2k+1}
  \dot{\tensor{U}{k}}{\tensor{U}{k}}
  \ .
\end{equation}

Finally, Eq.~\eqref{eq:H3} contains the quadratic Casimir operator of the special orthogonal group in $2l+1=7$ dimensions $\text{SO}(7)$.
In general, this operator is expressed as~\cite{Racah.1949,Judd.1963}
\begin{equation}
  \label{eq:H3c}
  \operator{C}_2(\text{SO}(2l+1)) =
  \frac{1}{2l-1}\,
  \sum_{\substack{k=1 \\ k\ \text{odd}}}^{2l-1}
  (2k+1)\,
  \dot{\tensor{U}{k}}{\tensor{U}{k}}
  \ .
\end{equation}

Within the \ameli\ package all three operators are implemented directly based on the scalar products of unit tensor operators as defined in Eq.~\eqref{eq:UU}.
Note that the general implementation of Casimirs operator of the orthogonal group in $2l+1$ dimensions may also be applied to the $d^N$ configuration in which the second order Coulomb operator~$\operator{H}_3$ is given by~\cite{Rajnak.1963}
\begin{equation}
  \operator{H}_3 = \alpha\operator{L}^2/\hbar^2 + \beta \operator{C}_2(\text{SO}(5))
  \ .
\end{equation}
For the sake of completeness it should be noted that the same operator in the $p^N$ configuration has only one term~\cite{Trees.1951,Rajnak.1963}:
\begin{equation}
  \operator{H}_3 = \alpha\operator{L}^2/\hbar^2
  \ .
\end{equation}

The effective three-electron Coulomb operator in second order~\cite{Judd.1966} involves six radial integrals $T^c$:
\begin{equation}
  \label{eq:H4}
  \operator{H}_4 =
  \!\!\!\sum_{c=2,3,4,6,7,8}\!\!\!
  T^c \operator{t}_c
  \ .
\end{equation}
Each of the operators $\operator{t}_c$ is a linear combination of triple scalar products:
\begin{multline}
  \label{eq:tc}
  \operator{t}_c =
  \!\sum_{k,k',k''}\!
  \langle kk'k''|c\rangle\,
  \sqrt{(2k+1)(2k'+1)(2k''+1)}\, \\
  \times 6 \!\sum_{h<i<j}\!
  \threedot{\tensor{u}{k}_h}{\tensor{u}{k'}_i}{\tensor{u}{k''}_j}
\end{multline}
where the three ranks $k$, $k'$, and $k''$ run over all even integer values from zero to $2l$ due to their origin from the Coulomb operator $\tensor{c}{k}$.
However, the value zero can be excluded, because a triple scalar product with scalar tensor factors would actually not be a three-electron operator.
Furthermore, the definition of the triple scalar product in Eq.~\eqref{eq:TripleDot} shows that it does not depend on the order of the three ranks and the \symiiij~symbol imposes the triangle condition on their values.
In total this means that the rank sum in Eq.~\eqref{eq:tc} needs to be evaluated for the nine triples 222, 224, 244, 246, 266, 444, 446, 466, and 666 only.

\begin{table*}[t]
  \caption{\label{tab:Judd-table} Vector coupling coefficient $\langle kk'k''|c\rangle$ with values compiled from Ref.~\onlinecite{Judd.1966}}
  \renewcommand{\arraystretch}{1.8}
\begin{tabularx}{\textwidth}{l >{\centering\arraybackslash}X >{\centering\arraybackslash}X >{\centering\arraybackslash}X >{\centering\arraybackslash}X >{\centering\arraybackslash}X >{\centering\arraybackslash}X >{\centering\arraybackslash}X >{\centering\arraybackslash}X >{\centering\arraybackslash}X}
\hline
$kk'k''$ & $c=1$ & 2 & 3 & 4 & 5 & 6 & 7 & 8 & 9 \\ \hline
222 & $-\sqrt{\frac{11}{1134}}$ & $\sqrt{\frac{605}{5292}}$ & $\sqrt{\frac{32761}{889056}}$ & $\sqrt{\frac{3575}{889056}}$ & $-\sqrt{\frac{17303}{396900}}$ & $-\sqrt{\frac{1573}{8232}}$ & $\sqrt{\frac{264407}{823200}}$ & $\sqrt{\frac{21879}{274400}}$ & $-\sqrt{\frac{46189}{231525}}$ \\
224 & $\sqrt{\frac{4}{189}}$ & $-\sqrt{\frac{6760}{43659}}$ & $\sqrt{\frac{33}{1372}}$ & $-\sqrt{\frac{325}{37044}}$ & $\sqrt{\frac{416}{33075}}$ & $-\sqrt{\frac{15028}{305613}}$ & $\sqrt{\frac{28717}{2778300}}$ & $-\sqrt{\frac{37349}{926100}}$ & $-\sqrt{\frac{8398}{694575}}$ \\
244 & $\sqrt{\frac{1}{847}}$ & $-\sqrt{\frac{1805}{391314}}$ & $-\sqrt{\frac{4}{33957}}$ & $-\sqrt{\frac{54925}{373527}}$ & $-\sqrt{\frac{117}{296450}}$ & $\sqrt{\frac{4693}{12326391}}$ & $-\sqrt{\frac{1273597}{28014525}}$ & $\sqrt{\frac{849524}{9338175}}$ & $-\sqrt{\frac{134368}{3112725}}$ \\
246 & $\sqrt{\frac{26}{3267}}$ & $-\sqrt{\frac{4160}{754677}}$ & $-\sqrt{\frac{13}{264}}$ & $\sqrt{\frac{625}{26136}}$ & $\sqrt{\frac{256}{571725}}$ & $\sqrt{\frac{1568}{107811}}$ & $\sqrt{\frac{841}{1960200}}$ & $-\sqrt{\frac{17}{653400}}$ & $-\sqrt{\frac{15827}{245025}}$ \\
444 & $-\sqrt{\frac{6877}{139755}}$ & $\sqrt{\frac{55016}{717409}}$ & $\sqrt{\frac{49972}{622545}}$ & $\sqrt{\frac{92480}{1369599}}$ & $\sqrt{\frac{178802}{978285}}$ & $-\sqrt{\frac{297680}{5021863}}$ & $-\sqrt{\frac{719104}{2282665}}$ & $-\sqrt{\frac{73644}{2282665}}$ & $-\sqrt{\frac{2584}{18865}}$ \\
446 & $\sqrt{\frac{117}{1331}}$ & $-\sqrt{\frac{195}{204974}}$ & $\sqrt{\frac{52}{1089}}$ & $\sqrt{\frac{529}{11979}}$ & $-\sqrt{\frac{2025}{18634}}$ & $-\sqrt{\frac{49}{395307}}$ & $-\sqrt{\frac{1369}{35937}}$ & $\sqrt{\frac{68}{11979}}$ & $0$ \\
266 & $\sqrt{\frac{2275}{19602}}$ & $\sqrt{\frac{1625}{143748}}$ & $\sqrt{\frac{325}{199584}}$ & $\sqrt{\frac{6889}{2195424}}$ & $\frac{71}{198}$ & $-\sqrt{\frac{1}{223608}}$ & $\sqrt{\frac{625}{81312}}$ & $\sqrt{\frac{1377}{27104}}$ & $\sqrt{\frac{323}{22869}}$ \\
466 & $\sqrt{\frac{12376}{179685}}$ & $\sqrt{\frac{88400}{1185921}}$ & $-\sqrt{\frac{442}{12705}}$ & $-\sqrt{\frac{10880}{251559}}$ & $-\sqrt{\frac{1088}{179685}}$ & $-\sqrt{\frac{174080}{8301447}}$ & $-\sqrt{\frac{8704}{3773385}}$ & $-\sqrt{\frac{103058}{1257795}}$ & $-\sqrt{\frac{19}{31185}}$ \\
666 & $\sqrt{\frac{4199}{539055}}$ & $\sqrt{\frac{29393}{790614}}$ & $\sqrt{\frac{205751}{784080}}$ & $-\sqrt{\frac{79135}{1724976}}$ & $\sqrt{\frac{2261}{1078110}}$ & $\sqrt{\frac{79135}{175692}}$ & $\sqrt{\frac{15827}{319440}}$ & $-\sqrt{\frac{8379}{106480}}$ & $-\sqrt{\frac{98}{1485}}$ \\
\hline
$WU$ & (000)(00) & (220)(22) & (222)(00) & (222)(40) & (400)(40) & (420)(22) & (420)(40) & (420)(42) & (600)(60) \\
\hline
\end{tabularx}
\end{table*}

The factor $\langle kk'k''|c\rangle$ was calculated by Judd et al.~\cite{Judd.1966} as vector coupling coefficient based on the symmetries of the Coulomb operator in second order.
Its values are given in Tab.~\ref{tab:Judd-table} together with the respective irreducible representations $W$ and $U$ of the rotational group in seven dimensions $\text{SO}(7)$ and the special group $\text{G}_2$ used by Judd.
Note that the full set contains nine indices~$c$, but Judd explains that only the six indices in Eq.~\eqref{eq:H4} actually relate to three-electron operators.
The other three indices result in operators which are proportional to linear combinations of the two-electron operators contained in $\operator{H}_1$.
In Subsection~\ref{sub:elementwise} we use this fact for testing the \ameli\ code.

The electrostatic spin-orbit interaction operator~$\operator{H}_6$ was introduced by Rajnak et al.~\cite{Rajnak.1964} and corrected by Judd et al.~\cite{Judd.1968} in their work on magnetic interactions.
The expressions later given by Goldschmidt~\cite{Goldschmidt.1978} fit well into this work,
except of her radial integrals $Q^k$, which are an unfortunate choice from an theoretical point of view, because they mix radial and angular parameters.
Therefore we use the parameters $P^k$ instead~\cite{Judd.1968}:
\begin{equation}
  \label{eq:H6}
  \operator{H}_6 = \sum_{k=2,4,6} P^k \operator{p}_k
  \ .
\end{equation}
They are related to $Q^k$ by
\begin{equation}
  \label{eq:Qk-Pk}
  Q^k = \frac{\cmat{k}^2}{6} P^k
  \ .
\end{equation}

Using this definition, the effective electrostatic spin-orbit operator given by Goldschmidt is~\cite{Goldschmidt.1978}
\begin{multline}
  \label{eq:H_ELSO}
  \operator{H}_\text{EL-SO} =
  -\frac{1}{3\hbar}
  \sum_{k\ \text{even}}
  P^k
  \cmat{k}^2
  \sqrt{\frac{l(l+1)(2l+1)}{2k+1}}
  \\
  \times
  \sum_{t\ \text{odd}}
  (2t+1)
  \vij{1}{k}{t}{l}{l}{l}
  \dot{\tensor{U}{k}}{\tensortop{T}{1t}{k}}
\end{multline}
containing the scalar product of the orbital unit tensor operator $\tensor{U}{k}$ and the coupled tensor operator~\cite{Goldschmidt.1978}
\begin{equation}
  \tensortop{T}{1t}{k} = \sum_{i}\cross{\tensor{s}{1}_i}{\tensor{u}{t}_i}{k}
  \ .
\end{equation}

Inserting this definition into Eq.~\eqref{eq:H_ELSO} results in a one-electron term which is obviously proportional to the first order spin-orbit interaction~$\operator{H}_2$ and a two-electron term, which delivers the angular part of the effective electrostatic spin-orbit operator from Eq.~\eqref{eq:H6}
\begin{multline}
  \label{eq:H6.1}
  \operator{p}_k =
  -\frac{2}{3\hbar}
  \cmat{k}^2
  \sqrt{\frac{l(l+1)(2l+1)}{2k+1}}
  \\
  \times
  \sum_{t\ \text{odd}}
  (2t+1)
  \vij{1}{k}{t}{l}{l}{l}
  \ \sum_{i<j}
  \bigdot{\tensor{u}{k}_i}{\cross{\tensor{s}{1}_j}{\tensor{u}{t}_j}{k}}
  \ .
\end{multline}

The \symiiij~symbol in the reduced matrix element of $\tensor{c}{k}$ restricts the rank $k$ to even values between 0 and $2l$ as usual, but since $\operator{p}_0$ is proportional to the spin-orbit operator, only the values $k=2,4,6$ are relevant. 
Furthermore, the odd rank~$t$ is restricted to $t=k\pm1$ by the triangle conditions of the \symvij~symbol.
A substitution of the spin operator by the unit tensor operator~$\tensor{t}{1}$ according to Eq.~\eqref{eq:def.S} gives
\begin{multline}
  \operator{p}_k =
  -\cmat{k}^2
  \sqrt{\frac{2l(l+1)(2l+1)}{3(2k+1)}}
  \\
  \times
  \Big[
  (2k-1)
  \vij{1}{k}{k-1}{l}{l}{l}
  \sum_{i<j}\bigdot{\tensor{u}{k}_i}{\cross{\tensor{t}{1}_j}{\tensor{u}{k-1}_j}{k}}
  \\
  +
  (2k+3)
  \vij{1}{k}{k+1}{l}{l}{l}
  \sum_{i<j}\bigdot{\tensor{u}{k}_i}{\cross{\tensor{t}{1}_j}{\tensor{u}{k+1}_j}{k}}
  \Big]
  \ .
\end{multline}

The lean structure of the \symvij~symbols in this expression results in rather simple algebraic expressions:
\begin{multline}
  \label{eq:6j.spec.m}
  \vij{1}{k}{k-1}{l}{l}{l} =
  \frac{(-1)^{2l+k}}{2}
  \\ \times
  \sqrt{\frac{(2l+k+1) (2l-k+1) k}{l (l+1) (2l+1) (2k-1) (2k+1)}}
\end{multline}
\begin{multline}
  \label{eq:6j.spec.p}
  \vij{1}{k}{k+1}{l}{l}{l} =
  \frac{(-1)^{2l+k+1}}{2}
  \\ \times
  \sqrt{\frac{(2l+k+2) (2l-k) (k+1)}{l (l+1) (2l+1) (2k+1) (2k+3)}}
  \ .
\end{multline}

We insert these and convert the scalar products to scalar tensor operators, which according to Eq.~\eqref{eq:ScalarProduct-2} gives another factor $\sqrt{2k+1}$:
\begin{multline}
  \operator{p}_k =
  -\cmat{k}^2
  \frac{1}{\sqrt{6(2k+1)}}
  \\ \times
  \Big[
  \sqrt{(2l+k+1) (2l-k+1) k (2k-1)}
  \\ \times
  \sum_{i<j}\bigelcross{\tensor{u}{k}_i}%
  {\cross{\tensor{u}{k-1}_j}{\tensor{t}{1}_j}{k}}{0}{0}
  \\ -
  \sqrt{(2l+k+2) (2l-k) (k+1) (2k+3)}
  \\ \times
  \sum_{i<j}\bigelcross{\tensor{u}{k}_i}%
  {\cross{\tensor{u}{k+1}_j}{\tensor{t}{1}_j}{k}}{0}{0}
  \Big]
  \ ,
\end{multline}
where we also reordered the mixed tensor products of the electron~$j$, which according to Eq.~\eqref{eq:ReOrder} does not change their signs.

In the final step we expand the tensor operator for electron~$i$ by a scalar unit tensor operator in the spin space, coming with a factor $\sqrt{2}$.
The expression used in the \ameli\ software is then
\begin{multline}
  \label{eq:pk}
  \operator{p}_k =
  -\cmat{k}^2
  \frac{1}{\sqrt{3(2k+1)}}
  \\ \times
  \Big[
  \sqrt{(2l+k+1) (2l-k+1) k (2k-1)}
  \\ \times
  \sum_{i<j}\bigelcross%
  {\cross{\tensor{u}{k}_i}{\tensor{t}{0}_i}{k}}%
  {\cross{\tensor{u}{k-1}_j}{\tensor{t}{1}_j}{k}}{0}{0}
  \\ -
  \sqrt{(2l+k+2) (2l-k) (k+1) (2k+3)}
  \\ \times
  \sum_{i<j}\bigelcross%
  {\cross{\tensor{u}{k}_i}{\tensor{t}{0}_i}{k}}%
  {\cross{\tensor{u}{k+1}_j}{\tensor{t}{1}_j}{k}}{0}{0}
  \Big]
  \ .
\end{multline}

\subsection{Crystal-field Hamiltonians}
\label{sub:crystal}

The approach of Wybourne~\cite{Wybourne.1965} to model the local field at the site of a lanthanide ion in a crystalline host uses the same spherical tensor operators~$\telement{C}{k}{q}$ that are employed to expand the intra-atomic Coulomb interaction~\cite{Smentek.1998,Fiorucci.2025}.
The crystal field Hamiltonian in Wybourne notation is written as
\begin{equation}
  \label{eq:cf}
  \operator{H}_\mathrm{cf} = \sum_{k} \sum_{q=-k}^{k} B^k_q \telement{C}{k}{q}
  \ ,
\end{equation}
where $B^k_q$ are the crystal field parameters.
In contrast to the purely radial integrals of the free-ion Hamiltonian, these crystal field parameters incorporate the local anisotropy of the host and are, in general, complex-valued.
The requirement that the Hamiltonian must be Hermitian to ensure real energy eigenvalues restricts these parameters to the relationship:
\begin{equation}
  \label{eq:Bkq}
  B^k_{-q} = (-1)^q B^{k*}_q
  \ .
\end{equation}
For local sites where the coordinate system can be chosen such that the crystal field parameters are strictly real-valued, the \ameli\ repository utilizes a condensed representation.
By exploiting the real-valued nature of the operator matrix elements within this framework, a convenient, self-adjoint crystal field operator is defined as:
\begin{equation}
  \label{eq:Hcf}
  \telement{\hat{C}}{k}{q} = 
  \begin{cases} 
    \telement{C}{k}{0} & \text{if } q = 0 \\
    \telement{C}{k}{q} + (-1)^q \telement{C}{k}{-q} & \text{if } q > 0 
  \end{cases}
  \ .
\end{equation}

\section{Classification of States}
\label{sec.coupling}

Adopting Slater determinants as basis states significantly streamlines the automated evaluation of Hamiltonian matrix elements for $f^N$ configurations.
In contrast, the traditional manual approach employs an $LS$-coupling scheme, where the spin and orbital angular momenta of $N$ electrons are coupled to a total spin~$\operator{S}$ and total orbital angular momentum~$\operator{L}$.
Because the full Hamiltonian is diagonal in neither the uncoupled product space nor the $LS$-space, the physical eigenstates of lanthanide systems are necessarily mixed.
They are formally described as states in intermediate coupling.

The diagonalization of the Hamiltonian yields a discrete set of eigenvalues and corresponding eigenvectors for these intermediate states.
Each eigenvalue defines an energy level, while the eigenvector components serve as expansion coefficients for the basis states used to construct the intermediate state.
Consequently, the eigenvectors generated in our framework represent linear combinations of product states, differing from the $LS$-state expansions conventional in the literature.
This represents a notable divergence from standard notation, as intermediate states are traditionally labeled according to the $LS$-component that provides the largest contribution to the expansion.

To resolve this discrepancy, we implement a linear transformation from the uncoupled product space to the $LS$-coupling representation, applied to each operator matrix immediately following its computation.
This transformation ensures that all numerical results and state assignments remain consistent with established literature and standard spectroscopic nomenclature.

\subsection{Transformation Matrix}
\label{sub:transform}

A state in $LS$-coupling is typically specified by the set of quantum numbers $\gamma S L J M$, where $J$ and $M$ denote the magnitude and magnetic quantum numbers of the total angular momentum operator, respectively.
In this representation, the total spin quantum number $S$ corresponds to the irreducible representation of the state under the special unitary group~$\text{SU}(2)$, while the total orbital angular momentum quantum number $L$ labels the irreducible representation under the special orthogonal group~$\text{SO}(3)$.

Strictly speaking, for an $N$-electron configuration, each group is an $N$-fold direct product, for example $\text{SU}(2)^N$ for the total spin.
However, this exponent is conventionally omitted for brevity, a notation we adopt in this work.

To distinguish between states characterized by identical $S$ and $L$ values, the additional label $\gamma$ is required.
Racah~\cite{Racah.1949} demonstrated that for an $l^N$ configuration, $\gamma$ can be largely replaced by irreducible representations of the state associated with a nested chain of symmetry subgroups.

Although mainly of mathematical and not physical relevance, the unambiguous classification of each $LS$-state using these additional quantum numbers turned out to be particularly helpful in the traditional calculation approach.
Because these labels are integral to the formal identity of each state, we must determine them concurrently with the physical observables $S$, $L$, $J$, and $M$.

From a mathematical perspective, the general linear group $\text{GL}(4l+2)$ comprises the most expansive set of transformations for non-relativistic electronic states.
However, many of these transformations lack physical significance, as they may result in states that violate fundamental physical symmetries. To isolate physically permissible transformations, Racah utilized a nested chain of symmetry groups~\cite{Racah.1949}:
\begin{multline}
  \label{eq:Groups-1}
  \text{GL}(4l+2)
  \supset \text{U}(4l+2) \\
  \supset \text{Sp}(4l+2)
 \supset \text{SU}_S(2) \times \text{SO}_L(2l+1)
  \ .
\end{multline}

This reduction leads to the direct product of the special unitary group $\text{SU}(2)$ for transformations in spin space and the special orthogonal group $\text{SO}(2l+1)$ for transformations in orbital space.

Within this framework, an $LS$-state is characterized by its irreducible representations: the quantum number $S$ under $\text{SU}(2)$ and the set of $l$ integers $W=(w_1\ldots w_l)$ under $\text{SO}(2l+1)$.
Alternatively, the seniority number $v$ can be employed, which labels the irreducible representations of the symplectic group $\text{Sp}(4l+2)$.
Consequently, utilizing either the pair $(S,W)$ or $(v,S)$ provides equal schemes for state classification.

Racah followed the chain of symmetry groups in the orbital space further according to~\cite{Racah.1949}
\begin{equation}
  \label{eq:Groups-2}
  \text{SO}(2l+1) \supset \text{G}_2 \supset \text{SO}(3)
  \ .
\end{equation}
The irreducible representation of a state with respect of the special group $\text{G}_2$ is the pair of integers $U=(u_1u_2)$ and $L$ with respect to the rotational group $\text{SO}(3)$ in three dimensions.

It turns out that this set of quantum numbers almost unambiguously identifies each $LS$-state of the configuration $f^N$. It is only for $N=5-9$ that some pairs of states remain unresolved. An additional quantum number $\tau$ is traditionally used to label these states with the values $A$ or $B$~\cite{Nielson.1963}.

Our goal is now to evaluate the expansion coefficients for the unitary transformation between the uncoupled product basis and the $LS$-coupling representation:
\begin{multline}
  \label{eq:LStrans}
  |SWU\tau LJM\rangle =
  \sum_{\alpha_1\alpha_2\ldots\alpha_N}
  |\{\alpha_1\alpha_2\ldots\alpha_N\}\rangle
  \\ \times
  \langle\alpha_1\alpha_2\ldots\alpha_N|SWU\tau LJM\rangle\,
  \ .
\end{multline}

This procedure is implemented by systematically descending the symmetry group chain, beginning with the $\text{SU}(2)$ spin symmetry.
Since the total spin operator $\operator{S}^2$ must be diagonal in the $LS$-coupling scheme, the first step involves computing the matrix elements of the dimensionless operator $\operator{S}^2/\hbar^2$ within the product space.
Using the angular basis states $|\Phi_i\rangle$, the matrix $M_1$ is defined as:
\begin{equation}
  \label{eq:M1}
  [M_1]_{ab} =
  \langle\Phi_a|\dot{\tensor{S}{1}}{\tensor{S}{1}}/\hbar^2|\Phi_b\rangle
\end{equation}
where the elements are evaluated according to Eq.~\eqref{eq:SS}.
The diagonalization of $M_1$ yields:
\begin{equation}
  \label{eq:M1s}
  \Lambda_1 = V_1^T M_1\,V_1
  \ ,
\end{equation}
where the diagonal matrix $\Lambda_1$ contains the eigenvalues $S(S+1)$, and $V_1$ represents the corresponding matrix of eigenvectors for the operator~$\operator{S}^2$.

In the subsequent computational steps, the columns of the transformation matrix $V_1$ must be ordered such that the corresponding eigenvalues in $\Lambda_1$ follow a consistent sorting scheme (usually ascending numerical order).
While most modern numerical libraries for eigendecomposition perform this sorting implicitly, it is a critical prerequisite for the systematic classification of states within the subgroup chain.

The generic eigendecomposition routines provided by the \texttt{SymPy} library are not suitable for the matrices in exact arithmetic encountered in this framework.
Consequently, rather than employing a global matrix diagonalization routine, \ameli\ utilizes a specialized, step-wise diagonalization algorithm.
For full algorithmic implementation details, the reader is referred to the source code within the \ameli\ software repository~\cite{Caspary.2026}.

The next stage in the symmetry reduction involves the group $\text{SO}(2l+1)$.
Specifically, for $f^N$ configurations, we utilize the quadratic Casimir operator of $\text{SO}(7)$ as given in Eq.~\eqref{eq:H3c}:
\begin{equation}
  \label{eq:M2}
  [M_2]_{ab} =
  \langle\Phi_a|\operator{C}_2(\text{SO}(7))|\Phi_b\rangle
  \ .
\end{equation}

The representation of this operator in the basis defined by the eigenvectors of $\operator{S}^2$ is obtained via the transformation:
\begin{equation}
  \label{eq:M2s}
  M'_2 = V_1^T M_2\,V_1
  \ .
\end{equation}

Because $\text{SO}(7)$ is a subgroup of the product space $\text{SU}(2) \times \text{SO}(7)$, the matrix $M'_2$ is necessarily block-diagonal with respect to the subspaces spanned by the distinct eigenvalues of $\operator{S}^2$.
Consequently, $M'_2$ possesses non-vanishing elements only within submatrices located along the principal diagonal.
Each of these submatrices corresponds to a specific $S(S+1)$ eigenvalue found in the diagonal elements of $\Lambda_1$.

In the subsequent step, each of these submatrices is individually diagonalized, with the resulting eigenvectors arranged according to sorted eigenvalues.
These local transformation matrices are then assembled into a global matrix, $V_2$, which maintains the same dimensionality and block-diagonal structure as $M'_2$.
The resulting diagonal matrix,
\begin{equation}
  \label{eq:M2ss}
  \Lambda_2 = V_2^TV_1^T M_2\,V_1V_2
\end{equation}
contains the eigenvalues of the quadratic Casimir operator $\operator{C}_2(\text{SO}(7))$.
These eigenvalues are uniquely determined by the irreducible representations~$W=(w_1 \ldots w_l)$ according to the eigenvalue equation~\cite{Racah.1949,Judd.1963}:
\begin{multline}
  \operator{C}_2(\text{SO}(2l+1))\, |W\rangle =
  \\ =
  \frac{1}{2(2l-1)}\sum_{i=1}^{l} w_i(w_i + 2l+1 -2i)\, |W\rangle
  \ .
\end{multline}
The complete set of irreducible representations $W$ for $f^N$ configurations, along with their corresponding eigenvalues, is provided in Tab.~\ref{tab:Eigen.WU}.
For computational implementation, scaling these eigenvalues by a factor of 5 is advantageous, as it maps the results onto a set of integers.

\begin{table}
    \caption{\label{tab:Eigen.WU} Irreducible representations $W$ and $U$ of $f^N$ states with respect to the symmetry groups $\text{SO}(7)$ and $\text{G}_2$ and the respective eigenvalues of the quadratic Casimir operators.}
    \renewcommand{\arraystretch}{1.1}
    \begin{tabular}{c|cc@{\hspace{2em}}c|cc@{\hspace{2em}}c|c}
    $W$ & $\operator{C}_2(\text{SO}(7))$
    && $U$ & $\operator{C}_2(\text{G}_2)$ 
    && $U$ & $\operator{C}_2(\text{G}_2)$ \\
    \cline{1-2}\cline{4-5}\cline{7-8}
    $(000)$ & $0$    && $(00)$ & $0$   && $(40)$ & $3$    \\
    $(100)$ & $3/5$  && $(10)$ & $1/2$ && $(41)$ & $15/4$ \\
    $(110)$ & $1$    && $(11)$ & $1$   && $(42)$ & $14/3$ \\
    $(111)$ & $6/5$  && $(20)$ & $7/6$ && $(43)$ & $23/4$ \\
    $(200)$ & $7/5$  && $(21)$ & $7/4$ && $(44)$ & $7$    \\
    $(210)$ & $9/5$  && $(22)$ & $5/2$ \\
    $(211)$ & $2$    && $(30)$ & $2$   \\
    $(220)$ & $12/5$ && $(31)$ & $8/3$ \\
    $(221)$ & $13/5$ && $(32)$ & $7/2$ \\
    $(222)$ & $3$    && $(33)$ & $9/2$ 
    \end{tabular}
\end{table}

In certain contexts, the seniority number $v$, which labels the irreducible representations of the symplectic group $\text{Sp}(4l+2)$, is used as an alternative to the $W$ notation.
For $l^N$ configurations, the pairs $(S,W)$ and $(v,S)$ exist in a bijective relationship, as previously explained.
This numerical correspondence is defined by the following relations~\cite{Racah.1949,Judd.1963,Wybourne.1965}:
\begin{align}
  v &= 2c + 2S \\
  d &= \min(2S, 2l+1 - v)
  \ ,
\end{align}
where $c$ and $d$ denote the multiplicities of the integers '2' and '1', respectively, within the sequence $w_1\ldots w_l$ with the remaining $l-c-d$ components $w_i$ being zero.

The next symmetry group in the reduction chain is the exceptional group $\text{G}_2$.
Accordingly, we consider the quadratic Casimir operator $\operator{C}_2(\text{G}_2)$ as defined in Eq.~\eqref{eq:H3b}:
\begin{equation}
  \label{eq:M3}
  [M_3]_{ab} =
  \langle\Phi_a|\operator{C}_2(\text{G}_2)|\Phi_b\rangle
  \ .
\end{equation}

The representation of this operator in the basis transformed by $V_1$ and $V_2$ is computed as:
\begin{equation}
  \label{eq:M3s}
  M'_3 = V_2^TV_1^T M_3\,V_1V_2
  \ .
\end{equation}

By virtue of the subgroup hierarchy, the matrix~$M'_3$ must be diagonal with respect to the subspaces defined by the eigenvalues of both $\operator{S}^2$ and $\operator{C}_2(\text{SO}(7))$.
Consequently, the block-diagonal structure of~$M'_3$ is a refinement of that observed in $M'_2$, where each submatrix now corresponds to a unique pair of irreducible representations $(S,W)$.

Each of these submatrices is individually diagonalized, with the resulting eigenvectors ordered by their eigenvalues.
These local transformations are then consolidated into a global matrix~$V_3$, which preserves the dimensionality and refined block structure of $M'_3$.
The resulting diagonal matrix,
\begin{equation}
  \label{eq:M3sss}
  \Lambda_3 = V_3^TV_2^TV_1^T M_3\,V_1V_2V_3
  \ ,
\end{equation}
contains the eigenvalues of $\operator{C}_2(\text{G}_2)$.
These are related to the irreducible representations~$U=(u_1u_2)$ via the eigenvalue equation~\cite{Racah.1949,Judd.1963}:
\begin{multline}
  \operator{C}_2(\text{G}_2)\, | U \rangle =
  \\ =
  \frac{1}{12} (u_1^2 + u_2^2 + u_1u_2 + 5u_1 + 4u_2)\, | U \rangle
  \ .
\end{multline}

The complete set of irreducible representations $U$ for $f^N$ configurations and their associated eigenvalues are tabulated in Tab.~\ref{tab:Eigen.WU}.
In numerical implementations, scaling these eigenvalues by a factor of 12 is advantageous, as it yields a set of pure integers.

The next stage in the symmetry reduction involves the $\text{SO}(3)$ group.
Following the established procedure, we diagonalize the dimensionless total orbital angular momentum operator~$\operator{L}^2/\hbar^2$, as given in Eq.~\eqref{eq:LL}, within the $SWU$ subspaces.
This yields the eigenvalues $L(L+1)$ and the corresponding transformation matrix~$V_4$.
Subsequently, the total angular momentum operator $\operator{J}^2/\hbar^2$, according to Eq.~\eqref{eq:JJ}, is diagonalized within the $SWUL$ subspaces, providing the eigenvalues $J(J+1)$ and the transformation matrix~$V_5$.

In the final step of the algorithm, the eigenvalues of the longitudinal angular momentum operator~$\operator{J}_z$ are determined by diagonalizing the rank-1 tensor component~$\telement{J}{1}{0}/\hbar$ from Eq.~\eqref{eq:J} within the $SWULJ$ subspaces.
This process identifies the magnetic quantum numbers~$M$ as eigenvalues and produces the final transformation matrix~$V_6$.

As previously noted, this successive diagonalization scheme leaves a small number of state pairs with identical $SWULJM$ quantum numbers unresolved in the $f^5$ through $f^9$ configurations.
To distinguish these degenerate states, we assign the ad hoc labels A and B as an auxiliary quantum number~$\tau$.

The culmination of this stepwise procedure is the construction of the total transformation matrix~$V$:
\begin{multline}
  V = V_1 V_2 V_3 V_4 V_5 V_6 =
  \\ =
  \langle\alpha_1\alpha_2\ldots\alpha_N|SWU\tau LJM\rangle\
  \ .
\end{multline}
In this representation, each column of V corresponds to a coupled $LS$-basis state $|SWU\tau LJM\rangle$, while each row corresponds to a specific uncoupled product state $|\{\alpha_1\alpha_2\ldots\alpha_N\}\rangle$.

While the labels $W$, $U$, and $\tau$ are primarily of mathematical rather than physical significance, they are frequently replaced by a single integer index to differentiate states with identical $S$ and $L$ values.
A prominent example is found in the comprehensive tables of states and matrix elements published by Nielson and Koster~\cite{Nielson.1963}, where states of the same $S$ and $L$ are indexed sequentially according to the lexicographic order of the $W$, $U$, and $\tau$ eigenvalues.

It should be emphasized that the specific chain of tensor operators described here is tailored to $f^N$ configurations.
However, analogous symmetry chains for other single-shell configurations, summarized in Tab.~\ref{tab:chains}, follow a nearly identical hierarchical logic~\cite{Judd.1963}.

\begin{table}
    \caption{\label{tab:chains} Chains of tensor operators representing different single-shell configurations~\cite{Judd.1963}.}
    \renewcommand{\arraystretch}{1.2}
    \begin{tabular}{|c|l|}
    \hline
    Configuration & Tensor Operators \\
    \hline
    $p^N$ & $\vector{S}^2$, $\vector{L}^2$, $\vector{J}^2$, $\velement{J}{0}$ \\
    $d^N$ & $\vector{S}^2$, $\operator{C}_2(\text{SO}(5))$, $\vector{L}^2$, $\vector{J}^2$, $\velement{J}{0}$ \\
    $f^N$ & $\vector{S}^2$, $\operator{C}_2(\text{SO}(7))$, $\operator{C}_2(\text{G}_2)$, $\vector{L}^2$, $\vector{J}^2$, $\velement{J}{0}$ \\
    \hline
    \end{tabular}
\end{table}

\subsection{Phase Consistency}
\label{sub:phase}

The transformation matrix defined above is highly effective for determining the $LS$-composition of intermediate-coupling states during energy-level fits of lanthanide ions, which is an important objective of this work.
However, the eigendecomposition of scalar operators inherently results in arbitrary phases for the column vectors.
This occurs because the phase of an eigenvector is not uniquely determined by the eigenvalue equation, leading to a random phase distribution across the resulting basis states.

To establish a uniform and computationally elegant sign convention within each $J$-multiplet, we utilize the standard angular momentum lowering operator $J_{-1}$.
This operator acts sequentially on the states within a given multiplet to propagate the phase from the stretched state $|\gamma J J\rangle$ acting as reference state.

For each $J$-multiplet, a subsequent state $|\gamma J M-1\rangle$ is generated by applying the lowering operator to the current state $|\gamma J M\rangle$ according to the standard relation:
\begin{equation}
  \label{eq:ladder-phase}
  J_{-1} |\gamma J M\rangle = \hbar \sqrt{\frac{J(J+1)-M(M-1)}{2}} \, |\gamma J M-1\rangle \ ,
\end{equation}
where $\gamma$ represents the collective quantum numbers $SWU\tau L$ required to uniquely identify the state.
The factor $1/\sqrt{2}$ distinguishes this spherical tensor expression from its counterpart for cartesian vector operators.

This process is applied iteratively, descending from $M=J$ down to $M=-J$ while determining all column vectors with synchronized relative signs within the $J$-multiplet completely and unambiguously.
Because this algorithm is executed entirely within an exact arithmetic framework, accumulation of numerical errors or degradation of floating-point precision is inherently prevented.

While this ladder-operator approach offers a direct and highly efficient route to phase alignment, it should be noted that using the Wigner-Eckart theorem as in Ref.~\onlinecite{Caspary.2002} for the same purpose is equivalent, albeit more complicated.
It utilizes the fact that inside a $J$-multiplet the reduced matrix element is a constant which is independent of the sign convention:
\begin{equation}
  \label{eq:Reduced-U}
  \langle \gamma J || \tensor{A}{k} || \gamma J \rangle = 
  (-1)^{M'-J}
  \frac{\langle \gamma J M' | \telement{A}{k}{M'-M} | \gamma J M \rangle}%
  {\iiij{J}{k}{J}{-M'}{M'-M}{M}}
  \ .
\end{equation}
The theorem thus provides a convenient independent tool for checking a given sign synchronization.

Upon completion of the phase-correction stage, the transformation matrix~$V$ is fully defined and standardized.
Any tensor matrix~$Q$ initially evaluated in the uncoupled product space of Slater determinants can be rigorously mapped into the $LS$-coupling representation via the similarity transformation:
\begin{equation}
  \label{eq:transform}
  Q' = V^T Q\, V
  \ .
\end{equation}

Consequently, the eigendecomposition of the Hamiltonian in the $LS$-basis $H' = V^T H\, V$, yields an energy spectrum identical to that of the original product-space Hamiltonian $H$.
However, the resulting eigenvectors now represent linear combinations of $LS$-coupled basis states, providing a direct and physically intuitive description of states in intermediate coupling.

It is important to emphasize that this specific approach to intermediate coupling is strictly valid for lanthanide ions in amorphous environments, where the effective rotational site symmetry precludes the coupling of states with differing $J$ quantum numbers.
Under these symmetry constraints, it is possible to employ Hamiltonian matrices constructed using only a single stretched state from each $J$-multiplet.
These reduced matrices are identified by the label \verb"SLJ" within the \ameli\ software package.
They significantly diminish the computational overhead required for energy-level fits in amorphous hosts.

In contrast, lanthanides in crystalline hosts generally reside in sites lacking full rotational symmetry.
In such cases, the crystal field induces $J$-mixing, meaning that states in intermediate coupling necessarily encompass $LS$-components with different $J$ values.
This requires a modification of the aforementioned procedure: the total Hamiltonian is constructed from interaction operators directly within the uncoupled product space.
Following diagonalization, the resulting eigenvectors, rather than the Hamiltonian itself, are then transformed into the $LS$-coupling representation.

\section{Data Management}
\label{sec.data}

In Section~\ref{sec.tensors}, we derived the analytical expressions required to compute the angular matrix elements for all spherical tensor operators within a given electron configuration.
These matrices contain constant elements that are uniquely determined by the configuration's symmetry.
The \ameli\ software package~\cite{Caspary.2026} evaluates these elements using exact arithmetic, representing them as signed roots of rational numbers to avoid the precision loss associated with floating-point approximations.

Because these constants are invariant once calculated, the \ameli\ source code was designed with a documentary philosophy.
Its implementation prioritizes structural clarity and exhaustive internal documentation over raw computational throughput.
The software uses Python and its exact arithmetic is based on the SymPy package. 

The primary output of this work is a comprehensive library of angular matrix elements archived in the Zenodo repository~\cite{Caspary.2026b}.
High-level applications should interface directly with these precomputed constant matrices rather than invoking \ameli\ code.
The repository provides dedicated datasets for each lanthanide configuration from $f^1$ to $f^{13}$ alongside a separate archive containing both current and legacy releases of the \ameli\ software package.
All datasets are organized under the Zenodo community AMELI, which serves as a centralized framework for the digital assets of this project.

Given the significant volume of these constants, we have implemented a robust data management strategy to ensure long-term utility and accessibility.
Our approach follows the established FAIR principles of research data management:
\begin{description}
\item[Findable] The tensor operator matrices and supplementary data are archived in the Zenodo repository~\cite{Caspary.2026b}, while the \ameli\ source code is hosted on GitHub~\cite{Caspary.2026} and release versions are stored on Zenodo as well.
The Zenodo resources are assigned a common persistent \gls{doi} to ensure long-term discoverability.
\item[Accessible] Data can be retrieved via standard web interfaces or programmatically through standardized \glspl{api}. Open access is guaranteed under the CC-BY-SA license for the datasets and the MIT license for the software.
\item[Interoperable] All data are stored in structured containers that include comprehensive metadata.
The used file formats ZIP, JSON, and HDF5 follow established industry standards for long-term scientific data preservation.
\item[Reusable] Documentation is provided at multiple levels to ensure long-term usability: within this manuscript, on the repository landing pages, and via embedded metadata in each individual data container, which typically represents a single matrix.
\end{description}

\section{Computational Workflow}
\label{sec.workflow}

The evaluation of matrix elements for a given electronic configuration within the \ameli\ software follows a linear pipeline.
Having established the theoretical framework for each segment in the Sections~\ref{sec.product} to \ref{sec.coupling}, we provide a detailed overview of this workflow below.

\subsection{Product States (module \texttt{\upshape config.py})}

The given electron configuration defines the pool of single electron states available for constructing the many-electron product states.
Each single electron state is characterized by the shell identification number $n$ and the standard quantum numbers $l$, $m_l$, $s$, and $m_s$.
For lanthanide ions, $l=3$ and $s=1/2$.
While \ameli\ supports arbitrary configurations, the pool for lanthanide ions follows the standard ordering defined in Eq.~\eqref{eq:StdOrder}.
Many-electron product states are then represented as lists of indices pointing to the single-electron pool.

The \ameli\ software generates both, the single-electron basis and the corresponding many-electron index lists as an initial step, archiving them in a dedicated data container \verb"config.zdc" for subsequent pipeline stages.
Consistent with all \ameli\ data structures, this container is enriched with extensive metadata to ensure interoperability and reusability in alignment with the FAIR data principles.

\subsection{Non-Zero Elements (module \texttt{\upshape product.py})}

In the subsequent workflow stage, \ameli\ identifies the lists of potentially non-zero matrix elements.
As discussed in Subsection~\ref{sub:Preparation}, these lists depend exclusively on the size of the single-electron pool and the particle-rank of the operator.
Consequently, the software generates three distinct lists for one-, two-, and three-electron operators, archiving them in the data containers \verb"product_{n}.zdc", where $n=1,2,3$ is the respective particle-rank.

Such a list, identified as \verb"indices" within the \ameli\ source code, contains the indices of the initial and final many-electron states along with the number of elementary matrix elements contributing to each full matrix element.
The specific parameters for these elementary elements are determined according to the rules detailed in Subsection~\ref{sub:Calculation}.
Each parameter set consists of $n$ single-electron indices extracted from the initial state, $n$ indices from the final state, and a sign flag representing the permutation parity of the respective elementary matrix element.
These parameter sets are consolidated into a comprehensive list named \verb"elements".

Given the combinatorial growth of the lists \verb"indices" and \verb"elements", which can easily exceed physical memory limits, \ameli\ bypasses memory-intensive operations by generating the datasets directly within an HDF5 file.
This allows for the management of massive datasets through memory-mapped I/O, ensuring the software remains performant even for complex configurations.

It is important to note that this stage of the workflow is entirely configuration-agnostic.
It does not depend on the quantum numbers of the single electron states in the pool.
The \ameli\ code also poses no limits on the particle-rank $n$.

\subsection{Unit Matrices (module \texttt{\upshape unit.py})}

The subsequent pipeline stage involves calculating the matrix elements for the mixed unit tensor operators acting on one, two, or three electrons.
In the \ameli\ code these matrices are labeled \verb"Unit_UT", \verb"Unit_UTUT", and \verb"Unit_UUU".

The evaluation of each potentially non-zero matrix element from the previously generated list \verb"indices" is reduced to a summation of elementary matrix elements as prescribed by Eqs.\ \eqref{eq:EvalOne}, \eqref{eq:EvalTwo}, and \eqref{eq:EvalThree}, depending on the respective particle-rank~$n$.
The parameters for these elementary operations are retrieved from the corresponding entries in the list \verb"elements".

It turns out that a vast majority of the potentially non-zero matrix elements are in fact identically zero.
One of the primary advantages of the exact arithmetic approach of the \ameli\ code is the unambiguous identification of these vanishing elements.
The code exploits the resulting high degree of sparsity by storing only non-vanishing elements and their coordinates according to the \gls{coo} standard for sparse matrices.
Furthermore, \ameli\ utilizes the inherent symmetry of the tensor operator matrices.
Rather than storing redundant values of matrix elements, it stores an index into a list of unique values, which is typically much smaller than the number of non-zero elements.

The algebraic definition of the \symiiij~symbol ensures that any matrix element~$v$ can be represented exactly as the signed square root of a rational number:
\begin{equation}
    v = (-1)^s \sqrt{\frac{n}{d}}
    \ .
\end{equation}
In all matrix data containers the value of a matrix element is stored as a triplet of unsigned integers: the sign flag $s$, the numerator $n$, and the denominator $d$.
In the rare event that these values exceed the 64-bit range of standard computational architectures, the integers are split bit-wise, allowing for exact reconstruction in an application software without precision loss.

While the current implementation is optimized for single-shell configurations like the lanthanides, the logic in the module \verb"unit.py" is fundamentally configuration-agnostic.
In multi-shell environments, the direct representation of the Coulomb tensor operator $\tensor{c}{k}$ via the unit tensor operator $\tensor{u}{k}$ in Eq.~\eqref{eq:def.C} is no longer valid.
However, the modular architecture of the software allows for the natural expansion of the module \verb"unit.py" to include more complex elementary mixed tensors, including $\tensor{c}{k}$, if required.

\subsection{Many-Electron Matrices (module \texttt{\upshape matrix.py})}

The mixed unit tensor operator matrices obtained from the previous step are of limited utility for direct practical applications.
However, they serve as the foundation for constructing the matrices of functional tensor operators.
This construction step consists exclusively of high-level matrix additions and multiplications, avoiding operations at the individual matrix-element level.
To enhance code readability, the \ameli\ package utilizes sparse matrix objects provided by the SymPy library.
All tensor operators currently supported by the software and provided by the \ameli\ repository are listed in Tab.~\ref{tab:matrices}, alongside their corresponding equations in this manuscript.

\begin{table*}[t]
  \caption{\label{tab:matrices} Unit, angular momentum, and perturbation Hamilton tensor operators provided by the \ameli\ repository. The file name of each matrix is derived from its given name by replacing the characters '\texttt{/}' and '\texttt{,}' by '\texttt{\_}'.}
  \renewcommand{\arraystretch}{1.1}
  \begin{tabularx}{\textwidth}{|c|c|X|c|}
    \hline
    Operator & Name & Description & Equation \\
    \hline
$\telement{U}{k}{q}$ & \verb"U/{k},{q}" & Component $q$ of the total unit tensor operator of rank $k$ in the orbital space & \eqref{eq:U} \\
$\telement{T}{k}{q}$ & \verb"T/{k},{q}" & Component $q$ of the total unit tensor operator of rank $k$ in the spin space & \eqref{eq:T} \\
$\dot{\tensor{U}{k}}{\tensor{U}{k}}$ & \verb"UU/{k}" & Squared total unit tensor operator of rank $k$ in the orbital space & \eqref{eq:UU} \\
$\dot{\tensor{T}{k}}{\tensor{T}{k}}$ & \verb"TT/{k}" &  Squared total unit tensor operator of rank $k$ in the spin space & \eqref{eq:TT} \\
$\dot{\tensor{U}{k}}{\tensor{T}{k}}$ & \verb"UT/{k}" & Scalar product of the total unit tensor operators in the orbital and spin spaces & \eqref{eq:UT} \\
    \hline
$\velement{L}{q} / \hbar$ & \verb"L/{q}" & Component $q$ of the total orbital angular momentum operator & \eqref{eq:L} \\
$\velement{S}{q} / \hbar$ & \verb"S/{q}" & Component $q$ of the total spin angular momentum operator & \eqref{eq:S} \\
$\velement{J}{q} / \hbar$ & \verb"J/{q}" & Component $q$ of the total angular momentum operator & \eqref{eq:J} \\
$\dot{\vector{L}}{\vector{L}} / \hbar^2$ & \verb"LL" & Squared total orbital angular momentum operator & \eqref{eq:LL} \\
$\dot{\vector{S}}{\vector{S}} / \hbar^2$ & \verb"SS" & Squared total spin angular momentum operator & \eqref{eq:SS} \\
$\dot{\vector{J}}{\vector{J}} / \hbar^2$ & \verb"JJ" & Squared total angular momentum operator & \eqref{eq:JJ} \\
$\dot{\vector{L}}{\vector{S}} / \hbar^2$ & \verb"LS" & Scalar product of the total orbital and spin angular momentum operators & \eqref{eq:LS} \\
    \hline
$\operator{f}_k$ & \verb"H1/{k}" & Coulomb first-order perturbation Hamiltonian of rank $k$ & \eqref{eq:fk-unit} \\
$\operator{z}$ & \verb"H2" & Spin-orbit first-order perturbation Hamiltonian & \eqref{eq:z} \\
$\operator{C}_2(\text{G}_2)$ & \verb"C2" & Casimir operator of the special group $\text{G}_2$ & \eqref{eq:H3b} \\
$\operator{C}_2(\text{SO}(2l+1))$ & \verb"CR" & Casimir operator of the special orthogonal (rotational) group in $2l+1$ dimensions & \eqref{eq:H3c} \\
$\operator{t}_c$ & \verb"H4/{c}" & Effective Coulomb second-order perturbation Hamiltonian & \eqref{eq:tc} \\
$\operator{m}_{k,ss}$ & \verb"Hss/{k}" & Spin-spin first-order perturbation Hamiltonian of rank $k$ & \eqref{eq:mkss} \\
$\operator{m}_{k,soo}$ & \verb"Hsoo/{k}" & Spin-other-orbit first-order perturbation Hamiltonian of rank $k$ & \eqref{eq:mksoo} \\
$\operator{m}_k$ & \verb"H5/{k}" & Spin-spin and spin-other-orbit first-order perturbation Hamiltonian of rank $k$ & \eqref{eq:mk} \\
$\operator{p}_k$ & \verb"H6/{k}" & Effective electrostatic spin-orbit second-order perturbation Hamiltonian of rank $k$ & \eqref{eq:pk} \\
    \hline
$\telement{C}{k}{q}$ & \verb"C/{k},{q}" & Component $q$ of the Wybourne crystal field operator of rank $k$ & \eqref{eq:ck} \\
$\telement{\hat{C}}{k}{q}$ & \verb"Hcf/{k},{q}" & Component $q$ of the crystal field operator of rank $k$ with real parameter $B^k_q$ & \eqref{eq:Hcf} \\
    \hline
  \end{tabularx}
\end{table*}

These matrices are stored in the same sparse format as the mixed unit tensor operator matrices.
Each data container includes a comprehensive set of states, incorporating electron pool and many-electron indices as described in the initial step.
This ensures that the containers are self-contained, allowing for the unambiguous identification of the product state associated with each row and column.
All product-state matrix containers are archived in the file \verb"product.zip" within the Zenodo repository~\cite{Caspary.2026b}.

It should be noted that the expressions used to evaluate these many-electron matrices are, in several respects, specific to single-shell configurations, such as the $f^N$ configurations of the lanthanides.
A representative example is the application of Eq.~\eqref{eq:reduce-uu-scalar-1} in scalar products.
For general multi-shell configurations the dominant many-electron Coulomb interaction Hamiltonians would be based on mixed unit tensor operators that explicitly include the tensor operator $\tensor{c}{k}$.
The most configuration-specific instances are the effective second-order Hamiltonians discussed in Subsection~\ref{sub:2ndOrder}, which are defined strictly for single-shell configurations.

\subsection{\texorpdfstring{$LS$}{LS}-Coupling (module \texttt{\upshape transform.py})}

The most computationally intensive phase of the workflow involves generating the transformation matrix from the product-state space to the $LS$-coupling basis, as detailed in Section~\ref{sec.coupling}.
Core of this procedure is the consecutive diagonalization of the product-state matrices of the tensor operators within increasingly smaller subspaces, following the symmetry chains specified in Tab.~\ref{tab:chains}.
The utilization of exact arithmetic within the \ameli\ package is particularly advantageous here, as it facilitates the unambiguous identification of degenerate subspaces with identical eigenvalues.

Following diagonalization, the phase adjustment step from Subsection~\ref{sub:phase} is applied to establish consistent signs of the states within each $J$-multiplet.

To ensure the data container \verb"transform.zdc" remains self-contained, it includes the transformation matrix along with the comprehensive lists of product and $LS$-states.
Product states, corresponding to the rows of the transformation matrix, are represented by the previously mentioned pool of single-electron states and their respective indices.
$LS$-states, corresponding to the columns, are characterized by the eigenvalues and irreducible representations of the respective operators.
Additionally, the code incorporates the seniority number~$v$, the number~$\tau$, and the term number for states with identical quantum numbers $LS$ but distinct $WU\tau$ classifications.
The latter follows the convention established by Nielson and Koster~\cite{Nielson.1963}.

The data container \verb"transform.zdc" is contained in the file \verb"product.zip" together with the product-state matrices within the Zenodo repository~\cite{Caspary.2026b}.
This file therefore contains all data required for energy-level fits of lanthanide ions in crystalline hosts including the transformation of the resulting intermediate states to the $LS$-classification scheme.

\subsection{\texorpdfstring{$LS$}{LS}-Matrices (module \texttt{\upshape matrix.py})}

In the final stage of the workflow, the \ameli\ package transforms all many-electron product-state matrices into the $LS$-coupling scheme using Eq.~\eqref{eq:transform}.
Although the use of exact arithmetic renders these transformations computationally demanding, particularly for near-half-filled shell configurations with high state densities, each matrix requires only a single transformation.
A significant advantage of exact arithmetic in this context is the preservation of matrix sparsity.
While floating-point transformations introduce rounding noise that can obscure null elements, exact arithmetic ensures that zero-valued elements in $LS$-coupling remain identically zero.

The resulting matrices within the full $LS$-state space are archived in the Zenodo repository~\cite{Caspary.2026b} in the file \verb"sljm.zip".
Unlike the product-state matrices, these $LS$-matrix data containers include the complete list of $LS$-states, including their eigenstates and irreducible representations, ensuring the unambiguous identification of every row and column.

For lanthanide ions embedded in amorphous materials, the random orientation of local environments typically justifies the assumption of overall rotational symmetry.
This structural disorder manifests as inhomogeneous broadening, resulting in continuous spectral features where individual $J$-levels merge.
Under these conditions, energy-level fitting and Judd-Ofelt calculations can be performed using significantly condensed matrices that consider only a single representative state for each $J$-multiplet.
To facilitate this, the \ameli\ package projects the full $LS$-space onto a basis of stretched states with $M=J$.
These reduced matrices are provided in the Zenodo repository~\cite{Caspary.2026b} within the file \texttt{slj.zip}, offering a computationally efficient alternative for energy-level fits in environments with rotational symmetry.

Within the framework of Judd-Ofelt theory, the intensities of radiative transitions between $J$-manifolds in amorphous materials are determined by the reduced matrix elements of electric and magnetic dipole operators.
To compute these values, the \ameli\ package utilizes the basis of stretched states in conjunction with Eq.~\eqref{eq:Reduced-U} based on the Wigner-Eckart theorem. 
The file \texttt{slj.zip} contains a comprehensive set of these reduced matrix elements in addition to the mentioned stretched state matrices.
Consequently, this file is the data source for energy-level fits and Judd-Ofelt calculations involving lanthanide ions in amorphous host materials.

It should be noted that synchronized phases of the $LS$-states within each $J$-multiplet according to the algorithm outlined in Subsection~\ref{sub:phase} are a mandatory precursor to these calculations to ensure the consistency of the resulting reduced matrix elements.

\section{Comparison with the Literature}
\label{sec.comparison}

The \ameli\ software package includes a comprehensive test suite designed to minimize the risk of erroneous results.
These tests are divided into three functional categories.

The first category consists of mathematical tests.
A failure in this category would indicate a fundamental error in the code.
Consequently, all such tests must be passed without exception.

The second category involves element-wise comparisons with previously published matrix elements.
Since \ameli\ is based on exact arithmetic, tests in this category are expected to reproduce literature values exactly.
Our validation against established datasets confirmed this, as the package successfully reproduced all referenced matrix elements without deviation.

The third category comprises application-based tests, which involve converting exact \ameli\ results for use in floating-point matrix operations.
Numerical comparisons with published data in this context require appropriate error margins.
Interestingly, some of these tests have revealed outliers that are most likely attributable to errors in the original publications.

\subsection{Mathematical Tests}

The most computationally intensive procedure within the \ameli\ package is the determination of the transformation matrix~$V$ from the product-state basis to the $LS$-coupling scheme. 
To ensure the integrity of this transformation, every matrix~$V$ must satisfy the orthonormality condition:
\begin{equation}
    V^TV = \ident
    \ .
\end{equation}
The use of exact arithmetic allows this test to be performed without the need for a numerical error margin.

In the second test, the following matrix product is evaluated for the matrix~$M$ of each operator in the symmetry chains listed in Table~\ref{tab:chains}:
\begin{equation}
    V^T M V = \Lambda
    \ .
\end{equation}

The resulting matrix~$\Lambda$ must be diagonal, and its diagonal elements must exactly reproduce the known eigenvalues of each $LS$-state.
Exact arithmetic is particularly advantageous here, as it guarantees that these eigenvalues are recovered as precise integer or half-integer values, as required by the underlying quantum mechanics.

The final test in this category involves verifying the identity of all $LS$-states by comparing their irreducible representations and term numbers with the values published by Nielson and Koster~\cite{Nielson.1963}.
This ensures that the state labeling of the \ameli\ package remains consistent with established group-theoretical conventions.

\subsection{Element-Wise Comparisons}
\label{sub:elementwise}

Judd published a comprehensive set of matrix elements for the $f^3$ configuration alongside the introduction of the effective three-electron second-order Coulomb interaction~\cite{Judd.1966}.
This publication includes not only the matrix elements for $c=2,3,4,6,7,8$ relevant to the perturbation operator~$\operator{H}_4$, but the complete set for $c=1\ldots9$.
Since the operator is independent of the total angular momentum~$J$, the full list contains only 21 elements for each value of $c$.
Judd reported these matrix elements as signed square roots of rational numbers, all of which are exactly reproduced by the \ameli\ software.

Ref.~\onlinecite{Judd.1966} also notes that the operator~$\operator{t}_9$ in all $f^N$ configurations is exactly zero, which is confirmed by the test suite.
Judd furthermore noted that the operator~$\operator{t}_5$ is proportional to the Coulomb operator~$\operator{e}_2$ in Racah notation and $\operator{t}_1$ is absorbed by $\operator{e}_0$ and $\operator{e}_1$.
The relation between the Coulomb interaction operators in Racah notation and the rank-based notation $\operator{f}_0$, $\operator{f}_2$, $\operator{f}_4$, and $\operator{f}_6$ used in this work is given by~\cite{Racah.1949}:
\begin{equation}\begin{split}
  \operator{e}_0 &= \operator{f}_0 \\
  \operator{e}_1 &= \frac{9}{7}\operator{f}_0 + \frac{75}{14}\operator{f}_2 +
                    \frac{99}{7}\operator{f}_4 + \frac{5577}{350}\operator{f}_6 \\
  \operator{e}_2 &= \frac{10725}{14}\operator{f}_2 -
                    \frac{12870}{7}\operator{f}_4 + \frac{5577}{10}\operator{f}_6
  \ .
\end{split}\end{equation}

The relation between the operators $\operator{t}_5$ and $\operator{e}_2$ is given by Judd~\cite{Judd.1966}:
\begin{equation}
\label{eq:t5}
    \operator{t}_5 = -\frac{N-2}{14 \sqrt{4290}}\, \operator{e}_2
\end{equation}
and the respective relation for the operator~$\operator{t}_1$ was identified by the author of this work as
\begin{equation}
    \operator{t}_1 = \sqrt{1155}\, (N-2) \left(\frac{\operator{e}_0}{245} -
      \frac{\operator{e}_1}{210}\right)
    \ .
\end{equation}
Both expressions are verified in exact arithmetic for all configurations from $f^3$ to $f^{11}$ by the \ameli\ test suite.

In conjunction with the introduction of the first-order spin-spin and spin-other-orbit interactions resembling the perturbation operator~$\operator{H}_5$, Judd et al.\ published the exact values of all of their matrix elements for the $f^2$ configuration~\cite{Judd.1968}.
The $J$-dependence of these scalar operators allows to reduce the matrix elements according to:
\begin{multline}
  \label{eq:reducedSL}
  \langle \gamma S L J | \operator{H} | \gamma' S' L' J' \rangle = 
  \delta(J,J') (-1)^{S'+L'+J+t}
  \\ \times
  \sqrt{2t+1} \vij{S'}{L'}{J}{L}{S}{t}
  \langle \gamma S L || \operator{H} || \gamma' S' L' \rangle
  \ ,
\end{multline}
where the common inner rank is $t=2$ in both orbital and spin space for the operator $\operator{H}_{ss}$, and $t=1$ for the operator $\operator{H}_{soo}$.
It should be noted that there is a typographical error in equation~(3) of Ref.~\onlinecite{Judd.1968}, where the prime on $L'$ is omitted in the phase exponent.
The publication lists all non-zero values of $(-1)^t\sqrt{2t+1}\langle \gamma SL || \operator{H} || \gamma'S'L' \rangle$, which the \ameli\ software reproduces exactly.

The same study includes the reduced matrix elements of the second-order spin-orbit interaction operator with $t=1$ and $\operator{H}=\operator{H}_6$ in Eq.~\eqref{eq:reducedSL} for the $f^2$ configuration.
These matrix elements are also reproduced exactly by \ameli.
While the original publication provides elements for ranks $k=0,2,4,6$, it is important to note that $k=0$ is excluded from the standard set of matrices for operator~$\operator{H}_6$.
As shown in Eq.~\eqref{eq:H6.1}, this operator is equivalent to the first-order spin-orbit interaction~\eqref{eq:z}.

The matrix elements for the nearly closed-shell configuration~$f^{12}$ are identical to those of $f^2$ for the spin-spin interaction operator~$\operator{H}_{ss}$.
Consequently, the \ameli\ software validates these interaction matrices against the same literature values.
The matrix elements for the spin-other-orbit interaction~$\operator{H}_{soo}$ of the configuration~$f^{12}$ were published by Carnall et al.~\cite{Carnall.1970}, also utilizing the reduction in Eq.~\eqref{eq:reducedSL}.
These authors further included the second-order spin-orbit operator~$\operator{H}_6$ for $f^{12}$.
The results generated by the \ameli\ software are identical to all of these published reduced matrix elements for the magnetic interactions of $f^{12}$.

As mentioned above, the second-order operator~$\operator{p}_0$ is proportional to the first-order spin-orbit operator~$\operator{z}$.
The comparison of Eq.~\eqref{eq:H6.1} and Eq.~\eqref{eq:z} reveals the relationship
\begin{equation}
\label{eq:p0}
    \operator{p}_0 = -\frac{N-1}{3}\, \operator{z}
    \ ,
\end{equation}
which is verified in exact arithmetic for all configurations from $f^2$ to $f^{12}$ by the \ameli\ test suite.

\subsection{Application-Based Tests}
\label{sub:application_tests}

The primary objective of this work is to provide a comprehensive library of matrix elements for the community utilizing lanthanide absorption spectra to characterize radiative transition intensities.
To this end, the perturbation Hamilton matrices generated by the \ameli\ software are extensively validated against published energy-level spectra.
These tests focus on studies that provide both a set of radial integrals and the corresponding calculated energy-level spectrum.

The test suite currently includes only datasets that treat each $J$-multiplet as degenerate, which is standard practice for amorphous or liquid materials.
Consequently, the tests utilize matrices from the \verb"slj.zip" dataset, which contains only the stretched states with $M=J$.
Unlike in crystalline hosts, the quantum number $J$ can be considered a good quantum number in these environments.
The tests use $J$ alongside the energy value to uniquely identify states when comparing results with literature energy levels.

\def\imgwidth{2}   
\def\imgheight{0.4}   
\def\lwidth{0.8pt}

\begin{table}[t]
  \caption{\label{tab:energy_test} Comparison of published energy-level spectra with results based on the \ameli\ package. The deviation histograms cover $\pm3.5$ times the resolution of the published energies. They are centered at zero and the upper line corresponds to \qty{100}{\percent}. An asterisk at the reference marks cases with extended set of radial integrals (see the text).}
    \begin{tabular}{|l|l|c|c|c|c|}
    \hline
    Ion & Host & Levels & Deviation & Outliers & Ref. \\
    \hline
    \hline
    $\mathrm{Pr^{3+}} (f^{2})$ & $\mathrm{LaCl_{3}}$ & 13 & \drawhistogram{{0.000, 0.000, 0.077, 0.923, 0.000, 0.000, 0.000}} & 0 & \onlinecite{Carnall.1968} \\
    $\mathrm{Pr^{3+}} (f^{2})$ & $\mathrm{LaF_{3}}$ & 13 & \drawhistogram{{0.000, 0.000, 0.000, 0.846, 0.154, 0.000, 0.000}} & 0 & \onlinecite{Carnall.1968} \\
    $\mathrm{Pr^{3+}} (f^{2})$ & $\mathrm{LaF_{3}}$ & 13 & \drawhistogram{{0.000, 0.000, 0.000, 0.923, 0.077, 0.000, 0.000}} & 0 & \onlinecite{Carnall.1969} \\
    $\mathrm{Pr^{3+}} (f^{2})$ & $\mathrm{LaF_{3}}$ & 13 & \drawhistogram{{0.000, 0.000, 0.000, 0.846, 0.154, 0.000, 0.000}} & 0 & \onlinecite{Carnall.1970}$^\ast$ \\
    $\mathrm{Pr^{3+}} (f^{2})$ & aq & 13 & \drawhistogram{{0.000, 0.000, 0.000, 1.000, 0.000, 0.000, 0.000}} & 0 & \onlinecite{Carnall.1968} \\
    $\mathrm{Pr^{3+}} (f^{2})$ & free ion & 13 & \drawhistogram{{0.000, 0.000, 0.000, 0.923, 0.077, 0.000, 0.000}} & 0 & \onlinecite{Carnall.1968} \\
    \hline
    $\mathrm{Nd^{3+}} (f^{3})$ & $\mathrm{LaCl_{3}}$ & 27 & \drawhistogram{{0.000, 0.000, 0.148, 0.630, 0.148, 0.037, 0.000}} & 1 & \onlinecite{Carnall.1968} \\
    $\mathrm{Nd^{3+}} (f^{3})$ & aq & 39 & \drawhistogram{{0.000, 0.000, 0.103, 0.692, 0.179, 0.000, 0.000}} & 1 & \onlinecite{Carnall.1968} \\
    \hline
    $\mathrm{Pm^{3+}} (f^{4})$ & aq & 49 & \drawhistogram{{0.000, 0.000, 0.020, 0.898, 0.082, 0.000, 0.000}} & 0 & \onlinecite{Carnall.1968} \\
    \hline
    $\mathrm{Sm^{3+}} (f^{5})$ & $\mathrm{LaCl_{3}}$ & 29 & \drawhistogram{{0.000, 0.000, 0.034, 0.862, 0.069, 0.000, 0.000}} & 1 & \onlinecite{Carnall.1968} \\
    $\mathrm{Sm^{3+}} (f^{5})$ & aq & 106 & \drawhistogram{{0.028, 0.009, 0.019, 0.000, 0.009, 0.009, 0.028}} & 95 & \onlinecite{Carnall.1968} \\
    \hline
    $\mathrm{Eu^{3+}} (f^{6})$ & $\mathrm{LaCl_{3}}$ & 29 & \drawhistogram{{0.000, 0.000, 0.069, 0.931, 0.000, 0.000, 0.000}} & 0 & \onlinecite{Carnall.1968d} \\
    $\mathrm{Eu^{3+}} (f^{6})$ & aq & 55 & \drawhistogram{{0.000, 0.018, 0.109, 0.782, 0.036, 0.000, 0.000}} & 3 & \onlinecite{Carnall.1968d} \\
    \hline
    $\mathrm{Gd^{3+}} (f^{7})$ & $\mathrm{GdCl_{3}}$ & 15 & \drawhistogram{{0.000, 0.000, 0.133, 0.867, 0.000, 0.000, 0.000}} & 0 & \onlinecite{Carnall.1968b} \\
    $\mathrm{Gd^{3+}} (f^{7})$ & aq & 21 & \drawhistogram{{0.000, 0.000, 0.095, 0.905, 0.000, 0.000, 0.000}} & 0 & \onlinecite{Carnall.1968b} \\
    \hline
    $\mathrm{Tb^{3+}} (f^{8})$ & $\mathrm{LaCl_{3}}$ & 9 & \drawhistogram{{0.000, 0.000, 0.222, 0.556, 0.000, 0.111, 0.000}} & 1 & \onlinecite{Carnall.1968c} \\
    $\mathrm{Tb^{3+}} (f^{8})$ & aq & 44 & \drawhistogram{{0.000, 0.000, 0.045, 0.818, 0.023, 0.000, 0.023}} & 4 & \onlinecite{Carnall.1968c} \\
    \hline
    $\mathrm{Dy^{3+}} (f^{9})$ & $\mathrm{LaCl_{3}}$ & 26 & \drawhistogram{{0.000, 0.000, 0.115, 0.769, 0.077, 0.000, 0.000}} & 1 & \onlinecite{Carnall.1968} \\
    $\mathrm{Dy^{3+}} (f^{9})$ & aq & 62 & \drawhistogram{{0.000, 0.000, 0.016, 0.984, 0.000, 0.000, 0.000}} & 0 & \onlinecite{Carnall.1968} \\
    \hline
    $\mathrm{Ho^{3+}} (f^{10})$ & $\mathrm{LaCl_{3}}$ & 38 & \drawhistogram{{0.000, 0.026, 0.026, 0.842, 0.079, 0.000, 0.000}} & 1 & \onlinecite{Carnall.1968} \\
    $\mathrm{Ho^{3+}} (f^{10})$ & aq & 57 & \drawhistogram{{0.000, 0.000, 0.070, 0.860, 0.070, 0.000, 0.000}} & 0 & \onlinecite{Carnall.1968} \\
    \hline
    $\mathrm{Er^{3+}} (f^{11})$ & $\mathrm{LaCl_{3}}$ & 22 & \drawhistogram{{0.000, 0.000, 0.227, 0.364, 0.045, 0.182, 0.045}} & 3 & \onlinecite{Carnall.1968} \\
    $\mathrm{Er^{3+}} (f^{11})$ & $\mathrm{LaF_{3}}$ & 24 & \drawhistogram{{0.000, 0.042, 0.292, 0.292, 0.125, 0.125, 0.083}} & 1 & \onlinecite{Carnall.1968} \\
    $\mathrm{Er^{3+}} (f^{11})$ & aq & 34 & \drawhistogram{{0.029, 0.088, 0.088, 0.412, 0.176, 0.088, 0.000}} & 4 & \onlinecite{Carnall.1968} \\
    $\mathrm{Er^{3+}} (f^{11})$ & free & 31 & \drawhistogram{{0.097, 0.000, 0.194, 0.452, 0.065, 0.065, 0.065}} & 2 & \onlinecite{Carnall.1968} \\
    \hline
    $\mathrm{Tm^{3+}} (f^{12})$ & $\mathrm{C_{2}H_{5}SO_{4}}$ & 13 & \drawhistogram{{0.000, 0.000, 0.077, 0.923, 0.000, 0.000, 0.000}} & 0 & \onlinecite{Carnall.1968} \\
    $\mathrm{Tm^{3+}} (f^{12})$ & $\mathrm{LaF_{3}}$ & 12 & \drawhistogram{{0.000, 0.083, 0.083, 0.667, 0.167, 0.000, 0.000}} & 0 & \onlinecite{Carnall.1970} \\
    $\mathrm{Tm^{3+}} (f^{12})$ & $\mathrm{LaF_{3}}$ & 12 & \drawhistogram{{0.000, 0.083, 0.000, 0.750, 0.000, 0.083, 0.000}} & 1 & \onlinecite{Carnall.1970}$^\ast$ \\
    $\mathrm{Tm^{3+}} (f^{12})$ & aq & 13 & \drawhistogram{{0.000, 0.000, 0.000, 1.000, 0.000, 0.000, 0.000}} & 0 & \onlinecite{Carnall.1968} \\
    \hline
  \end{tabular}
\end{table}

In their seminal series of four papers~\cite{Carnall.1968,Carnall.1968b,Carnall.1968c,Carnall.1968d}, Carnall et al.\ reported a complete set of measured energy levels and calculated radial integrals for all lanthanide aquo ions from $\mathrm{Pr^{3+}} (f^2)$ to $\mathrm{Tm^{3+}} (f^{12})$ in dilute acid solutions.
They compared these to lanthanides in other hosts, primarily $\mathrm{LaF_3}$ and $\mathrm{LaCl_3}$, and calculated radial integrals for first- and second-order Coulomb interactions $\operator{H}_1$ and $\operator{H}_3$ as well as first-order spin-orbit coupling~$\operator{H}_2$.
The \ameli\ test suite also incorporates data from two additional publications by Carnall~\cite{Carnall.1969,Carnall.1970}.
Ref.~\onlinecite{Carnall.1970} is of particular interest, as it includes radial integrals for first-order spin-spin and spin-other-orbit interactions~$\operator{H}_5$, as well as second-order electrostatic spin-orbit interactions~$\operator{H}_6$.

For Coulomb interactions, the referenced publications utilize the Racah parameter set $E^1$, $E^2$, $E^3$, which is related to the rank-based radial integrals $F^2$, $F^4$, $F^6$ used in this work by~\cite{Racah.1949}:
\begin{equation}\begin{split}
  F^2 &= \frac{75}{14} (E^1 + 143 E^2 + 11 E^3) \\
  F^4 &= \frac{99}{7} (E^1 - 130 E^2 + 4 E^3) \\
  F^6 &= \frac{5577}{750} (E^1 + 35 E^2 - 7 E^3)
  \ .
\end{split}\end{equation}

A similar conversion is required for the radial integrals of the second-order electrostatic spin-orbit interactions in Ref.~\onlinecite{Carnall.1970}, which use the parameters $P_2$, $P_4$, $P_6$ introduced by Judd~\cite{Judd.1968} to extract common factors.
The conversion to the radial integrals used in this work is given by~\cite{Judd.1968}:
\begin{equation}
    P^2 = 225 P_2 \quad
    P^4 = 1089 P_4 \quad
    P^6 = \frac{184041}{25} P_6
    \ .    
\end{equation}

Perturbation Hamiltonians were constructed as linear combinations according to Eq.~\eqref{eq:Hab}, utilizing published radial integrals and the corresponding angular matrices generated by \ameli.
The resulting energy matrices were diagonalized in floating-point double precision, and the calculated energy levels were compared individually to the literature results, as summarized in Table~\ref{tab:energy_test}.
References involving the more extensive set of radial integrals, including spin-spin, spin-other-orbit, and electrostatic spin-orbit interactions, are denoted with an asterisk.

Initially, the total angular momentum quantum number $J$ for each calculated level was extracted from the eigenvectors.
The test suite successfully confirmed the identity of these quantum numbers with those reported in the literature, with $\mathrm{Sm^{3+}\!\!:\!aq}$ as one specific exception discussed below.

For the numerical comparison, the difference between each calculated $J$-multiplet energy and the corresponding literature value was determined.
To account for arbitrary energy offsets across the spectra, these differences were shifted such that the average deviation was zero.
Table~\ref{tab:energy_test} presents the number of energy levels evaluated alongside a seven-bin deviation histogram.
The bin width corresponds to the literature's reporting resolution, which was \qty{1}{cm^{-1}} in all cases.
The gray upper line indicates \qty{100}{\percent} of the total set.

The vast majority of energy levels fall within the central bin, confirming they are essentially identical to the literature values.
A small fraction of levels contribute to the two adjacent bins.
However, given the limitations of 1960s-era single-precision arithmetic, these should also be considered successful matches.
In many instances, the outer bins remain empty or contain only a few elements, which can be attributed to the approximate diagonalization procedures of the era, such as basis truncation techniques necessitated by the limited core memory of early mainframe computers.

Certain datasets contain outliers that exceed the histogram range.
Beyond the aforementioned computational inaccuracies, such significant deviations likely result from typographical or transmission errors. which are common occurrences in an era when numerical data processing involved numerous manual steps.
Several such errors were sufficiently obvious to be corrected prior to the evaluation.

A notable example is the data in Table~II of Ref.~\onlinecite{Carnall.1970}, which provides measured and calculated energy levels for $\mathrm{Pr^{3+}}$ and $\mathrm{Tm^{3+}}$ simultaneously in a condensed format.
Although these ions share the same $LS$-terms, the terms appear in a different order for each ion.
This discrepancy is likely why the values for the $\mathrm{\prescript{1}{}{D}_2}$ and $\mathrm{\prescript{3}{}{P}_2}$ states seem to be swapped for $\mathrm{Pr^{3+}}$, as do the $\mathrm{\prescript{3}{}{F}_4}$ and $\mathrm{\prescript{3}{}{H}_4}$ states for $\mathrm{Tm^{3+}}$.
Notably, the latter inconsistency may also stem from a classic ambiguity regarding the Thulium ion: state labels can vary depending on whether they are based on extrapolation to vanishing spin-orbit coupling or on the dominant $LS$-component of the intermediate state. The \ameli\ package adopts the latter convention for its natural integration into the calculation framework.

Two additional cases of likely mislabeling were identified in Ref.~\onlinecite{Carnall.1968}.
The terms $\mathrm{\prescript{4}{}{F}_{7/2}}$ and $\mathrm{\prescript{4}{}{S}_{3/2}}$ seem to be exchanged for $\mathrm{Nd^{3+}\!\!:\!aq}$, as do $\mathrm{\prescript{6}{}{H}_{9/2}}$ and $\mathrm{\prescript{6}{}{F}_{11/2}}$ for $\mathrm{Nd^{3+}\!\!:\!LaCl_3}$.

While the majority of datasets showed excellent agreement with \ameli\ results, Tab.~\ref{tab:energy_test} highlights two significant anomalies.
First, the energy spectrum calculated for $\mathrm{Sm^{3+}\!\!:\!aq}$ using the radial integrals from Ref.~\onlinecite{Carnall.1968} differed fundamentally from the published values.
Interestingly, the squared reduced unit tensor matrix elements for the same ion-host combination were reproduced reasonably well, as shown below.
Given the alignment between the published calculated and measured levels, a transmission error in the published radial integrals is the most probable explanation.

A second, systematic observation was noted across all ion–host combinations involving the erbium ion.
While the squared reduced unit tensor matrix elements were again reproduced with high accuracy using the same radial integrals, the energy-level comparisons yielded notably flatter deviation histograms compared to the sharp distributions seen for other ions.
Although the overall agreement remains acceptable, this persistent broadening suggests a subtle discrepancy, possibly arising from a minor error in one of the perturbation matrices used in the original studies, rather than a random transmission error.

In summary, the comparison with established literature strongly validates the perturbation Hamiltonian matrices generated by the \ameli\ software and the underlying theoretical framework.

\begin{table}[t]
  \caption{\label{tab:reduced_test} Comparison of published squared reduced unit tensor matrix elements with results based on the \ameli\ package. The deviation histograms cover $\pm3.5$ times the resolution of the published values. They are centered at zero and the upper line corresponds to \qty{100}{\percent}.}
    \begin{tabular}{|l|l|c|c|c|c|}
    \hline
    Ion & Host & Values & Deviation & Outliers & Ref. \\
    \hline
    $\mathrm{Pr^{3+}} (f^{2})$ & aq & 36 & \drawhistogram{{0.000, 0.000, 0.056, 0.944, 0.000, 0.000, 0.000}} & 0 & \onlinecite{Carnall.1968} \\
    $\mathrm{Nd^{3+}} (f^{3})$ & aq & 114 & \drawhistogram{{0.000, 0.009, 0.018, 0.895, 0.079, 0.000, 0.000}} & 0 & \onlinecite{Carnall.1968} \\
    $\mathrm{Pm^{3+}} (f^{4})$ & aq & 144 & \drawhistogram{{0.000, 0.000, 0.000, 0.979, 0.021, 0.000, 0.000}} & 0 & \onlinecite{Carnall.1968} \\
    $\mathrm{Sm^{3+}} (f^{5})$ & aq & 315 & \drawhistogram{{0.000, 0.006, 0.048, 0.838, 0.076, 0.003, 0.003}} & 8 & \onlinecite{Carnall.1968} \\
    $\mathrm{Eu^{3+}} (f^{6})$ & aq & 162 & \drawhistogram{{0.000, 0.000, 0.000, 0.994, 0.006, 0.000, 0.000}} & 0 & \onlinecite{Carnall.1968d} \\
    $\mathrm{Gd^{3+}} (f^{7})$ & aq & 60 & \drawhistogram{{0.000, 0.000, 0.000, 1.000, 0.000, 0.000, 0.000}} & 0 & \onlinecite{Carnall.1968b} \\
    $\mathrm{Tb^{3+}} (f^{8})$ & aq & 129 & \drawhistogram{{0.000, 0.000, 0.023, 0.915, 0.047, 0.016, 0.000}} & 0 & \onlinecite{Carnall.1968c} \\
    $\mathrm{Dy^{3+}} (f^{9})$ & aq & 183 & \drawhistogram{{0.000, 0.022, 0.033, 0.852, 0.055, 0.000, 0.005}} & 6 & \onlinecite{Carnall.1968} \\
    $\mathrm{Ho^{3+}} (f^{10})$ & aq & 168 & \drawhistogram{{0.000, 0.000, 0.000, 0.958, 0.042, 0.000, 0.000}} & 0 & \onlinecite{Carnall.1968} \\
    $\mathrm{Er^{3+}} (f^{11})$ & aq & 99 & \drawhistogram{{0.000, 0.000, 0.061, 0.859, 0.071, 0.000, 0.000}} & 1 & \onlinecite{Carnall.1968} \\
    $\mathrm{Tm^{3+}} (f^{12})$ & aq & 36 & \drawhistogram{{0.000, 0.000, 0.000, 0.972, 0.028, 0.000, 0.000}} & 0 & \onlinecite{Carnall.1968} \\
    \hline
  \end{tabular}
\end{table}

In addition to the calculated energy levels, the seminal series on lanthanide aquo ions~\cite{Carnall.1968,Carnall.1968b,Carnall.1968c,Carnall.1968d} provides comprehensive lists of the squared reduced matrix elements for the unit tensor operators $\tensor{U}{2}$, $\tensor{U}{4}$, and $\tensor{U}{6}$ corresponding to ground-state absorption transitions.
The eigenvectors obtained from the diagonalization of the full perturbation Hamiltonian represent the state compositions in intermediate coupling, and these were utilized to calculate linear combinations of the reduced matrix elements of pure $LS$-states provided by the \ameli\ software.

As illustrated in Table~\ref{tab:reduced_test}, the resulting values match the published literature almost perfectly for all lanthanide ions, including the previously mentioned cases of $\mathrm{Sm^{3+}\!\!:\!aq}$ and $\mathrm{Er^{3+}\!\!:\!aq}$.
It should be noted that because three reduced matrix elements are associated with each energy level, the total number of data points in this comparison is significantly larger than in the energy-level validation.
The bin width in the resulting histograms is again defined by the numerical resolution of the published values, which is $0.0001$ in all instances.

\section{Application Examples}
\label{sec.examples}

Although \ameli\ itself is not an application software, the open-source Python package \texttt{YALIP}~\cite{Caspary.2026c} serves as a reference implementation for applications utilizing the \ameli\ repository.
In this section, we employ \texttt{YALIP} to demonstrate the utility of \ameli\ in two realistic scenarios.

\subsection{Energies and Radiative Transitions of \texorpdfstring{$J$}{J}-Multiplets}

The first example utilizes data published by Hehlen et~al.\ in their paper marking the 50th anniversary of Judd-Ofelt theory~\cite{Hehlen.2013}.
Among other results, the authors report on their proprietary lanthanide program and present corresponding calculations.
Their program accounts for the two most prominent perturbation Hamiltonians: the Coulomb interaction~$\operator{H}_1$ and the spin-orbit coupling~$\operator{H}_2$.
Specifically, the paper presents calculated barycenter energies for the first eleven $J$-multiplets of $\mathrm{Er^{3+}\!\!:\!LaCl_3}$, alongside the oscillator strengths of ground-state absorption transitions into these multiplets.

To reproduce the calculated energies and determine the omitted ones, we adopt the reported radial integrals: $F_2=\qty{433.22}{cm^{-1}}$, $F_4=\qty{66.887}{cm^{-1}}$, $F_6=\qty{7.2952}{cm^{-1}}$, and $\zeta=\qty{2385.9}{cm^{-1}}$.
\texttt{YALIP} utilizes the well-known relationships~\cite{Judd.1963}
\begin{equation}
    F^2 = 225 F_2 \quad
    F^4 = 1089 F_4 \quad
    F^6 = \frac{184041}{25} F_6
\end{equation}
to convert these Coulomb parameters into the respective Slater integrals: $F^2=\qty{97474.5}{cm^{-1}}$, $F^4=\qty{72839.9}{cm^{-1}}$, and $F^6=\qty{53704.6}{cm^{-1}}$.

The total Hamiltonian is constructed as a linear combination of the matrices \verb"H1/2", \verb"H1/4", \verb"H1/6", and \verb"H2" for the $f^{11}$ configuration from the \ameli\ repository, scaled by the factors $F^2$, $F^4$, $F^6$, and $\zeta$.
Even though the erbium ions reside within a crystalline host material in this scenario, the original publication neglects the Stark splitting effect induced by the crystal field.

Consequently, \texttt{YALIP} utilizes the matrices from the \verb"slj" folder within the \verb"slj.zip" container, which exclusively contains the matrix elements of stretched states. This container is chosen here to mirror the spherical symmetry approximation adopted in the original publication's calculations.
In general, it represents the appropriate choice for lanthanides in truly amorphous media.

\begin{table*}[p]
  \caption{\label{tab:example_1} Comparison of calculated energy levels $k^\mathrm{ref}$ and oscillator strengths $f^\mathrm{ref}$ from Ref.~\onlinecite{Hehlen.2013} for $\mathrm{Er^{3+}\!\!:\!LaCl_3}$ with the values $k^\mathrm{calc}$ and $f^\mathrm{calc}$ calculated via \texttt{YALIP}/\ameli, alongside the dominant $LS$-components of each energy level in intermediate coupling.}
  \renewcommand{\arraystretch}{1.35}
  \setlength{\tabcolsep}{6pt}
  \begin{tabular}{|rrr|rrrrr|l|}
  \hline
  \multicolumn{1}{|c}{$k^\mathrm{ref}$} &\multicolumn{1}{c}{$k^\mathrm{calc}$} &\multicolumn{1}{c|}{$\Delta_\mathrm{abs}$} &\multicolumn{1}{|c}{$f_{ed}^\mathrm{ref}$} &\multicolumn{1}{c}{$f_{md}^\mathrm{ref}$} &\multicolumn{1}{c}{$f_{ed}^\mathrm{calc}$} &\multicolumn{1}{c}{$f_{md}^\mathrm{calc}$} &\multicolumn{1}{c|}{$\Delta_\mathrm{rel}$} &\multicolumn{1}{|c|}{Intermediate State} \\
  \multicolumn{1}{|c}{\unit{cm^{-1}}} &\multicolumn{1}{c}{\unit{cm^{-1}}} &\multicolumn{1}{c|}{\unit{cm^{-1}}} &\multicolumn{1}{|c}{$10^{-8}$} &\multicolumn{1}{c}{$10^{-8}$} &\multicolumn{1}{c}{$10^{-8}$} &\multicolumn{1}{c}{$10^{-8}$} &\multicolumn{1}{c|}{\%} & \\
  \hline
  0 & -0.0 & 0.0 & 0 & 0 & 0 & 0 &  &
  $0.97 \prescript{4}{}{\mathrm{I}}_{15/2} + 0.03 \prescript{2}{}{\mathrm{K}}_{15/2}$ \\
  6527 & 6526.8 & 0.2 & 103.90 & 57.57 & 103.99 & 57.63 & -0.1 &
  $0.99 \prescript{4}{}{\mathrm{I}}_{13/2}$ \\
  10145 & 10145.2 & -0.2 & 52.70 & 0 & 52.59 & 0 & 0.2 &
  $0.82 \prescript{4}{}{\mathrm{I}}_{11/2} + 0.15 \prescript{2}{}{\mathrm{H}}(2)_{11/2} + 0.01 \prescript{4}{}{\mathrm{G}}_{11/2} \ldots$ \\
  12296 & 12296.0 & -0.0 & 57.33 & 0 & 56.94 & 0 & 0.7 &
  $0.50 \prescript{4}{}{\mathrm{I}}_{9/2} + 0.17 \prescript{2}{}{\mathrm{H}}(2)_{9/2} + 0.15 \prescript{4}{}{\mathrm{F}}_{9/2} \ldots$ \\
  15139 & 15139.1 & -0.1 & 264.10 & 0 & 263.53 & 0 & 0.2 &
  $0.57 \prescript{4}{}{\mathrm{F}}_{9/2} + 0.29 \prescript{4}{}{\mathrm{I}}_{9/2} + 0.08 \prescript{2}{}{\mathrm{G}}(1)_{9/2} \ldots$ \\
  18311 & 18311.0 & -0.0 & 35.02 & 0 & 34.87 & 0 & 0.4 &
  $0.69 \prescript{4}{}{\mathrm{S}}_{3/2} + 0.18 \prescript{2}{}{\mathrm{P}}_{3/2} + 0.08 \prescript{2}{}{\mathrm{D}}(1)_{3/2} \ldots$ \\
  19185 & 19185.3 & -0.3 & 1156.00 & 0 & 1163.62 & 0 & -0.7 &
  $0.47 \prescript{2}{}{\mathrm{H}}(2)_{11/2} + 0.36 \prescript{4}{}{\mathrm{G}}_{11/2} + 0.15 \prescript{4}{}{\mathrm{I}}_{11/2} \ldots$ \\
  20304 & 20303.8 & 0.2 & 189.70 & 0 & 188.49 & 0 & 0.6 &
  $0.92 \prescript{4}{}{\mathrm{F}}_{7/2} + 0.05 \prescript{2}{}{\mathrm{G}}(1)_{7/2} + 0.02 \prescript{2}{}{\mathrm{G}}(2)_{7/2}$ \\
  21949 & 21949.0 & 0.0 & 43.48 & 0 & 43.12 & 0 & 0.8 &
  $0.84 \prescript{4}{}{\mathrm{F}}_{5/2} + 0.13 \prescript{2}{}{\mathrm{D}}(1)_{5/2} + 0.02 \prescript{2}{}{\mathrm{D}}(2)_{5/2} \ldots$ \\
  22303 & 22303.1 & -0.1 & 25.05 & 0 & 24.77 & 0 & 1.1 &
  $0.63 \prescript{4}{}{\mathrm{F}}_{3/2} + 0.20 \prescript{2}{}{\mathrm{D}}(1)_{3/2} + 0.16 \prescript{4}{}{\mathrm{S}}_{3/2}$ \\
  24430 & 24429.9 & 0.1 & 61.42 & 0 & 61.30 & 0 & 0.2 &
  $0.24 \prescript{4}{}{\mathrm{F}}_{9/2} + 0.19 \prescript{2}{}{\mathrm{G}}(1)_{9/2} + 0.15 \prescript{2}{}{\mathrm{G}}(2)_{9/2} \ldots$ \\
  26415 & 26414.7 & 0.3 & 2159.00 & 0 & 2172.91 & 0 & -0.6 &
  $0.60 \prescript{4}{}{\mathrm{G}}_{11/2} + 0.26 \prescript{2}{}{\mathrm{H}}(2)_{11/2} + 0.11 \prescript{2}{}{\mathrm{H}}(1)_{11/2} \ldots$ \\
   & 27164.5 &  &  &  & 69.49 & 7.62 &  &
  $0.91 \prescript{2}{}{\mathrm{K}}_{15/2} + 0.06 \prescript{2}{}{\mathrm{L}}_{15/2} + 0.03 \prescript{4}{}{\mathrm{I}}_{15/2}$ \\
   & 27412.5 &  &  &  & 219.39 & 0 &  &
  $0.79 \prescript{4}{}{\mathrm{G}}_{9/2} + 0.15 \prescript{2}{}{\mathrm{H}}(2)_{9/2} + 0.05 \prescript{4}{}{\mathrm{I}}_{9/2} \ldots$ \\
   & 27925.0 &  &  &  & 44.96 & 0 &  &
  $0.42 \prescript{4}{}{\mathrm{G}}_{7/2} + 0.26 \prescript{2}{}{\mathrm{G}}(1)_{7/2} + 0.24 \prescript{2}{}{\mathrm{G}}(2)_{7/2} \ldots$ \\
   & 31560.0 &  &  &  & 5.54 & 0 &  &
  $0.36 \prescript{2}{}{\mathrm{P}}_{3/2} + 0.24 \prescript{4}{}{\mathrm{F}}_{3/2} + 0.21 \prescript{2}{}{\mathrm{D}}(1)_{3/2} \ldots$ \\
   & 32451.2 &  &  &  & 18.29 & 0.26 &  &
  $0.89 \prescript{2}{}{\mathrm{K}}_{13/2} + 0.10 \prescript{2}{}{\mathrm{I}}_{13/2}$ \\
   & 33321.5 &  &  &  & 0.92 & 0 &  &
  $0.92 \prescript{4}{}{\mathrm{G}}_{5/2} + 0.03 \prescript{2}{}{\mathrm{F}}(2)_{5/2} + 0.02 \prescript{2}{}{\mathrm{F}}(1)_{5/2} \ldots$ \\
   & 33334.4 &  &  &  & 0 & 0 &  &
  $0.92 \prescript{2}{}{\mathrm{P}}_{1/2} + 0.08 \prescript{4}{}{\mathrm{D}}_{1/2}$ \\
   & 33960.8 &  &  &  & 40.37 & 0 &  &
  $0.54 \prescript{4}{}{\mathrm{G}}_{7/2} + 0.28 \prescript{2}{}{\mathrm{G}}(1)_{7/2} + 0.15 \prescript{2}{}{\mathrm{G}}(2)_{7/2} \ldots$ \\
   & 34657.7 &  &  &  & 9.42 & 0 &  &
  $0.57 \prescript{2}{}{\mathrm{D}}(1)_{5/2} + 0.15 \prescript{2}{}{\mathrm{D}}(2)_{5/2} + 0.14 \prescript{4}{}{\mathrm{D}}_{5/2} \ldots$ \\
   & 36526.2 &  &  &  & 76.98 & 0 &  &
  $0.32 \prescript{2}{}{\mathrm{H}}(2)_{9/2} + 0.24 \prescript{2}{}{\mathrm{G}}(1)_{9/2} + 0.15 \prescript{2}{}{\mathrm{G}}(2)_{9/2} \ldots$ \\
   & 38536.4 &  &  &  & 21.68 & 0 &  &
  $0.45 \prescript{4}{}{\mathrm{D}}_{5/2} + 0.28 \prescript{2}{}{\mathrm{D}}(1)_{5/2} + 0.23 \prescript{2}{}{\mathrm{D}}(2)_{5/2} \ldots$ \\
   & 39082.0 &  &  &  & 2222.48 & 0 &  &
  $0.95 \prescript{4}{}{\mathrm{D}}_{7/2} + 0.02 \prescript{2}{}{\mathrm{F}}(1)_{7/2} + 0.01 \prescript{4}{}{\mathrm{G}}_{7/2} \ldots$ \\
   & 40116.2 &  &  &  & 83.29 & 0 &  &
  $0.66 \prescript{2}{}{\mathrm{I}}_{11/2} + 0.30 \prescript{2}{}{\mathrm{H}}(1)_{11/2} + 0.02 \prescript{2}{}{\mathrm{H}}(2)_{11/2} \ldots$ \\
   & 40471.4 &  &  &  & 53.23 & 0.03 &  &
  $1.00 \prescript{2}{}{\mathrm{L}}_{17/2}$ \\
   & 42169.6 &  &  &  & 2.45 & 0 &  &
  $0.59 \prescript{4}{}{\mathrm{D}}_{3/2} + 0.25 \prescript{2}{}{\mathrm{D}}(1)_{3/2} + 0.06 \prescript{2}{}{\mathrm{P}}_{3/2} \ldots$ \\
   & 42842.6 &  &  &  & 20.45 & 0.13 &  &
  $0.90 \prescript{2}{}{\mathrm{I}}_{13/2} + 0.10 \prescript{2}{}{\mathrm{K}}_{13/2}$ \\
   & 42914.7 &  &  &  & 0.03 & 0 &  &
  $0.37 \prescript{2}{}{\mathrm{P}}_{3/2} + 0.24 \prescript{2}{}{\mathrm{D}}(2)_{3/2} + 0.19 \prescript{2}{}{\mathrm{D}}(1)_{3/2} \ldots$ \\
   & 46752.8 &  &  &  & 2.74 & 0.02 &  &
  $0.94 \prescript{2}{}{\mathrm{L}}_{15/2} + 0.06 \prescript{2}{}{\mathrm{K}}_{15/2}$ \\
   & 46980.3 &  &  &  & 2.95 & 0 &  &
  $0.76 \prescript{2}{}{\mathrm{H}}(1)_{9/2} + 0.19 \prescript{2}{}{\mathrm{H}}(2)_{9/2} + 0.03 \prescript{2}{}{\mathrm{G}}(2)_{9/2} \ldots$ \\
   & 46980.5 &  &  &  & 0 & 0 &  &
  $0.92 \prescript{4}{}{\mathrm{D}}_{1/2} + 0.08 \prescript{2}{}{\mathrm{P}}_{1/2}$ \\
   & 48947.7 &  &  &  & 2.63 & 0 &  &
  $0.46 \prescript{2}{}{\mathrm{D}}(2)_{5/2} + 0.39 \prescript{4}{}{\mathrm{D}}_{5/2} + 0.08 \prescript{2}{}{\mathrm{F}}(2)_{5/2} \ldots$ \\
   & 50149.3 &  &  &  & 8.42 & 0 &  &
  $0.55 \prescript{2}{}{\mathrm{H}}(1)_{11/2} + 0.33 \prescript{2}{}{\mathrm{I}}_{11/2} + 0.10 \prescript{2}{}{\mathrm{H}}(2)_{11/2} \ldots$ \\
   & 54997.0 &  &  &  & 11.18 & 0 &  &
  $0.57 \prescript{2}{}{\mathrm{F}}(2)_{7/2} + 0.34 \prescript{2}{}{\mathrm{F}}(1)_{7/2} + 0.04 \prescript{4}{}{\mathrm{D}}_{7/2} \ldots$ \\
   & 55093.1 &  &  &  & 0.28 & 0 &  &
  $0.69 \prescript{2}{}{\mathrm{D}}(2)_{3/2} + 0.21 \prescript{4}{}{\mathrm{D}}_{3/2} + 0.06 \prescript{2}{}{\mathrm{D}}(1)_{3/2} \ldots$ \\
   & 63018.6 &  &  &  & 0.00 & 0 &  &
  $0.65 \prescript{2}{}{\mathrm{F}}(2)_{5/2} + 0.16 \prescript{2}{}{\mathrm{F}}(1)_{5/2} + 0.12 \prescript{2}{}{\mathrm{D}}(2)_{5/2} \ldots$ \\
   & 64605.1 &  &  &  & 1.49 & 0 &  &
  $0.58 \prescript{2}{}{\mathrm{G}}(2)_{7/2} + 0.39 \prescript{2}{}{\mathrm{G}}(1)_{7/2} + 0.03 \prescript{2}{}{\mathrm{F}}(2)_{7/2}$ \\
   & 68772.4 &  &  &  & 2.54 & 0 &  &
  $0.57 \prescript{2}{}{\mathrm{G}}(2)_{9/2} + 0.40 \prescript{2}{}{\mathrm{G}}(1)_{9/2} + 0.02 \prescript{2}{}{\mathrm{H}}(1)_{9/2} \ldots$ \\
   & 92919.2 &  &  &  & 0.00 & 0 &  &
  $0.78 \prescript{2}{}{\mathrm{F}}(1)_{5/2} + 0.22 \prescript{2}{}{\mathrm{F}}(2)_{5/2}$ \\
   & 96597.9 &  &  &  & 14.82 & 0 &  &
  $0.61 \prescript{2}{}{\mathrm{F}}(1)_{7/2} + 0.37 \prescript{2}{}{\mathrm{F}}(2)_{7/2} + 0.01 \prescript{2}{}{\mathrm{G}}(2)_{7/2} \ldots$ \\
  \hline
  \multicolumn{1}{|l}{RMS:} & & 0.2 & & & & & 0.6 & \\
  \hline
\end{tabular}
\end{table*}

The column $k^\mathrm{ref}$ in Table~\ref{tab:example_1} lists the published calculation results, while column $k^\mathrm{calc}$ displays the eigenvalues of the total Hamiltonian determined by \texttt{YALIP}.
In all instances, the absolute difference $\Delta_\mathrm{abs}$ between the two values is much smaller than the precision limit of the published data.

The eigenvectors of the total Hamiltonian define the composition of each state in intermediate coupling.
The final column in Table~\ref{tab:example_1} lists the respective $LS$-components in decreasing order of their weight factors, which correspond to the squared elements of the eigenvectors.

Evaluating the electric dipole oscillator strengths requires the provided Judd-Ofelt parameters: $\Omega_2=\qty{5.449}{pm^2}$, $\Omega_4=\qty{2.077}{pm^2}$, and $\Omega_6=\qty{0.6873}{pm^2}$.
According to Judd-Ofelt theory, the radiative oscillator strength of the transition $i \to j$ is expressed as~\cite{Judd.1962,Ofelt.1962,Krupke.1966}:
\begin{equation}
\label{eq:fed}
    f^\mathrm{ed}_{ij} =
    \frac{4 \pi m_e c k_{ij}}{3(2J_i+1)\hbar}\, \chi^\prime_\mathrm{ed}\!\!
    \sum\limits_{k=2,4,6}
    \!\!\Omega_k |\langle J_j\parallel \tensor{U}{k} \parallel J_i\rangle|^2
    \ ,
\end{equation}
where $m_e$ is the electron mass, $c$ the speed of light in vacuum, $k_{ij}$ the wavenumber difference between the initial and final states, and $\hbar$ the reduced Planck constant.

The local field correction factor for absorption transitions~\cite{Fowler.1962,Axe.1963}
\begin{equation}
  \chi^\prime_\mathrm{ed} = \frac{(n^2+2)^2}{9n}
\end{equation}
depends on the spectral refractive index $n(k_{ij})$, which is parametrized in Ref.~\onlinecite{Hehlen.2013} using the Sellmeier coefficients $B_1=0.62646$, $C^\prime_1=\qty{0.061295}{\micro\meter^2}$, $B_2=1.5212$, $C^\prime_2=\qty{0.0061087}{\micro\meter^2}$, $B_3=0.2465$, and $C^\prime_3=\qty{0.014086}{\micro\meter^2}$.
\texttt{YALIP} implements the Sellmeier equation using the alternative parameters $C_i=\sqrt{C^\prime_i}$:
\begin{equation}
    n(\lambda) = \sqrt{1 + \sum_i \frac{B_i \lambda^2}{\lambda^2-C_i^2}} \,,
\end{equation}
where the wavelength is given by $\lambda=1/k_{ij}$.

The magnetic dipole oscillator strength does not require empirical parameters and can be calculated directly from the intermediate coupling states~\cite{Condon.1935,Caspary.2002}:
\begin{equation}
\label{eq:fmd}
    f^\mathrm{md}_{ij} =
    \frac{\pi k_{ij} \hbar}{3(2J_i+1)m_e c}\, \chi^\prime_\mathrm{md}
    |\langle J_j\parallel \frac{\operator{L} + g_s \operator{S}}{\hbar} \parallel J_i\rangle|^2
    \ ,
\end{equation}
where $\chi^\prime_\mathrm{md}=n$ is the corresponding local field correction and $g_s$ the electron spin $g$-factor.

The matrices of the reduced elements \verb"U/{k}", \verb"L", and \verb"S" in $LS$-coupling are located in the \verb"slj_reduced" folder within the \verb"slj.zip" container of the \ameli\ repository. 
Because both the operator matrices in the \verb"slj" folder and the reduced operator matrices in the \verb"slj_reduced" folder use an identical basis configuration, where each $J$-multiplet is represented by a single row and column index, their transformation to intermediate coupling is perfectly isomorphic. 
To apply them in Eqs.~\eqref{eq:fed} and \eqref{eq:fmd}, each $LS$-coupled reduced matrix $R_{LS}$ can therefore be transformed to the intermediate coupling scheme directly via the eigenvector matrix $C$ obtained from the eigenvalue-decomposition of the total Hamiltonian:
\begin{equation}
  \label{eq:Ric}
  R_{IC} = C^T R_{LS}\, C
  \ .
\end{equation}

Columns $f_{ed}^\mathrm{ref}$ and $f_{md}^\mathrm{ref}$ in Table~\ref{tab:example_1} present the published calculation results for the electric and magnetic dipole oscillator strengths of ground-state absorption transitions, while columns $f_{ed}^\mathrm{calc}$ and $f_{md}^\mathrm{calc}$ display the values determined by \texttt{YALIP}.
For the level $\prescript{4}{}{I}_{11/2}$ the published electric dipole oscillator strength $0.5270$ is taken from the column 'total'.
The value of $0.5527$ in the 'ED' column should theoretically be identical, but it seems to contain a typesetting error.

The root-mean-square of the relative differences $\Delta_\mathrm{rel}$ for the total oscillator strengths is $\qty{0.6}{\percent}$, which is notably larger than the precision implied by the four significant digits of the published values.
However, the uniform distribution of these deviations indicates the absence of systematic errors. Instead, it suggests that the actual precision of the reference calculations is lower than their presentation would suggest.

Importantly, this discrepancy does not stem from the precision of the Judd-Ofelt fit itself, but rather from the accumulation of floating-point errors along the evaluation path, originating from the initial radial integrals and perturbation matrix elements, propagating through the eigenvectors, and compounding with the Judd-Ofelt parameters to yield the final oscillator strengths.
A rigorous cross-verification would require comparing the raw matrix elements generated by the software used in Ref.~\onlinecite{Hehlen.2013} with the exact-arithmetic values provided by the \ameli\ repository.

Regardless, it is remarkable that these extensive results are derived from only four radial integrals and three Judd-Ofelt parameters.
The essential code lines are:
\begin{verbatim}
from math import sqrt
from yalip import Levels, Coupling, Sellmeier

config = "f11"
coupling = Coupling.SLJ
radial = {"F_2": 433.22, "F_4": 66.887,
          "F_6": 7.2952, "H2": 2385.9}
omega = {"JO/2": 5.449, "JO/4": 2.077,
         "JO/6": 0.6873}
material = Sellmeier(
        0.62646, sqrt(0.061295),
        1.5212, sqrt(0.0061087),
        0.2465, sqrt(0.014086))

ion = Levels(config, coupling, radial,
             omega, material)
\end{verbatim}

The comprehensive Python script \texttt{example\_1.py} used to generate Table~\ref{tab:example_1} is contained in the \texttt{YALIP} software repository~\cite{Caspary.2026c}.

In summary, and in contrast to conventional workflows, this approach completely bypasses the need for tabulated lists of reduced matrix elements.
Instead, a compact set of radial parameters yields immediate access to every radiative transition property in both absorption and emission.
All derived physical quantities, such as level energies, state compositions, oscillator strengths, and radiative rates, are readily accessible as attributes or methods of the \verb"ion" object immediately following execution of the script above.

\subsection{Full Interaction Crystal Spectrum}

In the second example, we reproduce some foundational data published in 1978 by Carnall et~al.\ \cite{Carnall.1978} in a comprehensive Argonne National Laboratory report summarizing their detailed investigation of lanthanide ions doped into $\mathrm{LaF_3}$ crystals.
This work has since become a standard reference source for radial integrals, energy levels, and reduced matrix elements.
Because the semi-empirical theory of lanthanide spectra was already well-established at that time, the report provides radial integrals for all six types of perturbation Hamiltonians from the standard set.

Furthermore, for odd-electron configurations, the report supplies optimized crystal field parameters $B^2_0$, $B^4_0$, $B^6_0$, and $B^6_6$ (Wybourne notation) assuming a $\mathrm{D_{3h}}$ site symmetry.
While the authors provide evidence that the actual site symmetry of lanthanide ions in $\mathrm{LaF_3}$ is as low as $\mathrm{C_2}$, which would instead require 15 crystal field parameters~\cite{Morrison.1979,Carnall.1983}, the Kramers degeneracy inherent to configurations with an odd number of electrons renders $\mathrm{D_{3h}}$ an excellent approximation for these ions.

We select $\mathrm{Nd^{3+}\!\!:\!LaF_3}$ to demonstrate the capacity of \texttt{YALIP} and \ameli\ to replicate these foundational datasets.
A three-electron configuration is sufficiently large to encompass even the most complex perturbation Hamiltonian, $\operator{H}_4$, which contains six additional angular matrices.
Moreover, due to the odd electron count, the original publication provides a complete list of calculated energies for all 182 Kramers doublets.
Fortunately, this dataset is compact enough to allow the comprehensive comparison in Table~\ref{tab:example_2} to fit on a single printed page.
The Python script \texttt{example\_2.py} used to generate this table is contained in the \texttt{YALIP} software repository~\cite{Caspary.2026c}.

\begin{table}
  \caption{\label{tab:NdParameters} Radial integrals and crystal field parameters from Ref.~\onlinecite{Carnall.1978} for $\mathrm{Nd^{3+}\!\!:\!LaF_3}$ alongside the corresponding \texttt{YALIP}/\ameli\ matrix designations.}
  \setlength{\tabcolsep}{12pt}
  \begin{tabular}{|cSc|}
    \hline
    Parameter & \multicolumn{1}{l}{Value} & Matrix \\
    & \multicolumn{1}{l}{\unit{cm^{-1}}} & \\
    \hline
    $F^2$ & 73036 & \verb"H1/2" \\
    $F^4$ & 52624 & \verb"H1/4" \\
    $F^6$ & 35793 & \verb"H1/6" \\
    $\zeta$ & 884.9 & \verb"H2" \\
    $\alpha$ & 21.28 & \verb"H3/0" \\
    $\beta$ & -583 & \verb"H3/1" \\
    $\gamma$ & 1443 & \verb"H3/2" \\
    $T^2$ & 306 & \verb"H4/2" \\
    $T^3$ & 41 & \verb"H4/3" \\
    $T^4$ & 59 & \verb"H4/4" \\
    $T^6$ & -283 & \verb"H4/6" \\
    $T^7$ & 326 & \verb"H4/7" \\
    $T^8$ & 298 & \verb"H4/8" \\
    $M^0$ & 2.237 & \verb"H5/0" \\
    $M^2$ & 1.248 & \verb"H5/2" \\
    $M^4$ & 0.84 & \verb"H5/4" \\
    $P^2$ & 213 & \verb"H6/2" \\
    $P^4$ & 160 & \verb"H6/4" \\
    $P^6$ & 106.5 & \verb"H6/6" \\
    $B^2_0$ & 216 & \verb"Hcf/2,0" \\
    $B^4_0$ & 1225 & \verb"Hcf/4,0" \\
    $B^6_0$ & 1506 & \verb"Hcf/6,0" \\
    $B^6_6$ & 770 & \verb"Hcf/6,6" \\
    \hline
  \end{tabular}
\end{table}

In contrast to the four radial integrals evaluated in our first example, the total perturbation Hamiltonian here is a linear combination of 23 matrices scaled by the empirical factors listed in Table~\ref{tab:NdParameters}.
Consequently, this case serves as a rigorous test for the full spectrum of interaction operators hosted within the \ameli\ repository.

Since this example incorporates crystal field splitting, \texttt{YALIP} extracts the required matrices for the $f^3$ configuration from the \verb"product" folder within the repository's \verb"product.zip" archive and uses the transformation matrix from \verb"transform.zdc" to determine the $LS$-components in intermediate coupling using the algorithm explained at the end of Subsection~\ref{sub:phase}.

\begin{table*}[p]
  \caption{\label{tab:example_2} Comparison of calculated energy levels $k^\mathrm{ref}$ from Ref.~\onlinecite{Carnall.1978} for $\mathrm{Nd^{3+}\!\!:\!LaF_3}$ with the eigenvalues $k^\mathrm{calc}$ computed via \texttt{YALIP}/\ameli. 
  Discovered discrepancies in Ref.~\onlinecite{Carnall.1978}: the original text erroneously lists $\prescript{a}{}{M}=3/2$ and $\prescript{b}{}{k}^\mathrm{ref}=\qty{28634}{cm^{-1}}$.}
  \renewcommand{\arraystretch}{0.0}
  \setlength{\tabcolsep}{4pt}
  \begin{tabular}{|lrrrr|lrrrr|lrrrr|}
  \hline
  \multicolumn{2}{|c}{Level} & \multicolumn{1}{c}{$k^\mathrm{ref}$} & \multicolumn{1}{c}{$k^\mathrm{calc}$} & \multicolumn{1}{c|}{$\Delta_\mathrm{abs}$} & $\prescript{4}{}{\mathrm{G}}_{7/2}$ & $5/2$ & $19139$ & $19139.2$ & $-0.2$ & $\prescript{2}{}{\mathrm{L}}_{15/2}$ & $9/2$ & $30615$ & $30615.1$ & $-0.1$ \\[-0.09ex]
  \multicolumn{1}{|l}{$\prescript{2S+1}{}{L}_{J}$} & \multicolumn{1}{r}{$M$} & \multicolumn{1}{c}{\unit{cm^{-1}}} & \multicolumn{1}{c}{\unit{cm^{-1}}} & \multicolumn{1}{c|}{\unit{cm^{-1}}} & $\prescript{4}{}{\mathrm{G}}_{7/2}$ & $1/2$ & $19245$ & $19244.7$ & $0.3$ & $\prescript{4}{}{\mathrm{D}}_{7/2}$ & $7/2$ & $30646$ & $30646.7$ & $-0.7$ \\[-0.09ex]
  \cline{1-5}
  $\prescript{4}{}{\mathrm{I}}_{9/2}$ & $9/2$ & $3$ & $1.9$ & $1.1$ & $\prescript{4}{}{\mathrm{G}}_{7/2}$ & $3/2$ & $19271$ & $19271.0$ & $0.0$ & $\prescript{2}{}{\mathrm{L}}_{15/2}$ & $7/2$ & $30701$ & $30700.9$ & $0.1$ \\[-0.09ex]
  $\prescript{4}{}{\mathrm{I}}_{9/2}$ & $5/2$ & $38$ & $37.8$ & $0.2$ & $\prescript{4}{}{\mathrm{G}}_{7/2}$ & $7/2$ & $19324$ & $19324.1$ & $-0.1$ & $\prescript{4}{}{\mathrm{D}}_{7/2}$ & $3/2$ & $30712$ & $30712.3$ & $-0.3$ \\[-0.09ex]
  $\prescript{4}{}{\mathrm{I}}_{9/2}$ & $3/2$ & $142$ & $141.3$ & $0.7$ & $\prescript{2}{}{\mathrm{K}}_{13/2}$ & $13/2$ & $19567$ & $19566.7$ & $0.3$ & $\prescript{4}{}{\mathrm{D}}_{7/2}$ & $1/2$ & $30792$ & $30792.5$ & $-0.5$ \\[-0.09ex]
  $\prescript{4}{}{\mathrm{I}}_{9/2}$ & $1/2$ & $294$ & $293.7$ & $0.3$ & $\prescript{4}{}{\mathrm{G}}_{9/2}$ & $5/2$ & $19632$ & $19632.1$ & $-0.1$ & $\prescript{2}{}{\mathrm{I}}_{13/2}$ & $11/2$ & $30850$ & $30849.5$ & $0.5$ \\[-0.09ex]
  $\prescript{4}{}{\mathrm{I}}_{9/2}$ & $7/2$ & $500$ & $499.7$ & $0.3$ & $\prescript{2}{}{\mathrm{K}}_{13/2}$ & $1/2$ & $19645$ & $19645.2$ & $-0.2$ & $\prescript{2}{}{\mathrm{I}}_{13/2}$ & $9/2$ & $30895$ & $30894.8$ & $0.2$ \\[-0.09ex]
  $\prescript{4}{}{\mathrm{I}}_{11/2}$ & $11/2$ & $1963$ & $1962.7$ & $0.3$ & $\prescript{4}{}{\mathrm{G}}_{9/2}$ & $7/2$ & $19687$ & $19687.0$ & $-0.0$ & $\prescript{2}{}{\mathrm{I}}_{13/2}$ & $7/2$ & $30955$ & $30954.5$ & $0.5$ \\[-0.09ex]
  $\prescript{4}{}{\mathrm{I}}_{11/2}$ & $5/2$ & $2042$ & $2041.6$ & $0.4$ & $\prescript{4}{}{\mathrm{G}}_{9/2}$ & $9/2$ & $19693$ & $19693.0$ & $-0.0$ & $\prescript{2}{}{\mathrm{I}}_{13/2}$ & $13/2$ & $31002$ & $31001.7$ & $0.3$ \\[-0.09ex]
  $\prescript{4}{}{\mathrm{I}}_{11/2}$ & $3/2$ & $2075$ & $2074.9$ & $0.1$ & $\prescript{2}{}{\mathrm{K}}_{13/2}$ & $11/2$ & $19737$ & $19737.3$ & $-0.3$ & $\prescript{2}{}{\mathrm{I}}_{13/2}$ & $5/2$ & $31041$ & $31040.5$ & $0.5$ \\[-0.09ex]
  $\prescript{4}{}{\mathrm{I}}_{11/2}$ & $1/2$ & $2098$ & $2097.9$ & $0.1$ & $\prescript{4}{}{\mathrm{G}}_{9/2}$ & $3/2$ & $19738$ & $19737.7$ & $0.3$ & $\prescript{2}{}{\mathrm{I}}_{13/2}$ & $3/2$ & $31070$ & $31069.8$ & $0.2$ \\[-0.09ex]
  $\prescript{4}{}{\mathrm{I}}_{11/2}$ & $7/2$ & $2203$ & $2202.3$ & $0.7$ & $\prescript{2}{}{\mathrm{K}}_{13/2}$ & $3/2$ & $19790$ & $19789.6$ & $0.4$ & $\prescript{2}{}{\mathrm{I}}_{13/2}$ & $1/2$ & $31079$ & $31078.8$ & $0.2$ \\[-0.09ex]
  $\prescript{4}{}{\mathrm{I}}_{11/2}$ & $9/2$ & $2227$ & $2226.8$ & $0.2$ & $\prescript{4}{}{\mathrm{G}}_{9/2}$ & $1/2$ & $19845$ & $19844.5$ & $0.5$ & $\prescript{2}{}{\mathrm{L}}_{17/2}$ & $15/2$ & $31767$ & $31767.6$ & $-0.6$ \\[-0.09ex]
  $\prescript{4}{}{\mathrm{I}}_{13/2}$ & $13/2$ & $3901$ & $3901.0$ & $0.0$ & $\prescript{2}{}{\mathrm{K}}_{13/2}$ & $5/2$ & $19917$ & $19916.5$ & $0.5$ & $\prescript{2}{}{\mathrm{L}}_{17/2}$ & $17/2$ & $31836$ & $31836.3$ & $-0.3$ \\[-0.09ex]
  $\prescript{4}{}{\mathrm{I}}_{13/2}$ & $5/2$ & $3983$ & $3982.7$ & $0.3$ & $\prescript{2}{}{\mathrm{K}}_{13/2}$ & $9/2$ & $19927$ & $19927.1$ & $-0.1$ & $\prescript{2}{}{\mathrm{L}}_{17/2}$ & $1/2$ & $31926$ & $31926.4$ & $-0.4$ \\[-0.09ex]
  $\prescript{4}{}{\mathrm{I}}_{13/2}$ & $3/2$ & $4043$ & $4043.3$ & $-0.3$ & $\prescript{2}{}{\mathrm{K}}_{13/2}$ & $7/2$ & $19971$ & $19970.8$ & $0.2$ & $\prescript{2}{}{\mathrm{L}}_{17/2}$ & $3/2$ & $31968$ & $31968.3$ & $-0.3$ \\[-0.09ex]
  $\prescript{4}{}{\mathrm{I}}_{13/2}$ & $1/2$ & $4102$ & $4101.5$ & $0.5$ & $\prescript{2}{}{\mathrm{G}}(1)_{9/2}$ & $7/2$ & $21150$ & $21150.4$ & $-0.4$ & $\prescript{2}{}{\mathrm{L}}_{17/2}$ & $13/2$ & $31990$ & $31990.2$ & $-0.2$ \\[-0.09ex]
  $\prescript{4}{}{\mathrm{I}}_{13/2}$ & $7/2$ & $4126$ & $4125.6$ & $0.4$ & $\prescript{2}{}{\mathrm{G}}(1)_{9/2}$ & $3/2$ & $21183$ & $21183.4$ & $-0.4$ & $\prescript{2}{}{\mathrm{L}}_{17/2}$ & $5/2$ & $32013$ & $32012.9$ & $0.1$ \\[-0.09ex]
  $\prescript{4}{}{\mathrm{I}}_{13/2}$ & $9/2$ & $4205$ & $4204.9$ & $0.1$ & $\prescript{2}{}{\mathrm{G}}(1)_{9/2}$ & $7/2$ & $21199$ & $21199.4$ & $-0.4$ & $\prescript{2}{}{\mathrm{L}}_{17/2}$ & $11/2$ & $32048$ & $32048.0$ & $-0.0$ \\[-0.09ex]
  $\prescript{4}{}{\mathrm{I}}_{13/2}$ & $11/2$ & $4275$ & $4275.3$ & $-0.3$ & $\prescript{2}{}{\mathrm{G}}(1)_{9/2}$ & $1/2$ & $21235$ & $21235.5$ & $-0.5$ & $\prescript{2}{}{\mathrm{L}}_{17/2}$ & $9/2$ & $32093$ & $32093.3$ & $-0.3$ \\[-0.09ex]
  $\prescript{4}{}{\mathrm{I}}_{15/2}$ & $15/2$ & $5820$ & $5820.4$ & $-0.4$ & $\prescript{2}{}{\mathrm{G}}(1)_{9/2}$ & $9/2$ & $21267$ & $21267.5$ & $-0.5$ & $\prescript{2}{}{\mathrm{L}}_{17/2}$ & $7/2$ & $32126$ & $32126.6$ & $-0.6$ \\[-0.09ex]
  $\prescript{4}{}{\mathrm{I}}_{15/2}$ & $7/2$ & $5838$ & $5837.6$ & $0.4$ & $\prescript{2}{}{\mathrm{D}}(1)_{3/2}$ & $3/2$ & $21339$ & $21338.8$ & $0.2$ & $\prescript{2}{}{\mathrm{H}}(1)_{9/2}$ & $7/2$ & $33035$ & $33035.2$ & $-0.2$ \\[-0.09ex]
  $\prescript{4}{}{\mathrm{I}}_{15/2}$ & $9/2$ & $5997$ & $5997.4$ & $-0.4$ & $\prescript{2}{}{\mathrm{D}}(1)_{3/2}$ & $1/2$ & $21352$ & $21352.3$ & $-0.3$ & $\prescript{2}{}{\mathrm{H}}(1)_{9/2}$ & $1/2$ & $33137$ & $33137.0$ & $0.0$ \\[-0.09ex]
  $\prescript{4}{}{\mathrm{I}}_{15/2}$ & $5/2$ & $6171$ & $6170.5$ & $0.5$ & $\prescript{4}{}{\mathrm{G}}_{11/2}$ & $7/2$ & $21535$ & $21535.7$ & $-0.7$ & $\prescript{2}{}{\mathrm{H}}(1)_{9/2}$ & $9/2$ & $33168$ & $33168.3$ & $-0.3$ \\[-0.09ex]
  $\prescript{4}{}{\mathrm{I}}_{15/2}$ & $1/2$ & $6187$ & $6186.9$ & $0.1$ & $\prescript{2}{}{\mathrm{K}}_{15/2}$ & $15/2$ & $21620$ & $21619.5$ & $0.5$ & $\prescript{2}{}{\mathrm{H}}(1)_{9/2}$ & $5/2$ & $33226$ & $33225.7$ & $0.3$ \\[-0.09ex]
  $\prescript{4}{}{\mathrm{I}}_{15/2}$ & $3/2$ & $6293$ & $6292.4$ & $0.6$ & $\prescript{4}{}{\mathrm{G}}_{11/2}$ & $5/2$ & $21621$ & $21621.7$ & $-0.7$ & $\prescript{2}{}{\mathrm{H}}(1)_{9/2}$ & $3/2$ & $33258$ & $33257.9$ & $0.1$ \\[-0.09ex]
  $\prescript{4}{}{\mathrm{I}}_{15/2}$ & $11/2$ & $6420$ & $6420.3$ & $-0.3$ & $\prescript{4}{}{\mathrm{G}}_{11/2}$ & $9/2$ & $21731$ & $21731.0$ & $-0.0$ & $\prescript{2}{}{\mathrm{D}}(2)_{3/2}$ & $3/2$ & $33612$ & $33611.9$ & $0.1$ \\[-0.09ex]
  $\prescript{4}{}{\mathrm{I}}_{15/2}$ & $13/2$ & $6545$ & $6544.7$ & $0.3$ & $\prescript{4}{}{\mathrm{G}}_{11/2}$ & $11/2$ & $21774$ & $21774.2$ & $-0.2$ & $\prescript{2}{}{\mathrm{D}}(2)_{3/2}$ & $1/2$ & $33631$ & $33631.3$ & $-0.3$ \\[-0.09ex]
  $\prescript{4}{}{\mathrm{F}}_{3/2}$ & $1/2$ & $11596$ & $11595.4$ & $0.6$ & $\prescript{2}{}{\mathrm{K}}_{15/2}$ & $13/2$ & $21779$ & $21779.1$ & $-0.1$ & $\prescript{2}{}{\mathrm{H}}(1)_{11/2}$ & $9/2$ & $34274$ & $34274.0$ & $-0.0$ \\[-0.09ex]
  $\prescript{4}{}{\mathrm{F}}_{3/2}$ & $3/2$ & $11626$ & $11626.1$ & $-0.1$ & $\prescript{4}{}{\mathrm{G}}_{11/2}$ & $9/2$ & $21790$ & $21790.2$ & $-0.2$ & $\prescript{2}{}{\mathrm{H}}(1)_{11/2}$ & $1/2$ & $34374$ & $34374.1$ & $-0.1$ \\[-0.09ex]
  $\prescript{2}{}{\mathrm{H}}(2)_{9/2}$ & $7/2$ & $12585$ & $12584.0$ & $1.0$ & $\prescript{2}{}{\mathrm{K}}_{15/2}$ & $1/2$ & $21824$ & $21824.5$ & $-0.5$ & $\prescript{2}{}{\mathrm{D}}(2)_{5/2}$ & $5/2$ & $34445$ & $34445.0$ & $-0.0$ \\[-0.09ex]
  $\prescript{4}{}{\mathrm{F}}_{5/2}$ & $3/2$ & $12589$ & $12588.5$ & $0.5$ & $\prescript{2}{}{\mathrm{K}}_{15/2}$ & $5/2$ & $21827$ & $21827.0$ & $0.0$ & $\prescript{2}{}{\mathrm{H}}(1)_{11/2}$ & $7/2$ & $34519$ & $34519.0$ & $0.0$ \\[-0.09ex]
  $\prescript{4}{}{\mathrm{F}}_{5/2}$ & $1/2$ & $12630$ & $12629.1$ & $0.9$ & $\prescript{2}{}{\mathrm{K}}_{15/2}$ & $3/2$ & $21857$ & $21856.6$ & $0.4$ & $\prescript{2}{}{\mathrm{H}}(1)_{11/2}$ & $1/2$ & $34551$ & $34551.3$ & $-0.3$ \\[-0.09ex]
  $\prescript{4}{}{\mathrm{F}}_{5/2}$ & $5/2$ & $12678$ & $12677.7$ & $0.3$ & $\prescript{2}{}{\mathrm{K}}_{15/2}$ & $11/2$ & $21901$ & $21901.3$ & $-0.3$ & $\prescript{2}{}{\mathrm{H}}(1)_{11/2}$ & $3/2$ & $34573$ & $34573.8$ & $-0.8$ \\[-0.09ex]
  $\prescript{4}{}{\mathrm{F}}_{5/2}$ & $1/2$ & $12704$ & $12703.6$ & $0.4$ & $\prescript{2}{}{\mathrm{K}}_{15/2}$ & $9/2$ & $21931$ & $21931.0$ & $-0.0$ & $\prescript{2}{}{\mathrm{H}}(1)_{11/2}$ & $11/2$ & $34686$ & $34686.0$ & $-0.0$ \\[-0.09ex]
  $\prescript{2}{}{\mathrm{H}}(2)_{9/2}$ & $3/2$ & $12763$ & $12762.3$ & $0.7$ & $\prescript{2}{}{\mathrm{K}}_{15/2}$ & $7/2$ & $21946$ & $21946.1$ & $-0.1$ & $\prescript{2}{}{\mathrm{D}}(2)_{5/2}$ & $3/2$ & $34709$ & $34709.0$ & $-0.0$ \\[-0.09ex]
  $\prescript{2}{}{\mathrm{H}}(2)_{9/2}$ & $9/2$ & $12854$ & $12853.0$ & $1.0$ & $\prescript{4}{}{\mathrm{G}}_{11/2}$ & $1/2$ & $21995$ & $21995.6$ & $-0.6$ & $\prescript{2}{}{\mathrm{H}}(1)_{11/2}$ & $5/2$ & $34818$ & $34818.3$ & $-0.3$ \\[-0.09ex]
  $\prescript{2}{}{\mathrm{H}}(2)_{9/2}$ & $5/2$ & $12873$ & $12872.1$ & $0.9$ & $\prescript{2}{}{\mathrm{P}}_{1/2}$ & $1/2$ & $23455$ & $23454.6$ & $0.4$ & $\prescript{2}{}{\mathrm{F}}(2)_{5/2}$ & $5/2$ & $38723$ & $38723.2$ & $-0.2$ \\[-0.09ex]
  $\prescript{4}{}{\mathrm{F}}_{7/2}$ & $3/2$ & $13514$ & $13513.7$ & $0.3$ & $\prescript{2}{}{\mathrm{D}}(1)_{5/2}$ & $3/2$ & $23996$ & $23996.8$ & $-0.8$ & $\prescript{2}{}{\mathrm{F}}(2)_{5/2}$ & $1/2$ & $38778$ & $38778.3$ & $-0.3$ \\[-0.09ex]
  $\prescript{4}{}{\mathrm{F}}_{7/2}$ & $7/2$ & $13583$ & $13582.7$ & $0.3$ & $\prescript{2}{}{\mathrm{D}}(1)_{5/2}$ & $1/2$ & $23999$ & $23999.6$ & $-0.6$ & $\prescript{2}{}{\mathrm{F}}(2)_{5/2}$ & $3/2$ & $38815$ & $38815.9$ & $-0.9$ \\[-0.09ex]
  $\prescript{4}{}{\mathrm{F}}_{7/2}$ & $1/2$ & $13673$ & $13672.7$ & $0.3$ & $\prescript{2}{}{\mathrm{D}}(1)_{5/2}$ & $5/2$ & $24057$ & $24057.8$ & $-0.8$ & $\prescript{2}{}{\mathrm{F}}(2)_{7/2}$ & $7/2$ & $40113$ & $40113.8$ & $-0.8$ \\[-0.09ex]
  $\prescript{4}{}{\mathrm{S}}_{3/2}$ & $1/2$ & $13693$ & $13692.0$ & $1.0$ & $\prescript{2}{}{\mathrm{P}}_{3/2}$ & $1/2$ & $26394$ & $26393.7$ & $0.3$ & $\prescript{2}{}{\mathrm{F}}(2)_{7/2}$ & $3/2$ & $40126$ & $40126.3$ & $-0.3$ \\[-0.09ex]
  $\prescript{4}{}{\mathrm{S}}_{3/2}$ & $3/2$ & $13695$ & $13693.7$ & $1.3$ & $\prescript{2}{}{\mathrm{P}}_{3/2}$ & $3/2$ & $26416$ & $26416.5$ & $-0.5$ & $\prescript{2}{}{\mathrm{F}}(2)_{7/2}$ & $1/2$ & $40187$ & $40187.9$ & $-0.9$ \\[-0.09ex]
  $\prescript{4}{}{\mathrm{F}}_{7/2}$ & $5/2$ & $13711$ & $13711.3$ & $-0.3$ & $\prescript{4}{}{\mathrm{D}}_{3/2}$ & $1/2$ & $28361$ & $28361.1$ & $-0.1$ & $\prescript{2}{}{\mathrm{F}}(2)_{7/2}$ & $5/2$ & $40254$ & $40254.7$ & $-0.7$ \\[-0.09ex]
  $\prescript{4}{}{\mathrm{F}}_{9/2}$ & $1/2$ & $14847$ & $14847.5$ & $-0.5$ & $\prescript{4}{}{\mathrm{D}}_{3/2}$ & $3/2$ & $28369$ & $28369.3$ & $-0.3$ & $\prescript{2}{}{\mathrm{G}}(2)_{9/2}$ & $5/2$ & $47871$ & $47871.2$ & $-0.2$ \\[-0.09ex]
  $\prescript{4}{}{\mathrm{F}}_{9/2}$ & $9/2^a$ & $14861$ & $14861.1$ & $-0.1$ & $\prescript{4}{}{\mathrm{D}}_{5/2}$ & $5/2$ & $28495$ & $28495.0$ & $0.0$ & $\prescript{2}{}{\mathrm{G}}(2)_{9/2}$ & $9/2$ & $47888$ & $47887.6$ & $0.4$ \\[-0.09ex]
  $\prescript{4}{}{\mathrm{F}}_{9/2}$ & $3/2$ & $14886$ & $14886.2$ & $-0.2$ & $\prescript{4}{}{\mathrm{D}}_{5/2}$ & $1/2$ & $28528$ & $28528.9$ & $-0.9$ & $\prescript{2}{}{\mathrm{G}}(2)_{9/2}$ & $3/2$ & $47964$ & $47963.8$ & $0.2$ \\[-0.09ex]
  $\prescript{4}{}{\mathrm{F}}_{9/2}$ & $5/2$ & $14924$ & $14924.2$ & $-0.2$ & $\prescript{4}{}{\mathrm{D}}_{5/2}$ & $3/2$ & $28686$ & $28686.2$ & $-0.2$ & $\prescript{2}{}{\mathrm{G}}(2)_{9/2}$ & $7/2$ & $48006$ & $48006.0$ & $0.0$ \\[-0.09ex]
  $\prescript{4}{}{\mathrm{F}}_{9/2}$ & $7/2$ & $14957$ & $14956.6$ & $0.4$ & $\prescript{4}{}{\mathrm{D}}_{1/2}$ & $1/2$ & $28938$ & $28938.8$ & $-0.8$ & $\prescript{2}{}{\mathrm{G}}(2)_{9/2}$ & $1/2$ & $48055$ & $48055.2$ & $-0.2$ \\[-0.09ex]
  $\prescript{2}{}{\mathrm{H}}(2)_{11/2}$ & $5/2$ & $16028$ & $16028.4$ & $-0.4$ & $\prescript{2}{}{\mathrm{I}}_{11/2}$ & $9/2$ & $29459$ & $29458.7$ & $0.3$ & $\prescript{2}{}{\mathrm{G}}(2)_{7/2}$ & $7/2$ & $48861$ & $48861.6$ & $-0.6$ \\[-0.09ex]
  $\prescript{2}{}{\mathrm{H}}(2)_{11/2}$ & $11/2$ & $16046$ & $16045.7$ & $0.3$ & $\prescript{2}{}{\mathrm{I}}_{11/2}$ & $7/2$ & $29475$ & $29474.4$ & $0.6$ & $\prescript{2}{}{\mathrm{G}}(2)_{7/2}$ & $3/2$ & $48869$ & $48869.3$ & $-0.3$ \\[-0.09ex]
  $\prescript{2}{}{\mathrm{H}}(2)_{11/2}$ & $3/2$ & $16059$ & $16058.9$ & $0.1$ & $\prescript{2}{}{\mathrm{I}}_{11/2}$ & $5/2$ & $29565$ & $29564.7$ & $0.3$ & $\prescript{2}{}{\mathrm{G}}(2)_{7/2}$ & $5/2$ & $48979$ & $48979.4$ & $-0.4$ \\[-0.09ex]
  $\prescript{2}{}{\mathrm{H}}(2)_{11/2}$ & $7/2$ & $16060$ & $16059.9$ & $0.1$ & $\prescript{2}{}{\mathrm{I}}_{11/2}$ & $11/2$ & $29634^b$ & $29633.3$ & $0.7$ & $\prescript{2}{}{\mathrm{G}}(2)_{7/2}$ & $1/2$ & $49065$ & $49065.1$ & $-0.1$ \\[-0.09ex]
  $\prescript{2}{}{\mathrm{H}}(2)_{11/2}$ & $1/2$ & $16095$ & $16095.2$ & $-0.2$ & $\prescript{2}{}{\mathrm{I}}_{11/2}$ & $3/2$ & $29659$ & $29658.8$ & $0.2$ & $\prescript{2}{}{\mathrm{F}}(1)_{7/2}$ & $5/2$ & $66548$ & $66547.8$ & $0.2$ \\[-0.09ex]
  $\prescript{2}{}{\mathrm{H}}(2)_{11/2}$ & $9/2$ & $16140$ & $16139.9$ & $0.1$ & $\prescript{2}{}{\mathrm{I}}_{11/2}$ & $1/2$ & $29767$ & $29766.6$ & $0.4$ & $\prescript{2}{}{\mathrm{F}}(1)_{7/2}$ & $7/2$ & $66705$ & $66705.1$ & $-0.1$ \\[-0.09ex]
  $\prescript{4}{}{\mathrm{G}}_{5/2}$ & $3/2$ & $17308$ & $17307.8$ & $0.2$ & $\prescript{2}{}{\mathrm{L}}_{15/2}$ & $13/2$ & $30271$ & $30271.5$ & $-0.5$ & $\prescript{2}{}{\mathrm{F}}(1)_{7/2}$ & $3/2$ & $66793$ & $66793.0$ & $-0.0$ \\[-0.09ex]
  $\prescript{4}{}{\mathrm{G}}_{5/2}$ & $1/2$ & $17311$ & $17310.5$ & $0.5$ & $\prescript{2}{}{\mathrm{L}}_{15/2}$ & $15/2$ & $30346$ & $30346.6$ & $-0.6$ & $\prescript{2}{}{\mathrm{F}}(1)_{7/2}$ & $1/2$ & $66859$ & $66858.5$ & $0.5$ \\[-0.09ex]
  $\prescript{4}{}{\mathrm{G}}_{5/2}$ & $5/2$ & $17360$ & $17359.6$ & $0.4$ & $\prescript{2}{}{\mathrm{L}}_{15/2}$ & $1/2$ & $30411$ & $30411.0$ & $-0.0$ & $\prescript{2}{}{\mathrm{F}}(1)_{5/2}$ & $5/2$ & $67857$ & $67856.7$ & $0.3$ \\[-0.09ex]
  $\prescript{4}{}{\mathrm{G}}_{7/2}$ & $5/2$ & $17491$ & $17490.8$ & $0.2$ & $\prescript{2}{}{\mathrm{L}}_{15/2}$ & $3/2$ & $30451$ & $30451.4$ & $-0.4$ & $\prescript{2}{}{\mathrm{F}}(1)_{5/2}$ & $3/2$ & $67858$ & $67858.2$ & $-0.2$ \\[-0.09ex]
  $\prescript{2}{}{\mathrm{G}}(1)_{7/2}$ & $3/2$ & $17505$ & $17505.0$ & $0.0$ & $\prescript{2}{}{\mathrm{L}}_{15/2}$ & $11/2$ & $30533$ & $30533.0$ & $0.0$ & $\prescript{2}{}{\mathrm{F}}(1)_{5/2}$ & $1/2$ & $68075$ & $68075.2$ & $-0.2$ \\[-0.09ex]
  $\prescript{4}{}{\mathrm{G}}_{5/2}$ & $5/2$ & $17564$ & $17563.3$ & $0.7$ & $\prescript{2}{}{\mathrm{L}}_{15/2}$ & $5/2$ & $30534$ & $30534.3$ & $-0.3$ &  &  &  &  &  \\[-0.09ex]
  \cline{11-15}
  $\prescript{4}{}{\mathrm{G}}_{5/2}$ & $1/2$ & $17611$ & $17610.9$ & $0.1$ & $\prescript{4}{}{\mathrm{D}}_{7/2}$ & $5/2$ & $30602$ & $30603.0$ & $-1.0$ & \multicolumn{1}{l}{RMS:} &  &  &  & 0.4 \\[-0.09ex]
  \hline
\end{tabular}
\end{table*}

The column $k^\mathrm{ref}$ in Table~\ref{tab:example_2} records the published calculated values, while column $k^\mathrm{calc}$ displays the eigenvalues of the total Hamiltonian determined by \texttt{YALIP}.
The maximum absolute difference ($\Delta_\mathrm{abs}$) between the two datasets is $\qty{1.3}{cm^{-1}}$, which is compatible with the precision limit ($\qty{1}{cm^{-1}}$) of the original printed data.
The uniform, non-systematic distribution of these residuals indicates the absence of systematic errors.
The resulting root-mean-square deviation is merely $\qty{0.4}{cm^{-1}}$, confirming that the two calculations are functionally identical and that \texttt{YALIP} perfectly reproduces the historical dataset.

It should be noted that the absolute energy reference scale for any calculated eigenvalue spectrum is arbitrary.
For a rigorous comparison of two energy level sets or during an empirical fitting procedure, the eigenvalues must be globally shifted so that their mean difference is zero.
Under this condition, the root-mean-square of the residuals serves as a reliable metric for evaluating the equivalence of the datasets.

Table~\ref{tab:example_2} also details the quantum numbers corresponding to the dominant $LS$-component of each energy level in intermediate coupling.
Crucially, the inclusion of crystal field parameters induces significant $J$-mixing.
Consequently, none of the assigned values represent "good" quantum numbers.
Even $J$ varies across the sub-dominant $LS$-components of a given state.
Furthermore, because of Kramers degeneracy in odd-electron systems, each energy level is fundamentally a doublet.
The electronic composition of both states in the doublet is identical, differing only in the sign of the projection quantum number $M$ for each $LS$-component.
Because the sign of $M$ reported in the original literature carries no independent physical information, Table~\ref{tab:example_2} lists only its absolute magnitude.

Our implementation revealed two distinct numerical errors in Ref.~\onlinecite{Carnall.1978}, which are highlighted in Table~\ref{tab:example_2}.
First, one of the two reported $\prescript{4}{}{\mathrm{F}}_{9/2}$ states labeled with $M=3/2$ actually possesses $M=9/2$.
The quantum number assignments for the remaining 181 energy levels perfectly match the original literature, including the supplemental indices given in parentheses. These indexes differentiate terms sharing identical $S$ and $L$ quantum numbers, and \ameli\ strictly reproduces the conventional ordering established by Nielson and Koster~\cite{Nielson.1963}.

The second numerical error highlights a fascinating artifact of early computational physics.
The original publication lists the state $\prescript{2}{}{\mathrm{I}}_{11/2}$ ($M=11/2$) at $k^\mathrm{ref}=\qty{28634}{cm^{-1}}$, whereas our evaluation demonstrates that the mathematically correct value was $\qty{29634}{cm^{-1}}$.
While standard eigendecomposition subroutines intrinsically output sorted eigenvalues, a manual typographical error or a raw computational mistake of this magnitude would disrupt the strict monotonic ascending order of the final table.

However, due to the severe mainframe memory constraints of the late 1970s, the diagonalization of the $\mathrm{Nd^{3+}}$ Hamiltonian had to be executed sequentially using symmetry sub-blocks. 
The independent results from these blocks were subsequently merged and sorted by a master routine prior to printing. 
The fact that the original publication preserves the correct numerical sorting despite the $\qty{1000}{cm^{-1}}$ error strongly implies that the data corruption, likely an in-memory bit flip or a transmission error, occurred after block diagonalization but before the final master sort execution.
This subtle anomaly serves as a striking historical reminder of the computational architecture hurdles underlying atomic structure calculations half a century ago, executing tasks that require less than a second on modern hardware.

\section{Concluding Remarks}

The primary objective of this work is to provide exact matrix elements for energy level and transition strength calculations.
By providing a comprehensive, accessible dataset, the \ameli\ repository~\cite{Caspary.2026b} enables researchers to move beyond the common practice of relying on legacy lists of matrix elements published for specific host materials when determining radiative transition intensities in other media.
Furthermore, the \texttt{YALIP} package~\cite{Caspary.2026c} as a lean reference implementation demonstrates a direct interface to the \ameli\ repository and provides the core functionalities of a typical lanthanide application software.

To take full advantage of the \ameli\ matrices, one needs to extract as many energy levels as possible from an absorption spectrum and determine the set of radial integrals that minimizes the deviation between calculated and observed levels.
This procedure requires constructing the perturbation Hamiltonian as a linear combination according to Eq.~\eqref{eq:Hab}.
While modern programming environments provide robust routines for numerically diagonalizing such matrices, yielding eigenvalues for energy levels and eigenvectors for intermediate coupling weights, the inverse problem of fitting radial integrals to an experimental spectrum remains a significant challenge.

This fitting process requires a nonlinear, multidimensional optimization algorithm.
The \texttt{YALIP} package~\cite{Caspary.2026c} provides a foundational optimization framework based on the Levenberg–Marquardt algorithm.
However, because nonlinear fits are inherently sensitive to local minima, the resulting parameters should always be subjected to thorough plausibility checks.

A well-established path to achieving reliable results is a multi-stage energy level fitting procedure.
In the first stage, known radial parameters from the literature are adopted as an initial guess to optimize exclusively the radial integrals of the two most significant first-order interactions: the Coulomb interaction ($\operator{H}_1$) and the spin-orbit coupling ($\operator{H}_2$).
These refined parameters then serve as the initial guess for the subsequent stage, which incorporates the second-order Coulomb interactions ($\operator{H}_3$ and $\operator{H}_4$).
In the final stage, the user can evaluate whether including the magnetic interactions ($\operator{H}_5$ and $\operator{H}_6$) further improves the fit quality without overfitting.

Once a set of radial integrals is determined, either through a dedicated fit or by adopting established literature values, the subsequent steps toward calculating radiative intensities for absorption or emission are straightforward.
The eigenvectors of the perturbation Hamiltonian provide the weight factors necessary to express intermediate-coupling states as linear combinations of pure $LS$-states.
Furthermore, they are utilized to compute the electric and magnetic dipole operators in intermediate coupling as outlined in Section~\ref{sec.examples}.

For amorphous materials, the Judd-Ofelt theory provides a sophisticated yet convenient framework by utilizing effective spherical site symmetry to reduce the required radial integrals to three parameters: $\Omega_2$, $\Omega_4$, and $\Omega_6$.
These are used in conjunction with the reduced even-rank matrix elements of the unit tensor operators $\tensor{U}{k}$ provided by \ameli.

Unlike the complex nonlinear energy-level fit, determining the radial integrals of the electric dipole operator requires only a basic linear optimization step~\cite{Caspary.2002}.
This involves the computation of the pseudo-inverse of a low-dimensional matrix based on measured transition strengths in intermediate coupling and the constant angular matrix elements in $LS$-coupling.

The dipole parameters are typically used in basic expressions~\cite{Reisfeld.1975,Walsh.2006} to determine the absorption strength and emission branching ratio for any pair of states and the radiative lifetime of any state in intermediate coupling.
Examples are given in Section~\ref{sec.examples}.

In summary, the \ameli\ repository offers significant advantages even if one chooses to forgo the complexities of a full energy-level fit.
By providing independence from limited, host-specific lists of published matrix elements, the \ameli\ repository allows researchers to robustly determine the radiative properties of any electronic transition of a lanthanide ion based purely on a chosen set of radial integrals.

\section*{Conflict of Interest}
The author has no conflicts to disclose.

\begin{acknowledgments}
This work is funded by the Deutsche Forschungsgemeinschaft (DFG, German Research Foundation) with project ID 390833453 (Cluster of Excellence PhoenixD).
\end{acknowledgments}

\section*{Data Availability Statement}
The data from this work are openly available on Zenodo~\cite{Caspary.2026b} at \url{https://doi.org/10.5281/zenodo.20401467}.
The Zenodo community \ameli\ at \url{https://zenodo.org/communities/ameli}  should be used as access point to the latest version of all datasets.
The \ameli\ software~\cite{Caspary.2026} is open-source.
Its latest version is available on GitHub at \url{https://github.com/reincas/AMELI}.
The \texttt{YALIP} reference application~\cite{Caspary.2026c} used in Section~\ref{sec.examples} is also open-source and available on GitHub at \url{https://github.com/reincas/YALIP}.

\section*{References}
\bibliography{references.bib}

@article{Axe.1963,
 author = {Axe, John D.},
 year = {1963},
 title = {Radiative Transition Probabilities within 4 fn Configurations: The Fluorescence Spectrum of Europium Ethylsulfate},
 pages = {1154--1160},
 volume = {39},
 number = {5},
 issn = {0021-9606},
 journal = {The Journal of Chemical Physics},
 doi = {10.1063/1.1734405}
}

@book{Brik.2019,
 author = {Brik, Mikhail G. and {Ma Chong-Geng}},
 year = {2019},
 title = {Theoretical spectroscopy of transition metal and rare earth ions: From free state to crystal field},
 address = {Singapore},
 publisher = {{Jenny Stanford Publishing}},
 isbn = {978-0-429-27875-4},
}

@article{Carnall.1968,
 author = {Carnall, William T. and Fields, P. R. and Rajnak, K.},
 year = {1968},
 title = {Electronic Energy Levels in the Trivalent Lanthanide Aquo Ions. {I}. $\mathrm{Pr^{3+}}$, $\mathrm{Nd^{3+}}$, $\mathrm{Pm^{3+}}$, $\mathrm{Sm^{3+}}$, $\mathrm{Dy^{3+}}$, $\mathrm{Ho^{3+}}$, $\mathrm{Er^{3+}}$, and $\mathrm{Tm^{3+}}$},
 pages = {4424--4442},
 volume = {49},
 number = {10},
 issn = {0021-9606},
 journal = {The Journal of Chemical Physics},
 doi = {10.1063/1.1669893},
}

@article{Carnall.1968b,
 author = {Carnall, William T. and Fields, P. R. and Rajnak, K.},
 year = {1968},
 title = {Electronic Energy Levels of the Trivalent Lanthanide Aquo Ions. {II}. $\mathrm{Gd^{3+}}$},
 pages = {4443--4446},
 volume = {49},
 number = {10},
 issn = {0021-9606},
 journal = {The Journal of Chemical Physics},
 doi = {10.1063/1.1669894},
}

@article{Carnall.1968c,
 author = {Carnall, William T. and Fields, P. R. and Rajnak, K.},
 year = {1968},
 title = {Electronic Energy Levels of the Trivalent Lanthanide Aquo Ions. {III}. $\mathrm{Tb^{3+}}$},
 pages = {4447--4449},
 volume = {49},
 number = {10},
 issn = {0021-9606},
 journal = {The Journal of Chemical Physics},
 doi = {10.1063/1.1669895},
}

@article{Carnall.1968d,
 author = {Carnall, William. T. and Fields, P. R. and Rajnak, K.},
 year = {1968},
 title = {Electronic Energy Levels of the Trivalent Lanthanide Aquo Ions. {IV}. $\mathrm{Eu^{3+}}$},
 pages = {4450--4455},
 volume = {49},
 number = {10},
 issn = {0021-9606},
 journal = {The Journal of Chemical Physics},
 doi = {10.1063/1.1669896},
}

@article{Carnall.1969,
 author = {Carnall, William T. and Fields, P. R. and Sarup, R.},
 year = {1969},
 title = {$\vphantom{|}^1{S}$ Level of $\mathrm{Pr^{3+}}$ in Crystal Matrices and Energy-Level Parameters for the $4f^2$ Configuration of $\mathrm{Pr^{3+}}$ in $\mathrm{LaF_3}$},
 pages = {2587--2591},
 volume = {51},
 number = {6},
 issn = {0021-9606},
 journal = {The Journal of Chemical Physics},
 doi = {10.1063/1.1672382},
}

@article{Carnall.1970,
 author = {Carnall, William T. and Fields, P. R. and Morrison, J. and Sarup, R.},
 year = {1970},
 title = {Absorption Spectrum of $\mathrm{Tm^{3+}}$:$\mathrm{LaF_3}$},
 pages = {4054--4059},
 volume = {52},
 number = {8},
 issn = {0021-9606},
 journal = {The Journal of Chemical Physics},
 doi = {10.1063/1.1673608},
}

@book{Carnall.1978,
 author = {Carnall, W. T. and Crosswhite, H. and Crosswhite, H. M. and Crosswhite, Hannah},
 year = {1978},
 title = {Energy Level Structure and Transition Probabilities in the Spectra of the Trivalent Lanthanides in $\mathrm{LaF_3}$},
 address = {Illinois},
 volume = {ANL-78-XX-95},
 publisher = {{Argonne National Laboratory}},
 series = {Report},
 doi = {10.2172/6417825},
}

@article{Carnall.1983,
 author = {Carnall, W. T. and Crosswhite, H.},
 year = {1983},
 title = {Further interpretation of the spectra of $\mathrm{Pr^{3+}}$-$\mathrm{LaF_3}$ and $\mathrm{Tm^{3+}}$-$\mathrm{LaF_3}$},
 pages = {127--135},
 volume = {93},
 number = {1},
 issn = {00225088},
 journal = {Journal of the Less Common Metals},
 doi = {10.1016/0022-5088(83)90457-5},
}

@book{Caspary.2002,
 author = {Caspary, Reinhard},
 year = {2002},
 title = {Applied rare-earth spectroscopy for fiber laser optimization: Doctoral Dissertation},
 address = {Aachen},
 publisher = {Shaker},
 institution = {{Technische Universit{\"a}t Braunschweig}}
}

@dataset{Caspary.2005,
 author = {Caspary, Reinhard},
 year = {2005},
 title = {Lanthanide-0.3},
 version = {0.3},
 url = {https://github.com/reincas/Lanthanide-0.3},
 howpublished = "{GitHub}",
 note = "{V}. 0.3, {GitHub}",
}

@dataset{Caspary.2026,
 author = {Caspary, Reinhard},
 year = {2026},
 title = {{AMELI}: {Angular Matrix Elements of Lanthanide Ions}},
 version = {1.3.4},
 url = {https://github.com/reincas/AMELI},
 howpublished = "{GitHub}",
 note = "{V}. 1.3.4, {GitHub}"
}

@dataset{Caspary.2026b,
 author = {Caspary, Reinhard},
 year = {2026},
 title = {{AMELI}: {Angular Matrix Elements of Lanthanide Ions}},
 version = {1.3.4},
 doi = "https://doi.org/10.5281/zenodo.20401467",
 howpublished = "{Zenodo}",
 note = "{V}. 1.3.4, {Zenodo}"
}

@dataset{Caspary.2026c,
 author = {Caspary, Reinhard},
 year = {2026},
 title = {{YALIP}: {Yet Another Lanthanide Ion Package}},
 version = {0.9.4},
 url = {https://github.com/reincas/YALIP},
 howpublished = "{GitHub}",
 note = "{V}. 0.9.4, {GitHub}"
}

@article{Ciric.2019,
 author = {{\'C}iri{\'c}, Aleksandar and Stojadinovi{\'c}, Stevan and Sekuli{\'c}, Milica and Drami{\'c}anin, Miroslav D.},
 year = {2019},
 title = {{JOES}: An application software for {J}udd-{O}felt analysis from $\mathrm{Eu^{3+}}$ emission spectra},
 pages = {351--356},
 volume = {205},
 issn = {00222313},
 journal = {Journal of Luminescence},
 doi = {10.1016/j.jlumin.2018.09.048},
}

@book{CohenTannoudji.1977,
 year = {1977},
 title = {Quantum Mechanics, volume 1. 2nd edition,},
 address = {New York},
 urldate = {29.08.2025},
 publisher = {{Wiley {\&} Sons}},
 isbn = {978-3-527-82271-3},
 editor = {Cohen-Tannoudji, Claude and Diu, Bernard and Lalo{\"e}, Franck},
}

@book{CohenTannoudji.1977b,
 year = {1977},
 title = {Quantum Mechanics, volume 2. 2nd edition,},
 address = {New York},
 urldate = {29.08.2025},
 publisher = {{Wiley {\&} Sons}},
 isbn = {978-3527345540},
 editor = {Cohen-Tannoudji, Claude and Diu, Bernard and Lalo{\"e}, Franck},
}

@article{Condon.1935,
 author = {Condon, E. U. and {Shortley G. H.}},
 year = {1935},
 title = {The Theory of Atomic Spectra},
 pages = {57--59},
 volume = {83},
 number = {2142},
 issn = {0036-8075},
 journal = {Science},
 doi = {10.1126/science.83.2142.57.b},
}

@article{Dutra.2013,
 author = {Dutra, Jos{\'e} Diogo L. and Ferreira, Jos{\'e} Wesley and Rodrigues, Marcelo O. and Freire, Ricardo O.},
 year = {2013},
 title = {Theoretical methodologies for calculation of {J}udd-{O}felt intensity parameters of polyeuropium systems},
 pages = {14095--14099},
 volume = {117},
 number = {51},
 journal = {The journal of physical chemistry. A},
 doi = {10.1021/jp4098809},
}

@article{Dutra.2014,
 author = {Dutra, Jos{\'e} Diogo L. and Bispo, Thiago D. and Freire, Ricardo O.},
 year = {2014},
 title = {{LUMPAC} lanthanide luminescence software: Efficient and user friendly},
 pages = {772--775},
 volume = {35},
 number = {10},
 journal = {Journal of computational chemistry},
 doi = {10.1002/jcc.23542},
}

@article{Eckart.1930,
 author = {Eckart, Carl},
 year = {1930},
 title = {The Application of Group theory to the Quantum Dynamics of Monatomic Systems},
 pages = {305--380},
 volume = {2},
 number = {3},
 issn = {0034-6861},
 journal = {Reviews of Modern Physics},
 doi = {10.1103/RevModPhys.2.305},
}

@article{Edvardsson.2001,
 author = {Edvardsson, Sverker and {\AA}berg, Daniel},
 year = {2001},
 title = {An atomic program for energy levels of equivalent electrons: lanthanides and actinides},
 pages = {396--406},
 volume = {133},
 number = {2-3},
 issn = {00104655},
 journal = {Computer Physics Communications},
 doi = {10.1016/S0010-4655(00)00171-5},
}

@article{Fiorucci.2025,
 author = {Fiorucci, Letizia and Ravera, Enrico},
 year = {2025},
 title = {Not Just Another Crystal Field Software},
 pages = {e70063},
 volume = {46},
 number = {6},
 journal = {Journal of computational chemistry},
 doi = {10.1002/jcc.70063}
}

@article{Fowler.1962,
 author = {Fowler, W. Beall and Dexter, D. L.},
 year = {1962},
 title = {Relation between Absorption and Emission Probabilities in Luminescent Centers in Ionic Solids},
 pages = {2154--2165},
 volume = {128},
 number = {5},
 issn = {0031-899X},
 journal = {Physical Review},
 doi = {10.1103/PhysRev.128.2154}
}

@article{Freidzon.2018,
 author = {Freidzon, Alexandra Ya and Kurbatov, Ilia A. and Vovna, Vitaliy I.},
 year = {2018},
 title = {Ab initio calculation of energy levels of trivalent lanthanide ions},
 url = {https://pubs.rsc.org/en/content/articlehtml/2018/cp/c7cp08366a},
 pages = {14564--14577},
 volume = {20},
 number = {21},
 journal = {Physical chemistry chemical physics : PCCP},
 doi = {10.1039/C7CP08366A},
}

@incollection{Goldschmidt.1978,
 author = {Goldschmidt, Zippora B.},
 title = {Chapter 1: Atomic properties (free atom)},
 pages = {1--171},
 volume = {1},
 publisher = {North-Holland},
 isbn = {978-0-444-85020-1},
 series = {Handbook on the physics and chemistry of rare earths},
 editor = {Gschneidner, Karl A. and Eyring, LeRoy},
 booktitle = {Volume 1: Metals},
 year = {1978},
 address = {Amsterdam},
}

@article{Hansen.1996,
 author = {Hansen, J. E. and Judd, B. R. and Crosswhite, Hannah},
 year = {1996},
 title = {Matrix Elements of Scalar Three-Electron Operators for the Atomic f Shell},
 pages = {1--49},
 volume = {62},
 number = {1},
 issn = {0092-640X},
 journal = {Atomic Data and Nuclear Data Tables},
 doi = {10.1006/adnd.1996.0001}
}

@article{Hehlen.2013,
 author = {Hehlen, Markus P. and Brik, Mikhail G. and Kr{\"a}mer, Karl W.},
 year = {2013},
 title = {50th anniversary of the {J}udd--{O}felt theory: An experimentalist's view of the formalism and its application},
 pages = {221--239},
 volume = {136},
 issn = {00222313},
 journal = {Journal of Luminescence},
 doi = {10.1016/j.jlumin.2012.10.035},
}

@article{Horie.1953,
 author = {Horie, Hisashi},
 year = {1953},
 title = {Spin-spin and Spin-other-orbit Interactions},
 pages = {296--308},
 volume = {10},
 number = {3},
 issn = {0033-068X},
 journal = {Progress of Theoretical Physics},
 doi = {10.1143/PTP.10.296},
}

@article{Hrabovsky.2025,
 author = {Hrabovsky, Jan and Varak, Petr and Krystufek, Robin},
 year = {2025},
 title = {{LOMS.cz} computational platform for high-throughput classical and combinatorial {J}udd-{O}felt analysis and rare-earth spectroscopy},
 url = {https://www.nature.com/articles/s41598-025-13620-0},
 pages = {28945},
 volume = {15},
 number = {1},
 issn = {2045-2322},
 journal = {Scientific Reports},
 doi = {10.1038/s41598-025-13620-0},
}

@article{Judd.1962,
 author = {Judd, Brian R.},
 year = {1962},
 title = {Optical Absorption Intensities of Rare-Earth Ions},
 pages = {750--761},
 volume = {127},
 number = {3},
 issn = {0031-899X},
 journal = {Physical Review},
 doi = {10.1103/PhysRev.127.750},
}

@book{Judd.1963,
 author = {Judd, Brian R.},
 year = {1963},
 title = {Operator Techniques in Atomic Spectroscopy},
 address = {New York},
 urldate = {29.08.2025},
 publisher = {{McGraw-Hill Company}},
 institution = {{Advanced physics monograph series}},
}

@article{Judd.1966,
 author = {Judd, Brian R.},
 year = {1966},
 title = {Three-Particle Operators for Equivalent Electrons},
 pages = {4--14},
 volume = {141},
 number = {1},
 issn = {0031-899X},
 journal = {Physical Review},
 doi = {10.1103/PhysRev.141.4},
}

@article{Judd.1968,
 author = {Judd, Brian R. and Crosswhite, Henry Milton and Crosswhite, Hannah},
 year = {1968},
 title = {Intra-Atomic Magnetic Interactions for $f$ Electrons},
 pages = {130--138},
 volume = {169},
 issn = {0031-899X},
 journal = {Physical Review},
 doi = {10.1103/PhysRev.169.130},
}

@article{Judd.2003,
 author = {Judd, Brian R.},
 year = {2003},
 title = {Rare-earth intensity trials},
 pages = {885--890},
 volume = {101},
 number = {7},
 issn = {0026-8976},
 journal = {Molecular Physics},
 doi = {10.1080/0026897021000046690},
}

@article{Kotzian.1995,
 author = {Kotzian, Manfred and Fox, Thomas and Roesch, Notker},
 year = {1995},
 title = {Calculation of Electronic Spectra of Hydrated {Ln(III)} Ions within the {INDO/S-CI} Approach},
 pages = {600--605},
 volume = {99},
 number = {2},
 issn = {0092-7325},
 journal = {The Journal of Physical Chemistry},
 doi = {10.1021/j100002a023},
}

@article{Kozak.2005,
 author = {Kozak, M. M. and Goebel, D. and Caspary, R. and Kowalsky, W.},
 year = {2005},
 title = {Spectroscopic properties of thulium-doped zirconium fluoride and indium fluoride glasses},
 pages = {2009--2021},
 volume = {351},
 number = {24-26},
 issn = {0022-3093},
 journal = {Journal of Non-Crystalline Solids},
 doi = {10.1016/j.jnoncrysol.2005.05.008},
}

@article{Krupke.1966,
 author = {Krupke, William F.},
 year = {1966},
 title = {Optical Absorption and Fluorescence Intensities in Several Rare-Earth-Doped $\mathrm{Y_2O_3}$ and $\mathrm{LaF_3}$ Single Crystals},
 pages = {325--337},
 volume = {145},
 number = {1},
 issn = {0031-899X},
 journal = {Physical Review},
 doi = {10.1103/PhysRev.145.325},
}

@article{LizarazoFerro.2026,
 author = {Lizarazo-Ferro, Juan-David and Puel, Tharnier O. and Flatté, Michael E. and Zia, Rashid},
 year = {2026},
 title = {Refining spectroscopic calculations for trivalent lanthanide ions: A revised parametric Hamiltonian and open-source solution},
 volume = {113},
 number = {7},
 issn = {0556-2805},
 journal = {Physical Review B},
 doi = {10.1103/llzb-wrlg}
}

@article{Malta.2002,
 author = {Malta, O. L. and Batista, H. J. and Carlos, L. D.},
 year = {2002},
 title = {Overlap polarizability of a chemical bond: a scale of covalency and application to lanthanide compounds},
 pages = {21--30},
 volume = {282},
 number = {1},
 issn = {0301-0104},
 journal = {Chemical Physics},
 doi = {10.1016/s0301-0104(02)00631-6}
}

@book{Martin.1978,
 author = {Martin, W. C. and Zalubas, Romuald and Hagan, Lucy},
 year = {1978},
 title = {Atomic Energy Levels: The Rare-Earth Elements},
 url = {https://cir.nii.ac.jp/crid/1363388845421931008},
 address = {Gaithersburg, MD},
 volume = {60},
 publisher = {{National Bureau of Standards}},
 doi = {10.6028/nbs.nsrds.60},
}

@article{Marvin.1947,
 author = {Marvin, H. H.},
 year = {1947},
 title = {Mutual Magnetic Interactions of Electrons},
 pages = {102--110},
 volume = {71},
 number = {2},
 issn = {0031-899X},
 journal = {Physical Review},
 doi = {10.1103/PhysRev.71.102},
}

@article{Morrison.1979,
 author = {Morrison, C. A. and Leavitt, R. P.},
 year = {1979},
 title = {Crystal-field analysis of triply ionized rare earth ions in lanthanum trifluoride},
 pages = {2366--2374},
 volume = {71},
 number = {6},
 issn = {0021-9606},
 journal = {The Journal of Chemical Physics},
 doi = {10.1063/1.438641},
}

@article{Moura.2021,
 author = {{Moura Jr.}, Renaldo T. and {Carneiro Neto}, Albano N. and Aguiar, Eduardo C. and Santos-Jr., Carlos V. and de Lima, Ewerton M. and Faustino, Wagner M. and Teotonio, Ercules E.S. and Brito, Hermi F. and Felinto, Maria C.F.C. and Ferreira, Rute A.S. and Carlos, Luís D. and Longo, Ricardo L. and Malta, Oscar L.},
 year = {2021},
 title = {{JOYSpectra}: A web platform for luminescence of lanthanides},
 pages = {100080},
 volume = {11},
 issn = {2590-1478},
 journal = {Optical Materials: X},
 doi = {10.1016/j.omx.2021.100080},
}

@article{Naguleswaran.2003,
 author = {Naguleswaran, S. and Reid, M. F. and Stedman, G. E.},
 year = {2003},
 title = {Perturbation expansions and gauge choices in {J}udd-{O}felt theory},
 pages = {917--922},
 volume = {101},
 number = {7},
 issn = {0026-8976},
 journal = {Molecular Physics},
 doi = {10.1080/0026897021000046753},
}

@book{Nielson.1963,
 author = {Nielson, C. W. and Koster, G. F.},
 year = {1963},
 title = {Spectroscopic coefficients for the $p^n$, $d^n$, and $f^n$ configurations},
 address = {Cambridge, MA},
 publisher = {{The M. I. T. Press}},
}

@article{Ofelt.1962,
 author = {Ofelt, G. S.},
 year = {1962},
 title = {Intensities of Crystal Spectra of Rare-Earth Ions},
 pages = {511--520},
 volume = {37},
 number = {3},
 issn = {0021-9606},
 journal = {The Journal of Chemical Physics},
 doi = {10.1063/1.1701366},
}

@article{Ofelt.2003,
 author = {Ofelt, George S.},
 year = {2003},
 title = {Reflections on the development of the {J}udd-{O}felt theory},
 pages = {891--892},
 volume = {101},
 number = {7},
 issn = {0026-8976},
 journal = {Molecular Physics},
 doi = {10.1080/0026897021000046708},
}

@article{Peijzel.2005,
 author = {Peijzel, P. S. and Meijerink, A. and Wegh, R. T. and Reid, M. F. and Burdick, G. W.},
 year = {2005},
 title = {A complete {$4f^N$} energy level diagram for all trivalent lanthanide ions},
 pages = {448--453},
 volume = {178},
 number = {2},
 issn = {00224596},
 journal = {Journal of Solid State Chemistry},
 doi = {10.1016/j.jssc.2004.07.046},
}

@article{Racah.1942,
 author = {Racah, Giulio},
 year = {1942},
 title = {Theory of Complex Spectra. {I}},
 pages = {186--197},
 volume = {61},
 number = {3-4},
 issn = {0031-899X},
 journal = {Physical Review},
 doi = {10.1103/PhysRev.61.186},
}

@article{Racah.1942b,
 author = {Racah, Giulio},
 year = {1942},
 title = {Theory of Complex Spectra. {II}},
 pages = {438--462},
 volume = {62},
 number = {9-10},
 issn = {0031-899X},
 journal = {Physical Review},
 doi = {10.1103/physrev.62.438},
}

@article{Racah.1943,
 author = {Racah, Giulio},
 year = {1943},
 title = {Theory of Complex Spectra. {III}},
 pages = {367--382},
 volume = {63},
 number = {9-10},
 issn = {0031-899X},
 journal = {Physical Review},
 doi = {10.1103/physrev.63.367},
}

@article{Racah.1949,
 author = {Racah, Giulio},
 year = {1949},
 title = {Theory of Complex Spectra. {IV}},
 pages = {1352--1365},
 volume = {76},
 number = {9},
 issn = {0031-899X},
 journal = {Physical Review},
 doi = {10.1103/physrev.76.1352},
}

@article{Racah.1967,
 author = {Racah, Giulio and Stein, J.},
 year = {1967},
 title = {Effective Electrostatic Interactions in $l^{N}$ Configurations},
 pages = {58--64},
 volume = {156},
 number = {1},
 issn = {0031-899X},
 journal = {Physical Review},
 doi = {10.1103/PhysRev.156.58},
}

@article{Rajnak.1963,
 author = {Rajnak, Katheryn and Wybourne, Brian G.},
 year = {1963},
 title = {Configuration Interaction Effects in $l^{N}$ Configurations},
 pages = {280--290},
 volume = {132},
 number = {1},
 issn = {0031-899X},
 journal = {Physical Review},
 doi = {10.1103/PhysRev.132.280},
}

@article{Rajnak.1964,
 author = {Rajnak, Katheryn and Wybourne, Brian G.},
 year = {1964},
 title = {Electrostatically Correlated Spin-Orbit Interactions in $l^{N}$-Type Configurations},
 pages = {A596-A600},
 volume = {134},
 number = {3A},
 issn = {0031-899X},
 journal = {Physical Review},
 doi = {10.1103/PhysRev.134.A596},
}

@article{Reid.1983,
 author = {Reid, Michael F. and Richardson, F. S.},
 year = {1983},
 title = {Electric dipole intensity parameters for lanthanide $4f\to4f$ transitions},
 url = {https://pubs.aip.org/aip/jcp/article/79/12/5735/87711},
 pages = {5735--5742},
 volume = {79},
 number = {12},
 issn = {0021-9606},
 journal = {The Journal of Chemical Physics},
 doi = {10.1063/1.445760},
}

@article{Reid.1984,
 author = {Reid, Michael F. and Richardson, F. S.},
 year = {1984},
 title = {Lanthanide $4f\to4f$ electric dipole intensity theory},
 pages = {3579--3586},
 volume = {88},
 number = {16},
 issn = {0092-7325},
 journal = {The Journal of Physical Chemistry},
 doi = {10.1021/j150660a041},
}

@incollection{Reid.2005,
 author = {Reid, M. F.},
 title = {Transition Intensities},
 pages = {95--129},
 volume = {83},
 publisher = {{Tsinghua Univ. Press} and Springer},
 isbn = {978-3-540-23886-7},
 series = {Springer series in materials science},
 editor = {Hull, Robert and Parisi, J{\"u}rgen and Osgood, R. M. and Warlimont, Hans and Liu, Guokui and Jacquier, Bernard},
 booktitle = {Spectroscopic properties of rare earths in optical materials},
 year = {2005},
 address = {Beijing and Berlin and Heidelberg and New York},
 doi = {10.1007/3-540-28209-2_2},
}

@incollection{Reisfeld.1975,
 author = {Reisfeld, Renata},
 title = {Radiative and non-radiative transitions of rare-earth ions in glasses},
 pages = {123--175},
 publisher = {Springer},
 isbn = {9783540072683},
 series = {Structure and bonding},
 editor = {Dunitz, Jack D.},
 booktitle = {Rare earths},
 year = {1975},
 address = {Berlin},
 doi = {10.1007/bfb0116557},
}

@article{Sa.2000,
 author = {de S{\'a}, G.F and Malta, O.L and de {Mello Doneg{\'a}}, C. and Simas, A.M and Longo, R.L and Santa-Cruz, P.A and {Da Silva}, E.F},
 year = {2000},
 title = {Spectroscopic properties and design of highly luminescent lanthanide coordination complexes},
 pages = {165--195},
 volume = {196},
 number = {1},
 issn = {0010-8545},
 journal = {Coordination Chemistry Reviews},
 doi = {10.1016/s0010-8545(99)00054-5}
}

@article{Slater.1929,
 author = {Slater, John C.},
 year = {1929},
 title = {The Theory of Complex Spectra},
 pages = {1293--1322},
 volume = {34},
 number = {10},
 issn = {0031-899X},
 journal = {Physical Review},
 doi = {10.1103/PhysRev.34.1293},
}

@book{Slater.1960,
 author = {Slater, John C.},
 year = {1960},
 title = {Quantum Theory of Atomic Structure},
 address = {New York},
 volume = {1},
 publisher = {McGraw-Hill},
 isbn = {0070580405},
 series = {International series in pure and applied physics}
}

@book{Slater.1960b,
 author = {Slater, John C.},
 year = {1960},
 title = {Quantum Theory of Atomic Structure},
 address = {New York},
 volume = {2},
 publisher = {McGraw-Hill},
 series = {International series in pure and applied physics}
}

@article{Smentek.1998,
 author = {Smentek, Lidia},
 year = {1998},
 title = {Theoretical description of the spectroscopic properties of rare earth ions in crystals},
 pages = {155--237},
 volume = {297},
 number = {4},
 issn = {0370-1573},
 journal = {Physics Reports},
 doi = {10.1016/s0370-1573(97)00077-x},
}

@article{Smentek.2003,
 author = {Smentek, Lidia},
 year = {2003},
 title = {{J}udd-{O}felt theory: past, present and future},
 pages = {893--897},
 volume = {101},
 number = {7},
 issn = {0026-8976},
 journal = {Molecular Physics},
 doi = {10.1080/0026897021000046717},
}

@article{Trees.1951,
 author = {Trees, R. E.},
 year = {1951},
 title = {Configuration Interaction in {Mn II}},
 pages = {756--760},
 volume = {83},
 number = {4},
 issn = {0031-899X},
 journal = {Physical Review},
 doi = {10.1103/PhysRev.83.756},
}

@incollection{Walsh.2006,
 author = {Walsh, Brian M.},
 title = {Judd-{O}felt theory: principles and practices},
 pages = {403--433},
 publisher = {Springer},
 isbn = {9781402047886},
 series = {NATO science series Series 2, Mathematics, physics and chemistry},
 editor = {{Di Bartolo}, Baldassare and Forte, Ottavio},
 booktitle = {Advances in spectroscopy for lasers and sensing},
 year = {2006},
 address = {Dordrecht},
 doi = {10.1007/1-4020-4789-4_21},
}

@book{Wigner.1959,
 author = {Wigner, Eugene},
 year = {1959},
 title = {Group Theory: And its Application to the Quantum Mechanics of Atomic Spectra},
 address = {Burlington},
 publisher = {{Elsevier Science}},
 isbn = {9780323152785}
}

@book{Wybourne.1965,
 author = {Wybourne, Brian G.},
 year = {1965},
 title = {Spectroscopic Properties of Rare Earths},
 address = {New York},
 publisher = {{Interscience Publishers}},
 isbn = {9780470965078},
}

\end{document}